\documentclass[sigconf, nonacm, review=false]{acmart}
\usepackage{graphicx}
\usepackage{balance}  
\usepackage{multirow}
\usepackage{bm}
\usepackage{makecell}
\usepackage{enumitem}
\usepackage{subcaption}
\usepackage{listings}
\usepackage{xcolor}
\usepackage[linesnumbered,vlined,ruled]{algorithm2e}

\newcommand\vldbdoi{xxxxxxxxxxxxxxxx}

\newcommand\vldbavailabilityurl{}

\newcommand{\sun}[1]{{\color{black}{#1}}\xspace}

\newcommand{\HBS}[1]{{\textcolor{black}{ #1}}}

\definecolor{codegreen}{rgb}{0,0.6,0}
\definecolor{codegray}{rgb}{0.5,0.5,0.5}
\definecolor{codepurple}{rgb}{0.58,0,0.82}
\definecolor{backcolour}{rgb}{0.95,0.95,0.92}

\lstdefinestyle{mystyle}{
    backgroundcolor=\color{backcolour},   
    commentstyle=\color{codegreen},
    keywordstyle=\color{magenta},
    numberstyle=\tiny\color{codegray},
    stringstyle=\color{codepurple},
    basicstyle=\ttfamily\footnotesize,
    breakatwhitespace=false,         
    breaklines=true,                 
    captionpos=b,                    
    keepspaces=true,                 
    numbers=left,                    
    numbersep=5pt,                  
    showspaces=false,                
    showstringspaces=false,
    showtabs=false,                  
    tabsize=2
}

\begin{document}
\title{ThunderRW: An In-Memory Graph Random Walk Engine}

\author{Shixuan Sun}
\affiliation{\institution{National University of Singapore}}
\email{sunsx@comp.nus.edu.sg}

\author{Yuhang Chen}
\affiliation{\institution{National University of Singapore}}
\email{yuhangc@comp.nus.edu.sg}

\author{Shengliang Lu}
\affiliation{\institution{National University of Singapore}}
\email{lusl@comp.nus.edu.sg}

\author{Bingsheng He}
\affiliation{\institution{National University of Singapore}}
\email{hebs@comp.nus.edu.sg}

\author{Yuchen Li}
\affiliation{\institution{Singapore Management University}}
\email{yuchenli@smu.edu.sg}

\begin{abstract}

As random walk is a powerful tool in many graph processing, mining and learning applications, this paper proposes an efficient in-memory random walk engine named \textbf{ThunderRW}. Compared with existing parallel systems on improving the performance of a single graph operation, ThunderRW supports massive parallel random walks. The core design of ThunderRW is motivated by our profiling results: common RW algorithms have as high as 73.1\% CPU pipeline slots stalled due to irregular memory access, which suffers significantly more memory stalls than the conventional graph workloads such as BFS and SSSP. To improve the memory efficiency, we first design a generic step-centric programming model named Gather-Move-Update to abstract different RW algorithms. Based on the programming model, we develop the step interleaving technique to hide memory access latency by switching the executions of different random walk queries. In our experiments, we use four representative RW algorithms including PPR, DeepWalk, Node2Vec and MetaPath to demonstrate the efficiency and programming flexibility of ThunderRW. Experimental results show that ThunderRW outperforms state-of-the-art approaches by an order of magnitude, and the step interleaving technique significantly reduces the CPU pipeline stall from 73.1\% to 15.0\%.

\end{abstract}

\maketitle



\section{Introduction} \label{sec:introduction}

\sun{Random walk (RW) is an effective tool for extracting relationships between entities in a graph, and is widely used in
many applications such as \emph{Personalized PageRank} (PPR) \cite{page1999pagerank}, \emph{SimRank} \cite{jeh2002simrank},
\emph{Random Walk Domination} \cite{li2014random}, \emph{Graphlet Concentration} (GC) \cite{prvzulj2007biological},
\emph{Network Community Profiling} (NCP) \cite{fortunato2016community}, \emph{DeepWalk} \cite{perozzi2014deepwalk}
and \emph{Node2Vec} \cite{grover2016node2vec}. For graph analysis tasks such as GC and NCP, RW queries generally dominate the
cost \cite{prvzulj2007biological,fortunato2016community}. Even for graph representation learning, the cost of sampling RW
is non-trivial, for example, a naive implementation of Node2Vec takes more than eight hours on the \emph{twitter} graph in our
experiments. Moreover, increasing the number of RW queries can improve the effectiveness of RW algorithms \cite{grover2016node2vec,prvzulj2007biological}.
Therefore, accelerating RW queries is an important problem.}

RW algorithms generally follow the execution paradigm illustrated
in Algorithm \ref{algo:common_paradigm}, which consists of massive RW queries.
Each query $Q$ starts from a given source vertex. At each step, $Q$ moves to a neighbour of the
current residing vertex at random, and repeats this process until satisfying a specific termination condition,
e.g., a target length is reached (Lines 2-5). Despite that RW algorithms follow a similar execution paradigm, there are
quite some variants of RW algorithms, which can differ significantly in neighbor selections (see Section~\ref{sec:random_walk}).
Encouraged by the success of in-memory graph processing engines~\cite{shun2013ligra,nguyen2013lightweight,zhang2018graphit,sundaram2015graphmat},
there have been some recent systems designed specifically for RW algorithms, including C-SAW \cite{pandey2020c}, GraphWalker \cite{wang2020graphwalker}
and KnightKing \cite{yang2019knightking}. They focus on accelerators, disk-based or distributed settings,
without specially optimizing in-memory execution of RW queries. \sun{However, with the rapid development of hardwares,
modern servers equip with hundred gigabytes, even several terabytes memory, which empowers in-memory processing of
graphs with hundred billions of edges. This covers many real-world graphs in applications \cite{dhulipalaprovably}. As such,
this paper studies the design and implementation of an in-memory graph engine for RW algorithms.}

\setlength{\textfloatsep}{0pt}
\begin{algorithm}[t]
	\caption{Execution Paradigm of RW algorithms}
	\label{algo:common_paradigm}
	\footnotesize
	\SetKwRepeat{Do}{do}{while}
	 \KwIn{a graph $G$ and a set $\mathbb{Q}$ of random walk queries\;}
	 \KwOut{the walk sequences of each query in $\mathbb{Q}$\;}
	 \ForEach{$Q \in \mathbb{Q}$}{
	    \Do{Terminate($Q$) is false}{
	        Select a neighbor of the current residing vertex $Q.cur$ at random\;
	        Add the selected vertex to $Q$\;
	    }
	 }
	 \KwRet $\mathbb{Q}$\;
\end{algorithm}


To crystallize the performance factors for in-memory RW executions, we conduct profiling studies on RW algorithms in comparison with conventional workloads of a single graph operation like BFS and SSSP (see Section \ref{sec:workload_profiling}). Our profiling results show that
common RW algorithms have as high as 73.1\% CPU pipeline slots stalled due to irregular memory access, which suffers significantly more memory stalls than the conventional workloads. Consequently, the CPUs frequently wait on the high-latency access to the main memory, which becomes the major performance bottleneck. Besides, we observe that the sampling methods such as \emph{inverse transformation sampling} \cite{marsaglia1963generating}, \emph{alias sampling} \cite{walker1977efficient} and \emph{rejection sampling} \cite{robert2013monte} have significant varying performance on different RW algorithms
\HBS{(with the difference as much as 6 times)}. Thus, it requires non-trivial and significant engineering efforts to develop any efficient RW algorithms considering the cache stall bottleneck, as well as parallelization and the choice of sampling methods. 

\sun{In this paper, we propose \textbf{ThunderRW}, a generic and efficient in-memory RW framework.
We employ a \emph{step-centric} programming model abstracting the computation from the local view of moving one step of a walker.
Users implement their RW algorithms by "thinking like a walker" in user-defined functions (UDF). The framework applies UDFs to each query
and parallelizes the execution by regarding a step of a query as a task unit. Furthermore, ThunderRW provides variant sampling methods
so that users can select an appropriate one based on the characteristics of workloads.
Built upon the step-centric programming model, we propose the \emph{step interleaving} technique to resolve the cache stalls
caused by irregular memory access with \emph{software prefetching} \cite{lee2012prefetching}.
As modern CPUs can process multiple memory access requests simultaneously \cite{williams2009roofline},
the core idea of step interleaving is to hide memory access latency by issuing multiple outstanding memory
accesses, which exploits \emph{memory level parallelism} \cite{beamer2015locality} among different RW queries.}

We demonstrate the generality and programming flexibility of ThunderRW by showcasing four representative
RW algorithms including \HBS{PPR~\cite{page1999pagerank}, DeepWalk \cite{perozzi2014deepwalk}, Node2Vec \cite{grover2016node2vec} and MetaPath~\cite{sun2013mining}}. \sun{We conduct extensive experiments with
twelve real-world graphs. Experiment results show that (1) ThunderRW runs 8.6-3333.1X faster
than the naive implementation in popular open-source packages; (2) ThunderRW provides speedups of \sun{1.7-14.6X} over the state-of-the-art
frameworks including GraphWalker \cite{wang2020graphwalker} and KnightKing \cite{yang2019knightking} running on the same machine;
and (3) the step interleaving technique significantly reduces the memory stalls from \sun{73.1\%} to \sun{15.0\%}.}

\section{Background and Related Work} \label{sec:background}

\subsection{Preliminary} \label{sec:preliminaries}

This paper focuses on the directed graph $G = (V, E)$ where $V$ is a set of vertices and $E$ is a set of edges.
\HBS{An undirected graph can be supported by representing each undirected edge with two directed edges with the same two vertexes in our system.}
Given a vertex $v \in V$, $N_v$ denotes the neighbors of $v$, i.e., $\{v'| e(v, v') \in E\}$ where $e(v, v')$
represents the edge between $v$ and $v'$. The degree $d_v$ denotes the number of neighbors of $v$. $E_v$ is the
set of edges adjacent to $v$, i.e., $\{e(v, v')| v' \in N_v\}$. Given
$e \in E$ (resp. $v \in V$), $w_e$ and $l_e$ (resp. $w_v$ and $l_v$) represent its weight and label, respectively.
Given $G$, a RW $Q$ is a stochastic process on $G$, which consists of a sequence of adjacent vertices.
$Q[i]$ is the $i$th vertex in the sequence where $i$ starts from 0. $Q.cur$ is the current residing vertex of $Q$.
$|Q|$ is the number of vertices in $Q$.
Suppose that $Q.cur$ is $v$. Given $e \in E_v$, we call the probability of $e$ being selected
the \emph{transition probability}, which is represented by $p(e)$. Then, the neighbor selection is equivalent
to sampling from the discrete probability distribution $P = \{p(e)\}$ where $e \in E_v$. Specifically,
it is to pick an element $h$ from $E_v$ based on the distribution of $P$, i.e., $P[h = e] = p(e)$.
For example, if the relative chance of $e$ being selected is proportional to the edge weight $w_e$, then
$p(e) = \hat{w}_e$ is the normalized probability where $\hat{w}_e = \frac{w_e}{\sum_{e' \in E_v}w_{e'}}$.

\subsection{Random Walk based Algorithms} \label{sec:random_walk}

RW algorithms generally follow the execution paradigm in Algorithm \ref{algo:common_paradigm}.
They mainly differ in the neighbor selection step. We first categorize them into \emph{unbiased}
and \emph{biased} based on the transition probability properties. Unbiased RW selects each edge $e \in E_v$ with the same probability where $v = Q.cur$,
while \HBS{the transition probability is nonuniform for biased RWs}, e.g., depending on the edge weight. We further classify the biased
RWs into \emph{static} and \emph{dynamic}. If the transition probability is determined before execution, then RW is static. Otherwise, it is dynamic, which is affected by states of RW queries. 
In the following, we introduce four representative RW algorithms \HBS{that have been used in many applications}.

\textbf{PPR (Personalized PageRank)} \cite{page1999pagerank} assigns a score to each vertex $v'$ in the graph from the personalized view of a given
source $v$, which describes how much $v$ is interested in (or similar to) $v'$. A common solution for this problem is to start a number of
RW queries from $v$, which have a fixed termination probability at each step, and approximately calculates the scores based on the
distribution of the end vertices of random walk queries \cite{liu2016powerwalk,fogaras2005towards}. The algorithms generally set RW queries as unbiased \cite{lofgren2015efficient}.

\textbf{DeepWalk} \cite{perozzi2014deepwalk} is a graph embedding technique widely used in machine learning.
It is developed based on the SkipGram model \cite{mikolov2013efficient}. For each vertex,
it starts a specified number of RW queries with a target length to generate embeddings. The original DeepWalk is unbiased,
while the recent work \cite{cochez2017biased} extends it to consider the edge weight, which becomes biased (static) random walk.

\textbf{Node2Vec} \cite{grover2016node2vec} is a popular graph embedding technique based on the second-order random walk. 
Different from DeepWalk, its transition probability depends on the last vertex visited. Suppose that $Q.cur$ is $v$.
Equation \ref{eq:n2v} describes the transition probability of selecting the edge $e(v, v')$ where $u$ is
the last vertex visited, $dist(v', u)$ is the distance between $v'$ and $u$, and $a$ and $b$ are two hyperparameters
controlling the random walk behaviour. Node2Vec is dynamic because the transition probability relies on the states of queries.
Moreover, It can take the edge weight into the consideration by multiplying $p(e)$ with $w_e$.

\begingroup
\setlength\abovedisplayskip{1pt}
\setlength\belowdisplayskip{1pt}
\begin{equation} \label{eq:n2v}
  p(e(v, v')) =
    \begin{cases}
      \frac{1}{a} & \text{if $dist(v', u) = 0$},\\
      1 & \text{if $dist(v', u) = 1$}, \\
      \frac{1}{b} & \text{if $dist(v', u) = 2$}.
    \end{cases}       
\end{equation}
\endgroup

\textbf{MetaPath}~\cite{sun2013mining} is a powerful tool to extract semantics information from heterogeneous information networks,
and is widely used in machine learning tasks such as natural language processing \cite{lao2011random,lv2019adapting}.
The RW queries are associated with a \emph{meta-path schema} $H$, which defines the pattern of the walk paths
based on the edge type, e.g., "write->publish->mention". Let $H[i]$ be the $i$th label in $H$. At each step, the RW query only considers the
edges $e \in E_v$ where $v = Q.cur$ such that $l_e$ is equal to $H[|Q|]$. In other words, if $l_e \neq H[|Q|]$,
then $p(e) = 0$. Thus, the transition probability depends on the states of the RW, and MetaPath is dynamic.

\subsection{Sampling Methods} \label{sec:sampling_methods}

Sampling from a discrete probability distribution $P = \{p_0, p_1,...,p_{n - 1}\}$ is to select an element $h$
from $\{0,1,...,n-1\}$ based on $P$ (i.e., $P[h = i] = p_i$). In this paper, we focus on five sampling techniques,
including \emph{naive sampling}, \emph{inverse transformation sampling} \cite{marsaglia1963generating},
\emph{alias sampling} \cite{walker1977efficient}, \emph{rejection sampling} \cite{robert2013monte} and a special case of \emph{rejection sampling} \cite{yang2019knightking} because they are efficient and widely used \cite{schwarz2011darts,wang2020graphwalker,pandey2020c,yang2019knightking,shao2020memory}.
Naive sampling only works on the uniform discrete distribution, while the other four can handle non-uniform and select the element $h$ in two phases:
\emph{initialization}, which preprocesses the distribution $P$, and \emph{generation}, which picks an element on the basis of the initialization result.
Please refer to \cite{schwarz2011darts} for the details. In the following, we briefly introduce the sampling methods in the context of this paper, i.e.,
selecting an edge from $E_v$ based on the transition probability distribution $P$ where $v = Q.cur$.

\textbf{Naive sampling (\texttt{NAIVE}).} This method generates a uniform random integer number $x$ in the range $[0, d_v)$
and picks $E_v[x]$, which is the $x$th element in $E_v$.
It only works on the uniform discrete distribution. The time and space complexities are both $O(1)$.

\textbf{Inverse transformation sampling (\texttt{ITS}).} The initialization phase of \texttt{ITS}
computes the \emph{cumulative distribution function} of $P$ as follows: $P' = \{p'_i = \sum_{j = 0} ^ {i}p_{j}\}$ where $0 \leqslant i < d_v$.
After that, the generation phase first generates a uniform real number $x$ in $[0, p'_{d_v - 1})$, then uses a binary
search to find the smallest index $i$ such that $x < p_{i}'$, and finally selects $E_v[i]$. The time complexity
of the initialization is $O(d_v)$, and that of the generation is $O(\log d_v)$. As \texttt{ITS} needs to store $P'$,
the space complexity is $O(d_v)$.

\textbf{Alias sampling (\texttt{ALIAS}).} The initialization phase
builds two tables: the \emph{probability table} $H$, and the \emph{alias table} $A$. Both of them have
$d_v$ values. $H[i]$ and $A[i]$ represent the $i$th value of $H$ and $A$, respectively.
Given $0 \leqslant i < d_v$, $A[i]$ is a bucket containing one or two elements from $\{0, 1,...,d_v - 1\}$,
which are denoted by $A[i].first$ and $A[i].second$, respectively. $H[i]$ is the probability selecting $A[i].first$.
If $A[i]$ has only one element, then $A[i].second$ is $null$ and $H[i]$ is equal to 1. The generation
phase first generates a uniform integer number $x$ in $[0, d_v)$ and then retrieves $H[x]$ and $A[x]$. Next,
it generates a uniform real number $y$ in $[0, 1)$. If $y < H[x]$, then picks $e(v, A[x].first)$.
Otherwise, the edge selected is $e(v,A[x].second)$. The time complexity of initialization is $O(d_v)$
and that of generation is $O(1)$. The space complexity is $O(d_v)$.

\textbf{Rejection sampling (\texttt{REJ}).} The initialization phase of \texttt{REJ} gets $p ^ * = \max_{p \in P} p$.
The generation phase can be viewed as throwing darts on a rectangle dartboard until hitting the target area.
Specifically, it has two steps: (1) generate a uniform integer number $x$ in $[0, d_v)$ and a uniform real number
$y$ in $[0, p^*)$ (i.e., the dart is thrown at the position $(x, y)$); and (2) if $y < p_x$,
then select $E_v[x]$ (i.e., hit the target area); otherwise, repeat Step (1). The time complexity of initialization
is $O(d_v)$, and that of generation is $O(\mathbb{E})$ where $\mathbb{E} = \frac{d_v \times p^*}{\sum_{p \in P}p}$
(i.e., the area of the rectangle board divides the target area). Based on the computation method of $\mathbb{E}$,
we can get that $1 \leqslant \mathbb{E} \leqslant d_v$. The space complexity is $O(1)$.

\textbf{A special case of \texttt{REJ} (\texttt{O-REJ}).} A special case of \texttt{REJ} is that we
can set a value $p ^ * \geqslant \max_{p \in P} p$ without the initialization phase, but of keeping
$\mathbb{E} = \frac{d_v \times p^*}{\sum_{p \in P}p}$ is close to $\frac{d_v \times \max_{p \in P}}{\sum_{p \in P}p}$.
For example, set $p ^ *$ to $\max \{1, \frac{1}{a}, \frac{1}{b}\}$ for Node2Vec \cite{yang2019knightking}.
The generation phase is the same as \texttt{REJ}. Therefore, the time complexity is $O(\mathbb{E})$ where $\mathbb{E} = \frac{d_v \times p^*}{\sum_{p \in P}p}$
and $p ^ *$ is specified by users. The space complexity is $O(1)$.

\sun{In existing works, unbiased random walks (e.g., PPR \cite{page1999pagerank} and unweighted DeepWalk \cite{perozzi2014deepwalk}) adopt
\texttt{NAIVE} sampling. In contrast, biased random walks (e.g., weighted DeepWalk \cite{ye2019improved,dai2018adversarial},
Node2Vec \cite{grover2016node2vec} and MetaPath \cite{fu2017hin2vec,hu2018leveraging})
use \texttt{ALIAS} sampling because the time complexity of the generation phase is $O(1)$.
C-SAW \cite{pandey2020c} adopts \texttt{ITS} to utilize the parallel computation capability of GPUs to calculate the prefix sum.
KnightKing \cite{yang2019knightking} uses \texttt{O-REJ} to avoid scanning neighbors of $Q.cur$ to reduce the network communication cost.}

\subsection{Related Work} \label{sec:related_work}

\textbf{Graph computing frameworks.} There are a number of generic graph computing frameworks working on different computation environments,
for example, (1) Single Machine (CPUs): GraphChi \cite{kyrola2012graphchi}, Ligra \cite{shun2013ligra}, Graphene \cite{liu2017graphene}, and
GraphSoft \cite{jun2018grafboost}; (2) GPUs: Medusa \cite{zhong2013medusa}, CuSha \cite{khorasani2014cusha} and Gunrock \cite{wang2016gunrock};
and (3) Distributed Environment: Pregel \cite{malewicz2010pregel}, GraphLab \cite{low2012distributed},
PowerGraph \cite{gonzalez2012powergraph}, GraphX \cite{gonzalez2014graphx}, Blogel \cite{yan2014blogel}, Gemini \cite{zhu2016gemini},
and Grapes \cite{fan2018parallelizing}. They usually adopt vertex- or edge-centric model, and are highly optimized for a single graph operation.
In contrast, ForkGraph \cite{lu2021cache} targets at graph algorithms consisting of concurrent graph queries, for example, betweenness centrality.
However, all of them focus on traditional graph query operations such as BFS and SSSP without considering RW workloads.
\HBS{That motivates the development of engines specially optimized for RW~\cite{yang2019knightking, pandey2020c, wang2020graphwalker}.}

\sun{\textbf{Random walk frameworks.} In contrast to graph computing frameworks abstracting the computation from
the view of the graph data, existing RW frameworks adopt the \emph{walker-centric} model, which regards
each query as the parallel task. KnightKing \cite{yang2019knightking} is a distributed
framework. It adopts the BSP model that moves a step for all queries
at each iteration until all queries complete. To reduce data transfers in network, it utilizes \texttt{O-REJ} sampling to
avoid scanning $E_v$ where $v = Q.cur$. It exposes an API for users to set a suitable upper
bound for the edge transition probability for each edge adjacent to $Q.cur$. Unfortunately, we find that this design introduces
an implicit constraint on RW algorithms: a suitable upper bound must be determined without
looping over $E_v$. This works well for Node2Vec by setting the upper bound as $\max{\{1.0/a, 1.0, 1.0/b\}}$
according to Equation \ref{eq:n2v}. However, it cannot
handle MetaPath because the transition probability of each $e \in E_v$ can be zero because of
the label filter. Another limitation is that KnightKing can
suffer the tail problem since it moves a step for all queries at an iteration, whereas queries can have variant lengths.}

\sun{C-SAW \cite{pandey2020c} is a framework on GPUs. It adopts the BSP model as well. To utilize the parallel computation capability in
the many-core architecture, C-SAW uses \texttt{ITS} sampling in computation. Particularly, for all random walk types including
unbiased, static and dynamic, C-SAW first conducts a prefix sum on the
transition probability of edges adjacent to $Q.cur$, and then selects an edge. Consequently, it incurs high overhead for unbiased and
static random walks. Moreover, C-SAW cannot support random walks with variant lengths (e.g., PPR) since such RW queries can degrade the
utilization of GPUs. Additionally, Node2Vec is not supported by C-SAW, because C-SAW does not support
the distance verification on GPUs.}

\sun{GraphWalker \cite{wang2020graphwalker} is an I/O efficient framework on a single machine.
For a graph that cannot reside in memory, GraphWalker divides it into a set of partitions, and focuses on optimizing the scheduling
of loading partitions into memory to reduce the number of I/Os. Specifically, for each partition, GraphWalker records
the number of queries residing in it, and the scheduler prioritizes partitions with
more queries. Given a partition $G'$ in memory, GraphWalker adopts the ASP model to execute queries in it. It assigns a query $Q$ to
each worker (i.e., a thread), and executes it independently until $Q$ completes or jumps out $G'$.
Once all queries in $G'$ complete or leave $G'$, GraphWalker swaps it out, and reads the partition with most queries in disk.
It repeats this process till all queries complete. GraphWalker supports unbiased RW only.}

\sun{This paper focuses on accelerating the in-memory execution of RW queries. ThunderRW abstracts the computation of RW algorithms
from the perspective of queries as well to exploit the parallelism in RW algorithms, but takes the \emph{step-centric} model,
which regards one step of a query as the task unit and factors one step into the gather-move-update operations to empower the
step interleaving technique. Moreover, ThunderRW supports all the five sampling methods in Section \ref{sec:sampling_methods} so
that users can adopt an appropriate sampling method given a specific workload. ThunderRW
supports all the four RW-algorithms in Section \ref{sec:random_walk}, which demonstrates its programming flexibility over other RW frameworks.}

\textbf{RW algorithm optimization.} Due to the importance of the RW-based
applications, a variety of algorithm-specific optimizations have been proposed for
different RW applications, e.g., PPR \cite{wang2017fora,shi2019realtime,lofgren2014fast,wei2018topppr,guo2017parallel},
Node2Vec \cite{zhou2018efficient} and second-order random walks \cite{shao2020memory}. In contrast,
we aim to design a generic and efficient random walk framework on which users can easily implement different
kinds of random walk applications. Thus, the algorithm-specific optimizations are beyond the scope of this paper.

\textbf{Prefetching in databases.} Our step-interleaving techniques are inspired by the prefetching techniques in query processing of databases. As the performance gap between main memory and CPU widens, prefetching has been an effective means to improve database performance. There have been studies applying prefetching to B-tree index~\cite{10.1145/376284.375688} and hash joins~\cite{chen2007improving, 6544839, 10.14778/1687553.1687564, 10.14778/2735703.2735704}. Hash joins are probably the most widely studied operator for prefetching. The group prefetching (GP) and software
pipeline prefetching (SPP) \cite{chen2007improving} are the classic prefetching technique for hash joins, which rearrange a sequence of operations in a loop to several
stages and execute all queries stage by stage in batch. However, GP and SPP cannot efficiently handle queries with
irregular access patterns, for example a binary search performs three searches to find the target value,
while the other one needs four times. To resolve the problem, AMAC \cite{kocberber2015asynchronous} proposes
to execute the stages of each query asynchronously by explicitly maintaining the states of each stage. However,
AMAC incurs more overhead than GP and SPP, especially when there are a number of stages
because it needs to maintain the states of each stage. As in the context of random walk, there is a lack of a model to abstract stages from a
sequence of operations and model their dependency relationships to guide the implementation.

\section{Motivations} \label{sec:workload_profiling}

In this section, we study the profiling results to assess the performance bottlenecks of in-memory computation
of RW algorithms. Specifically, we execute RW queries with different sampling methods and
examine the hardware utilization with the \emph{top-down microarchitecture
analysis method} (TMAM). In the following, we first introduce TMAM and then present the profiling results.

\textbf{Top-down analysis method (TMAM) \cite{coorporation2016intel}.} TMAM is a simplified and intuitive model for
identifying the performance bottlenecks in out-of-order CPUs. It uses
the \emph{pipeline slot} to represent the hardware resources required to process the micro-operations (uOps). In
a cycle, a pipeline slot is either empty (\emph{stalled}) or filled with a uOp. The execution stall is caused by the \emph{front-end}
or the \emph{back-end} part of the pipeline. \HBS{Specifically, the back-end cannot accept new operations due to the lack of required resources. It can be
    further split into \emph{memory bound}, which represents the stall caused by the memory subsystem, and
    \emph{core bound}, which reflects the stall incurred by the unavailable execution units.} When the slot is filled with a uOp, it will be classified as \emph{retiring} if
the uOp eventually retires (Otherwise, the slot is categorized as \emph{bad speculation}). 
We use Intel Vtune Profiler to measure the percentage of pipeline slots in each category (retiring, bad speculation, front-end bound, memory bound and core bound) in our experiments.

\subsection{Observations}

\begin{table}[t]
\footnotesize
    \setlength{\abovecaptionskip}{0pt}
    \setlength{\belowcaptionskip}{0pt}
\caption{Comparison of pipeline slot breakdown and memory bandwidth (the total value of read and write) between traditional graph algorithms and RW algorithms.}
\label{tab:comparison_breakdown}
\resizebox{0.48\textwidth}{!}{
\begin{tabular}{c|c|c|c|c|c|c}
\hline
\textbf{Method} & \textbf{\begin{tabular}[c]{@{}c@{}}Front\\ End\end{tabular}} & \textbf{\begin{tabular}[c]{@{}c@{}}Bad\\ Spec\end{tabular}} & \textbf{Core} & \textbf{Memory} & \textbf{Retiring} & \textbf{\begin{tabular}[c]{@{}c@{}}Memory\\ Bandwidth\end{tabular}} \\ \hline\hline
BFS             & 11.6\%                                                       & 9.1\%                                                       & 20.8\%        & 40.6\%          & 18.0\%            & 51.7 GB/s                                                            \\ \hline
SSSP            & 9.1\%                                                        & 12.5\%                                                      & 24.9\%        & 36.9\%          & 16.6\%            & 38.2 GB/s                                                            \\ \hline
\sun{PPR}             & \sun{0.6\%}     & \sun{0.7\%}     & \sun{15.8\%}              & \sun{\textbf{73.1\%}}   & \sun{9.7\%}                        & \sun{\textbf{1.4 GB/s}} \\ \hline       
DeepWalk        & 1.0\%                                                        & 3.9\%                                                       & 16.7\%        & \textbf{69.7\%}          & 8.7\%             & \textbf{5.6 GB/s}                                                             \\ \hline
\sun{Node2Vec}        & \sun{11.5\%}     & \sun{22.1\%}      &   \sun{24.3\%}       &    \sun{28.1\%}     & \sun{14.1\%}        & \sun{17.1 GB/s} \\ \hline
\sun{MetaPath}        & \sun{6.2\%}      & \sun{7.5\%}       &   \sun{29.7\%}       &    \sun{33.9\%}     & \sun{22.7\%}        & \sun{9.9 GB/s} \\ \hline
\end{tabular}
}
\end{table}

\sun{\textbf{Varying random walk workloads.} We first evaluate the four RW algorithms in Section \ref{sec:random_walk}.
Specifically, we set PPR as unbiased, and configure the termination probability as 0.2.
For DeepWalk and Node2Vec, we set the target length as 80. The transition probability of DeepWalk
is the edge weight, and that of Node2Vec is calculated based on Equation \ref{eq:n2v} where $a=2$ and $b=0.5$.
The schema length of MetaPath is 5, and we generate it by randomly choosing five labels from the edge label set.
PPR starts $|V|$ queries from a given vertex, and the others start a query from each vertex in $V$. 
Following existing studies \cite{page1999pagerank,cochez2017biased,grover2016node2vec,sun2013mining} (as well as popular open-source packages
\footnote{\url{https://github.com/aditya-grover/node2vec}, Last accessed on 2021/03/20} \footnote{\url{https://github.com/GraphSAINT/GraphSAINT}, Last accessed on 2021/03/20.}), we use \texttt{NAIVE} sampling for PPR,
while \texttt{ALIAS} sampling for the others. Moreover, we build alias tables for DeepWalk in a preprocessing phase to accelerate the execution
of queries. However, this method is prohibitively expensive for high order RW due the the exponential memory consumption 
\cite{yang2019knightking,shao2020memory}. For example, the space complexity of such an index for Node2Vec, which is second-order, is
$O(\sum_{v\in V}d_{v} ^ 2)$, and it can consume more than 1000 TB space for \emph{twitter}. As such, we compute the transition probability
and perform the initialization of \texttt{ALIAS} in run time. To compare the performance characteristics with RW algorithms,
we evaluate BFS and SSSP, which are two conventional graph algorithms. We develop RW algorithms without any frameworks, whereas
implementing BFS and SSSP with Ligra \cite{shun2013ligra}.}

Table \ref{tab:comparison_breakdown} presents the experiment results on \emph{livejournal}, the details of which are listed
in Table \ref{tab:datasets}. RW queries randomly visit nodes on the graph that leads to a massive number of random memory accesses.
Consequently, as high as 73.1\% pipeline slots of PPR and DeepWalk
are stalled due to memory access. In contrast, the memory bound of BFS and SSSP
is less than 45\%, which demonstrates much better cache locality than that of PPR and DeepWalk.
Due to the large proportional of memory stalls, the retiring of PPR and DeepWalk is less than 10\%.
Furthermore, we measure the memory bandwidth utilization of these algorithms. Our benchmark shows
that the max memory bandwidth of our test bed is 60 GB/s. As shown in the table, \HBS{
the bandwidth utilization of BFS and SSSP are rather high (86.2\% and 63.6\%, respectively), while that of PPR and DeepWalk
is very low (2.3\% and 9.3\%, respectively).}

\sun{Compared with PPR and DeepWalk, Node2Vec and MetaPath exhibit different characteristics. The memory bound is lower than
PPR and DeepWalk, whereas the retirement and bandwidth are much higher. To achieve more insights, we first examine the execution time breakdown
on computing the transition probability (denoted by \emph{compute $p(e)$}), and the initialization
and generation phases of sampling an edge (denoted by \emph{Init} and \emph{Gen}, respectively), and then analyze the complexity of these operations
at a step.}

\begin{table}[t]
\footnotesize
    \setlength{\abovecaptionskip}{0pt}
    \setlength{\belowcaptionskip}{0pt}
\caption{\sun{Comparison of execution time breakdown and the time complexity per step among RW algorithms where $v = Q.cur$
and $u$ is the last vertex of $Q$.}}
\label{tab:time_breakdown}
\resizebox{0.48\textwidth}{!}{
\begin{tabular}{c|c|c|c|c|c|c}
\hline
\multirow{3}{*}{\textbf{Method}} & \multicolumn{3}{c|}{\textbf{Time Breakdown}}                                                                                & \multicolumn{3}{c}{\textbf{Complexity per Step}}                                                                           \\ \cline{2-7} 
                                 & \multirow{2}{*}{\textbf{\begin{tabular}[c]{@{}c@{}}Compute\\ $p(e)$\end{tabular}}} & \multicolumn{2}{c|}{\textbf{Sampling}} & \multirow{2}{*}{\textbf{\begin{tabular}[c]{@{}c@{}}Compute\\ $p(e)$\end{tabular}}} & \multicolumn{2}{c}{\textbf{Sampling}} \\ \cline{3-4} \cline{6-7} 
                                 &                                                                                    & \textbf{Init}      & \textbf{Gen}      &                                                                                    & \textbf{Init}      & \textbf{Gen}      \\ \hline\hline
PPR                              & \textit{N/A}                                                                       & \textit{N/A}       & 100\%             & \textit{N/A}                                                                       & \textit{N/A}       & $O(1)$            \\ \hline
DeepWalk                         & \textit{N/A}                                                                       & \textit{N/A}       & 100\%             & \textit{N/A}                                                                       & \textit{N/A}       & $O(1)$            \\ \hline
Node2Vec                         & 89.9\%                                                                             & 9.9\%              & 0.2\%             & $O(d_v \times \log d_{u})$                                                         & $O(d_v)$           & $O(1)$            \\ \hline
MetaPath                         & 29.0\%                                                                             & 69.9\%             & 1.1\%             & $O(d_v)$                                                                           & $O(d_v)$           & $O(1)$            \\ \hline
\end{tabular}
}
\end{table}

\sun{Table \ref{tab:time_breakdown} lists the results. PPR and DeepWalk are static, and they only need to sample an edge and move
$Q$ along it in run time. In contrast, Node2Vec and MetaPath are dynamic, and they first compute the transition probability for each $e \in E_v$
where $v = Q.cur$, and then sample an edge. Consequently, the cost on \emph{Gen} is neglected as shown in Table \ref{tab:time_breakdown}.
Moreover, the memory bound is much lower than static RWs in Table \ref{tab:comparison_breakdown} since the computation scans $E_v$ in
a continuous manner. Given $e \in E_v$ and $u$ is the last vertex of $Q$, the complexity of computing $p(e)$ in Node2Vec is $O(\log d_u)$ because the 
distance check in Equation \ref{eq:n2v} is implemented by a binary search. However, MetaPath computes $p(e)$ with a simple label filter. 
As a result, computing $p(e)$ accounts for around 90\% of the execution time in Node2Vec, whereas \emph{Init} dominates the cost in MetaPath.}

\sun{\textbf{Observation 1.} \emph{The in-memory computation of common RW algorithms suffers severe performance
issues due to memory stalls caused by cache misses and under-utilizes the memory bandwidth. For high order RW algorithms,
computing $p(e)$ and initializing the auxiliary data structure for sampling dominate the in-memory computation cost, and their complexities
are determined by the RW algorithm and the selected sampling method, respectively.}}

\textbf{Varying sampling methods and RW types.} We further examine the performance of sampling methods.
We \HBS{continue to develop a micro benchmark that executes} $10 ^ 7$ RW queries each of which starts from a vertex randomly selected from the graph. The target length is 80.
We evaluate three types of RW queries as discussed in Section~\ref{sec:random_walk}: \textbf{unbiased}, \textbf{static} and \textbf{dynamic}. 
For unbiased RW, we first perform the initialization
phase of sampling methods for the neighbor set of each vertex in a \emph{preprocessing} step. We then use the generation phase of a
sampling method to select a neighbor of $Q.cur$ in execution. For static RW, we evaluate queries with the same process as that of unbiased. 
The only difference is that the edge weight is used to set the transition probability for static RW whereas the transition probability in unbiased RW is the default uniform. 
For dynamic RW, we set the edge weight as the transition probability, while performing the initialization phase for the neighbor set of $Q.cur$
in execution because the transition probability of dynamic RW varies during the computation.

\begin{figure}[t]\small
    \setlength{\abovecaptionskip}{0pt}
    \setlength{\belowcaptionskip}{0pt}
    \captionsetup[subfigure]{aboveskip=0pt,belowskip=0pt}
    \centering
    \begin{subfigure}[t]{0.155\textwidth}
        \centering
        \includegraphics[scale=0.23]{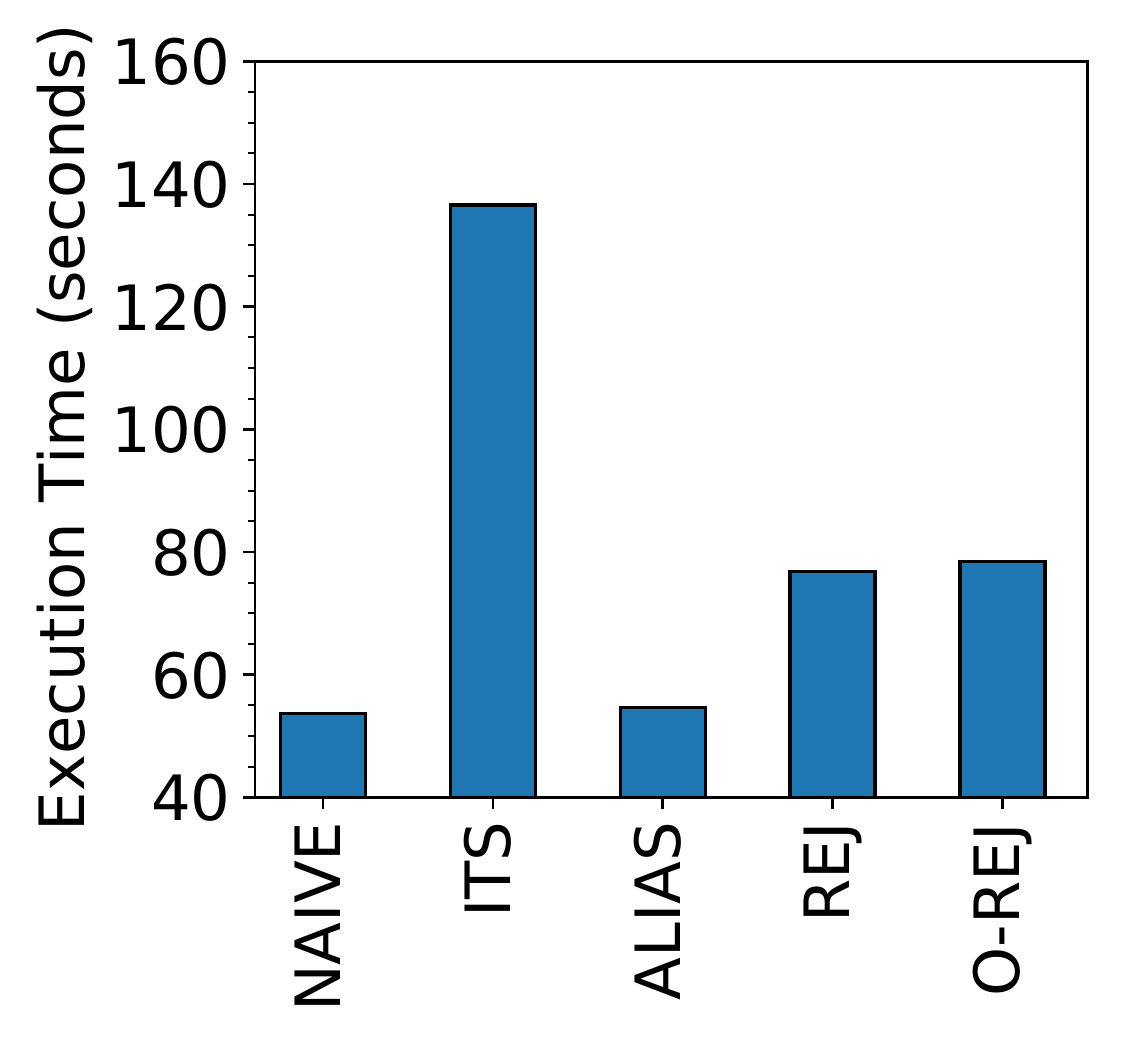}
        \caption{Unbiased.}
        \label{fig:unbiased}
    \end{subfigure}
    \begin{subfigure}[t]{0.155\textwidth}
        \centering
        \includegraphics[scale=0.23]{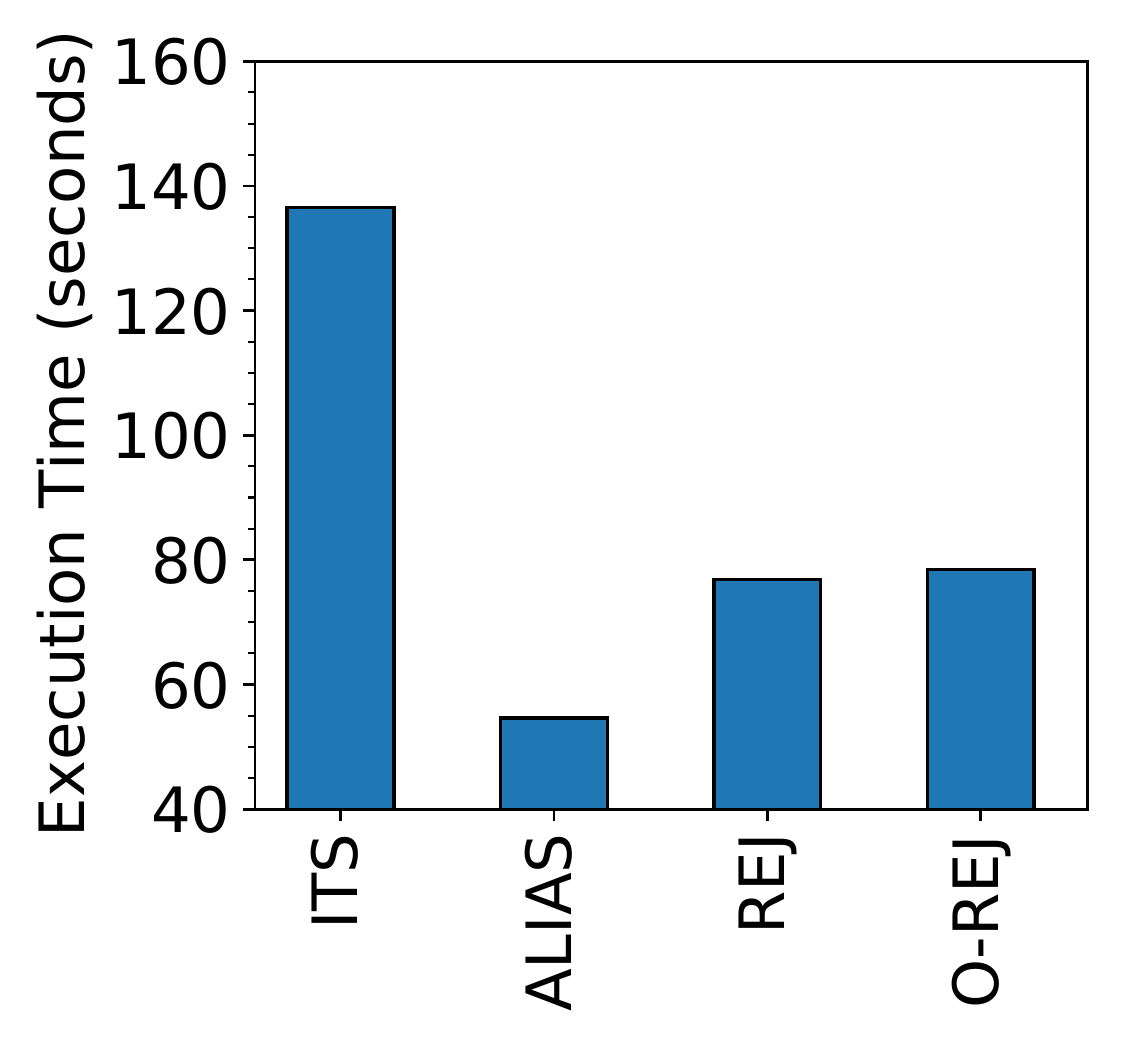}
        \caption{Static.}
        \label{fig:static}
    \end{subfigure}
    \begin{subfigure}[t]{0.155\textwidth}
        \centering
        \includegraphics[scale=0.23]{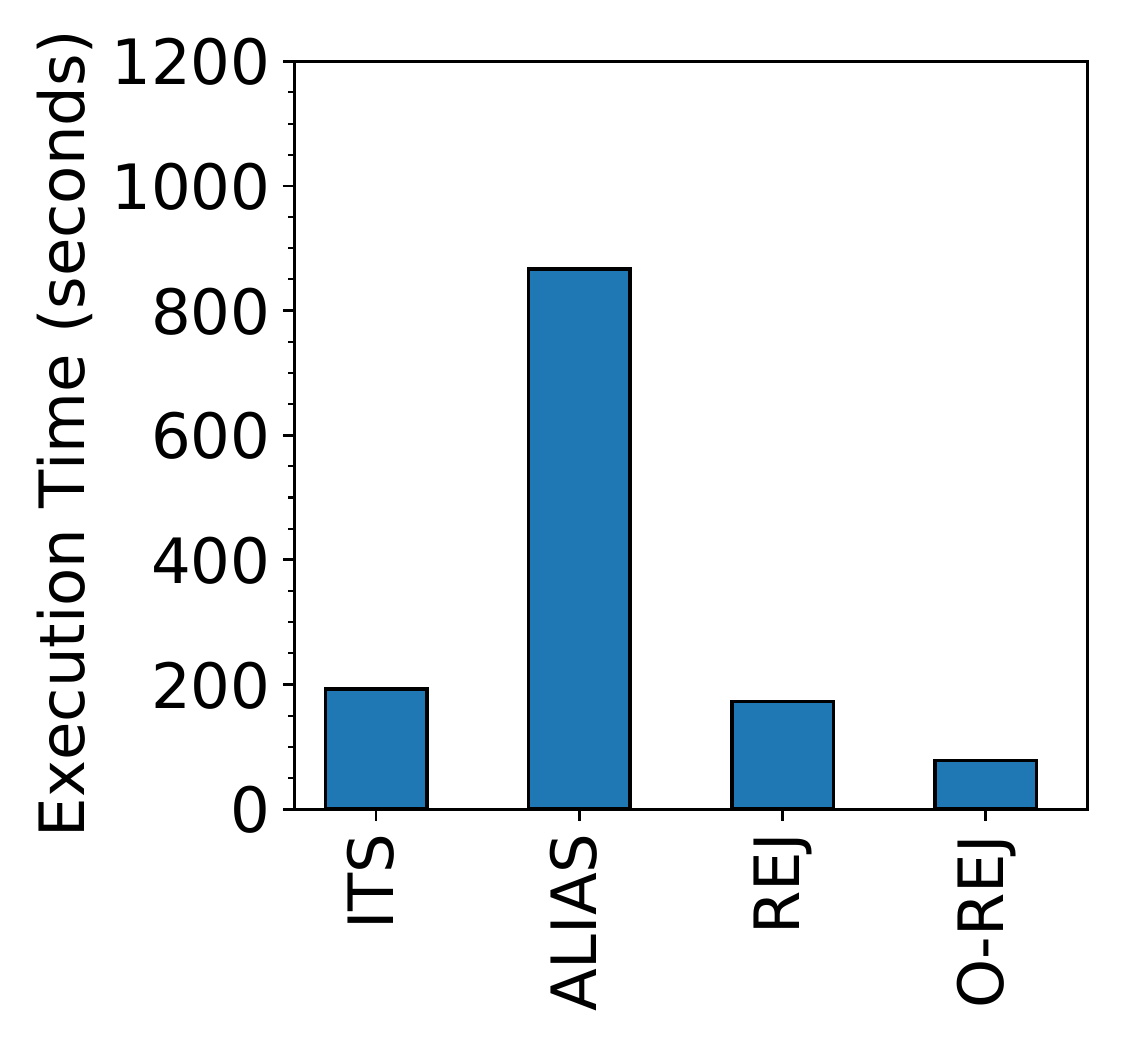}
        \caption{Dynamic.}
        \label{fig:dynamic}
    \end{subfigure}
    \caption{Effectiveness of sampling methods.}
    \label{fig:sampling_method_effectiveness}
\end{figure}

Figure \ref{fig:sampling_method_effectiveness} presents the experiment results of the sequential execution with
variant sampling methods on different RW types. \HBS{We have the following findings. First, the \texttt{NAIVE} sampling method performs the best on unbiased RW as
it has no initialization phase. Second, among static methods, the \texttt{ALIAS} sampling method outperforms others because its generation phase has lower time complexity. However, \texttt{ALIAS} runs much slower than other
methods on dynamic RW since its initialization cost is high in practice. Third, \texttt{O-REJ} performs well
on dynamic RW since it does not have the initialization phase. Fourth, we can observe that
the cost of evaluating dynamic RW is significantly expensive than that of unbiased and static RW
because of the initialization phase (if exists) at each step.}

\textbf{Observation 2.} \emph{Sampling methods have an important impact on the performance and
no sampling method can dominate on all cases. Generally, dynamic RW is expensive than unbiased and static RW.}

\subsection{System Implications}

\sun{Based on the profiling results, we can categorize the cost of evaluating RW queries into two classes, that of computing
$p(e)$ and that of sampling an edge. As the former is determined by the RW algorithms (i.e., algorithm-specific), our framework
targets at accelerating the latter operation.} Moreover, we have the following implications for the design and implementation of ThunderRW.
First, we need to develop mechanisms to reduce the cache stalls. Our profiling
results show that in-memory computation of common RW algorithms suffer severe performance issues due to
the irregular memory accesses. None of previous random walk frameworks 
\cite{pandey2020c,wang2020graphwalker,yang2019knightking}
address the problem. On the other hand, there are massive queries in random walk workloads, but
the memory bandwidth is under-utilized.
Inspired by previous work on accelerating multiple index lookups in database systems with prefetching
\cite{chen2007improving,kocberber2015asynchronous}, there are opportunities for prefetching and
interleaving executions among different queries.

Second, there is a need to support multiple sampling methods. However, existing frameworks support
one sampling method only and generally regard all RW as dynamic (e.g., C-SAW), while (1) the sampling method has an
important impact on the performance and none of them can dominate on all cases; and (2) the cost of evaluating dynamic
RW is generally much more expensive than that of unbiased and static RW.

\section{ThunderRW Abstraction} \label{sec:computation_model}

In this section, we present the abstraction of the computation in ThunderRW.

\subsection{Step-centric Model} \label{sec:step_centric_model}


To abstract the computation of RW algorithms, we propose the \emph{step-centric} model in this paper.
We observe that RW algorithms are built upon a number of RW queries rather than a single query.
In spite of limited intra-query parallelism, there is abundant inter-query parallelism in RW-algorithms
as each RW query can be executed independently. Therefore, our step-centric
model abstracts the computation of RW algorithms from the perspective of queries to exploit the inter-query parallelism.

Specifically, we model the computation from the local view of moving one step of a query $Q$.
Then, we abstract a step of $Q$ into the \texttt{Gather}-\texttt{Move}-\texttt{Update} (GMU) operations
to characterize the common structure of RW algorithms. With the step-centric model,
users develop RW algorithms by "thinking like a walker". They focus on defining functions setting
the transition probability of $e \in E_v$ and updating states of $Q$ at each
step, while the framework facilitates applying \HBS{user-defined} step-oriented functions to RW
queries. 

\subsection{Step-centric Programming} \label{sec:api}

\setlength{\textfloatsep}{0pt}
\begin{algorithm}[t]
	\caption{ThunderRW Framework}
	\label{algo:ThunderRW_framework}
	\footnotesize
	\SetKwProg{func}{Function}{}{}
	\SetKwFunction{Gather}{Gather}
	\SetKwFunction{Move}{Move}
	\SetKwRepeat{Do}{do}{while}
	 \KwIn{a graph $G$ and a set $\mathbb{Q}$ of random walk queries\;}
	 \KwOut{the walk sequences of each query in $\mathbb{Q}$\;}
	 \ForEach{$Q \in \mathbb{Q}$}{
	    \Do{$stop$ is false}{
	        $C \leftarrow$ \Gather{$G$, $Q$, \textbf{Weight}}\;
	        $e \leftarrow$ \Move{$G$, $Q$, $C$}\;
	        $stop \leftarrow$ \emph{\textbf{Update}}($Q$, $e$)\;
	    }
	 }
	 \KwRet $\mathbb{Q}$\;
	 
	\func{\Gather{$G, Q$, \textbf{Weight}}}{
	    $C\leftarrow \{\}$\;
	    \ForEach{$e \in E_{Q.cur}$}{
	        Add \emph{\textbf{Weight}}($Q$, $e$) to $C$;
	    }
	    $C\leftarrow$ execute initialization phase of a given sampling method on $C$\; 
	    \KwRet $C$\;
	}
	
	\func{\Move{$G, Q, C$}}{
	    Select an edge $e(Q.cur, v) \in E_{Q.cur}$ based on $C$ and add $v$ to $Q$\;
	    \KwRet $e(Q.cur, v)$\;
	}
\end{algorithm}

\textbf{Framework.} Algorithm \ref{algo:ThunderRW_framework} gives
an overview of ThunderRW.
Lines 1-6 execute each query one-by-one. Lines 3-5 factor one step
into three functions based on the step-centric model. \texttt{Gather}
collects the transition probabilities of edges adjacent to $Q.cur$.
It loops over $E_{Q.cur}$, applies \texttt{Weight},
a user-defined function, to each edge $e$ and add the transition
probability of $e$ to $C$ (Lines 10-11). Then, Line 12 executes
the initialization phase of a given sampling method to update $C$.
\texttt{Move} picks an edge based on $C$ and moves $Q$ along the
selected edge (Lines 14-16). \sun{As random memory accesses in the \emph{system space}
(i.e., the framework excluding user-defined functions) are mainly in \texttt{Move},
we apply step-interleaving techniques to optimize its performance (see Section~\ref{sec:step_interleaving}).}
Finally, Line 5 invokes \texttt{Update}, a user-defined function, to update states of $Q$ based on the
movement. The return value of \texttt{Update} decides whether $Q$ should be 
terminated.

The framework described in Algorithm \ref{algo:ThunderRW_framework}
can \HBS{support} unbiased, static and dynamic RW with different
sampling methods. Furthermore, we optimize the execution flow of ThunderRW
based on the RW type and the selected sampling method.
The transition probability of static RW is fixed during
the execution. In that case, ThunderRW omits the \texttt{Gather}
operation but introducing a preprocessing step to reduce the runtime cost, which obtains transition 
probabilities \HBS{in the system initialization}. Algorithm 
\ref{algo:preprocessing} presents the preprocessing for static RW.
Given a vertex $v$, Lines 3-4 loop over each edge $e$ in $E_v$ and apply the
\texttt{Weight} function to $e$ to obtain the transition probability. As
the probability does not rely on a query, we set $Q$ as \emph{null}. After
that, Lines 5-6 perform the initialization phase of a given sampling method
on $C_v$ and store $C_v$ for the usage in the query execution. As such,
we can load $C_{Q.cur}$ directly without \texttt{Gather} for static RW in Algorithm \ref{algo:ThunderRW_framework}.

\setlength{\textfloatsep}{0pt}
\begin{algorithm}[t]
	\caption{Preprocessing for Static Random Walk}
	\label{algo:preprocessing}
	\footnotesize
	 \KwIn{a graph $G$\;}
	 \KwOut{the transition probabilities $C_v$ on $E_v$ for each vertex $v$\;}
	 \ForEach{$v \in V(G)$}{
	    $C_v \leftarrow \{\}$\;
	    \ForEach{$e \in E_{v}$}{
	        Add \emph{\textbf{Weight}}($null$, $e$) to $C_v$;
	    }
	    $C_v\leftarrow$ execute initialization phase of a given sampling method on $C_v$\;
	    Store $C_v$ for the usage in query execution.
	 }
\end{algorithm}

Moreover, the \texttt{NAIVE} and \texttt{O-REJ} sampling methods have no
initialization phase as discussed in Section \ref{sec:sampling_methods}.
Hence, we do not need to collect the transition probability for initialization. As such, ThunderRW skips both the preprocessing
step and the \texttt{Gather} operation in the execution if 
\texttt{NAIVE} or \texttt{O-REJ} is used. 

\textbf{Application Programming Interfaces (APIs).} \sun{ThunderRW provides two
kinds of APIs, which include hyperparameters and user-defined functions.
Users develop their RW algorithms in two steps. Firstly, set the RW type and
the sampling method via hyperparameters \texttt{walker\_type} and \texttt{sampling\_method}, respectively.
Secondly, define the \texttt{Weight} and \texttt{Update} functions.} The
\texttt{Weight} function specifies the relative chance of an edge being
selected. The \texttt{Update} function modifies states of $Q$
given the selected edge. If its return value is \emph{true}, then
the framework terminates $Q$. Otherwise, $Q$ continues walking on $G$.
When using \texttt{O-REJ}, users need to implement the
\texttt{MaxWeight} function to set the maximum value of the transition probability.
We present an example in the following.

\begin{lstlisting} [language=C++,label={list:node2vec},numbers=none,mathescape=true,caption=Node2Vec sample code.]
WalkerType walker_type = WalkerType::Dynamic;
SamplingMethod sampling_method = SamplingMethod::O-REJ;
double Weight(Walker Q, Edge e) {
    if (Q.length == 0) return max(1.0 / a, 1.0, 1.0 / b);
    else if (e.dst == Q.prev) return 1.0 / a;
    else if (IsNeighbor(e.dst, Q.prev)) return 1.0;
    else return 1.0 / b;
}
bool Update(Walker Q, Edge e) {
    return Q.length == target_length;
}
double MaxWeight() {
    return max(1.0 / a, 1.0, 1.0 / b);
}
\end{lstlisting}

\begingroup
\setlength\abovedisplayskip{1pt}
\setlength\belowdisplayskip{1pt}
\begin{example} \label{exmp:n2v}
    List \ref{list:node2vec} shows the sample code of Node2Vec,
    which is dynamic. As the maximum value can be easily
    determined by the parameters $a$ and $b$, we use \texttt{O-REJ} to
    avoid scanning each edge adjacent to $Q.cur$ at each step. Thus, we
    set \texttt{sampling\_method} to \texttt{O-REJ} and implement
    \texttt{MaxWeight}. The \texttt{Weight} function is configured
    based on Equation \ref{eq:n2v}. Once the length of
    $Q$ meets the target length, we terminate it.
\end{example}
\endgroup

ThunderRW applies user-defined functions to RW queries,
and evaluates the queries based on RW type and selected
sampling method in parallel. Thus, users can easily implement customized RW algorithms with ThunderRW, which significantly reduces
the engineering effort. For example, users write only around ten lines of code to implement Node2Vec as shown in Example \ref{exmp:n2v}.

\textbf{Parallelization.} RW algorithms contain massive random walk queries
each of which can be completed independently and rapidly. Therefore, ThunderRW adopts
the static scheduling method to keep load balancing among workers. Specifically,
we regard each thread as a worker and evenly assign $\mathbb{Q}$ to the workers.
A worker independently executes the assigned queries with
Algorithm \ref{algo:ThunderRW_framework}. Our experiment results show
that the simple scheduling method achieves good performance.

\subsection{Analysis} \label{sec:analysis}

In this subsection, we analyze the space and time cost of
Algorithm \ref{algo:ThunderRW_framework} on different RW types with variant sampling methods. As the
cost of \texttt{Weight} and \texttt{Update} is determined by users' implementation, we assume their cost
is a constant value for the ease of analysis.

\begin{table}[t]
\footnotesize
    \setlength{\abovecaptionskip}{0pt}
    \setlength{\belowcaptionskip}{0pt}
\caption{The time complexity of ThunderRW on different random walk types with variant sampling methods}
\label{tab:summary_time_complexity}
\resizebox{0.48\textwidth}{!}{
\begin{tabular}{c|c|c|c}
\hline
\textbf{Method} & \textbf{Unbiased}                & \textbf{Static}                    & \textbf{Dynamic}                   \\ \hline\hline
NAIVE             & $O(T)$           & \textit{N/A}                       & \textit{N/A}                       \\ \hline
ITS               & $O(|E| + T \times \log d_{avg})$  & \textit{Same as unbiased}  & $O(T \times (d_{avg} + \log d_{avg}))$ \\ \hline
ALIAS             & $O(|E| + T)$ & \textit{Same as unbiased}       & $O(T \times (d_{avg} + 1))$                          \\ \hline
REJ               & $O(|E| + T \times \mathbb{E})$         &  \textit{Same as unbiased}                   & $O(T \times (d_{avg} + \mathbb{E}))$                \\ \hline
O-REJ             & $O(T \times \mathbb{E})$           & \emph{Same as unbiased}                    & \textit{Same as unbiased}                    \\ \hline
\end{tabular}
}
\end{table}

\textbf{Space.} \sun{The space for storing the graph is $O(|E| + |V|)$,
and that for maintaining the output is $O(\sum_{Q \in \mathbb{Q}}|Q|)$.}
\texttt{Gather} in Algorithm \ref{algo:ThunderRW_framework}
requires $O(d_{max})$ space to store $C$ where $d_{max}$ is the max degree
value of $G$. Suppose that ThunderRW has $n$ threads. Then, the memory
cost is $O(n \times d_{max})$. When there is a preprocessing step,
the memory cost of \texttt{ITS} and \texttt{ALIAS} is $O(|E|)$,
while that of \texttt{REJ} is $O(|V|)$ based on the analysis in Section
\ref{sec:sampling_methods}.

\textbf{Time.} Given a sampling method, $\alpha$ and $\beta$ denote
the cost of its initialization phase and generation phase, respectively.
Let $d_{avg}$ represent the average degree of $G$. Thus, the cost
of \texttt{Gather} in Algorithm \ref{algo:ThunderRW_framework} is
$d_{avg} + \alpha$, and that of \texttt{Move} is $\beta$. For static RW, the preprocessing cost is $\sum_{v \in V}(d_v + \alpha)$,
while the cost of processing one step is $\beta$ as it does not conduct
\texttt{Gather} during execution. From Section \ref{sec:sampling_methods}
we can get the value of $\alpha$ and $\beta$ for the sampling methods.
Support that $T = \sum_{Q \in \mathbb{Q}} |Q|$, which is the total
number of steps of all queries. Table \ref{tab:summary_time_complexity}
summarizes the time complexity on different RW types with
variant sampling methods.

As shown in the table, \texttt{NAIVE} supports unbiased RW only.
For \texttt{ITS}, \texttt{ALIAS} and \texttt{REJ}, the cost on unbiased
and static RW consists of the preprocessing cost and the execution
cost. Because RW algorithms can have massive RW queries with a long length, the execution cost is generally much more expensive than
the preprocessing cost. As \texttt{O-REJ} has no initialization phase,
it neither performs the preprocessing for unbiased and static RW
nor executes \texttt{Gather} for dynamic RW. Thus, the time complexity is
the same for the three RW types.

\textbf{Recommendation.} From the analysis, we have the following guidelines for setting sampling methods for users:
(1) \texttt{NAIVE} is the best sampling method for unbiased RW;
(2) \texttt{ALIAS} is a good choice for static RW since
the execution time is generally longer than the preprocessing time; and (3)
if we can set a reasonable max value for the transition probability,
then use \texttt{O-REJ} for dynamic RW. Users can easily tell
the RW type based on the properties of transition probability.
To further ease the programming efforts, we set the default sampling method
of unbiased, static and dynamic RW to \texttt{NAIVE},
\texttt{ALIAS} and \texttt{ITS}, respectively. We use \texttt{ITS}
instead of \texttt{ALIAS} for dynamic RW because the initialization
cost of \texttt{ALIAS} at each step is much more than that of \texttt{ITS}
in practice. If users can set a good max value for the transition probability,
then they can select \texttt{O-REJ} for dynamic RW.

\section{Step-Interleaving} \label{sec:step_interleaving}

In this section, we present the step interleaving technique,
which reduces the pipeline stall caused by random memory accesses.

\subsection{General Idea} \label{sec:step_interleaving_general_idea}

Based on the step-centric model, ThunderRW processes a step of a query
$Q$ with the GMU operations. \sun{According to the profiling results in Section \ref{sec:workload_profiling},
there can be two main sources for random memory accesses under the model. First, the \texttt{Move} operation
picks an edge randomly and moves $Q$ along the selected edge. Second, the operations in user-defined functions
can introduce cache misses, for example, the distance check operation in Node2Vec. As operations in the user space (i.e., user-defined functions)
are determined by RW algorithms, and can be very flexible, we target at memory issues incurred by the system (i.e., the \texttt{Move} operation).}
Motivated by the profiling result, we propose to use the
software prefectching technique \cite{lee2012prefetching} to accelerate
in-memory computation of ThunderRW. However, a step of a query $Q$ does not
have enough computation workload to hide memory access latency
because steps of $Q$ have dependency relationship.
Therefore, we propose to hide memory access latency via
executing steps of different queries alternately.

Specifically, given a sequence of operations in \texttt{Move},
we decompose them into multiple stages such that the computation of a stage
consumes the data generated by previous stages and
it retrieves the data for the subsequent stages if necessary. We execute
a group of queries simultaneously. Once a stage of a query $Q$
completes, we switch to stages of other queries in the group.
We resume the execution of $Q$ \HBS{when stages of other queries complete.}
In such a way we hide the memory access latency in a single query and
keep CPUs busy. We call this approach \emph{step interleaving}.

\begin{figure}[h]\small
    \setlength{\abovecaptionskip}{5pt}
    \setlength{\belowcaptionskip}{0pt}
    \centering
    \includegraphics[scale=0.49]{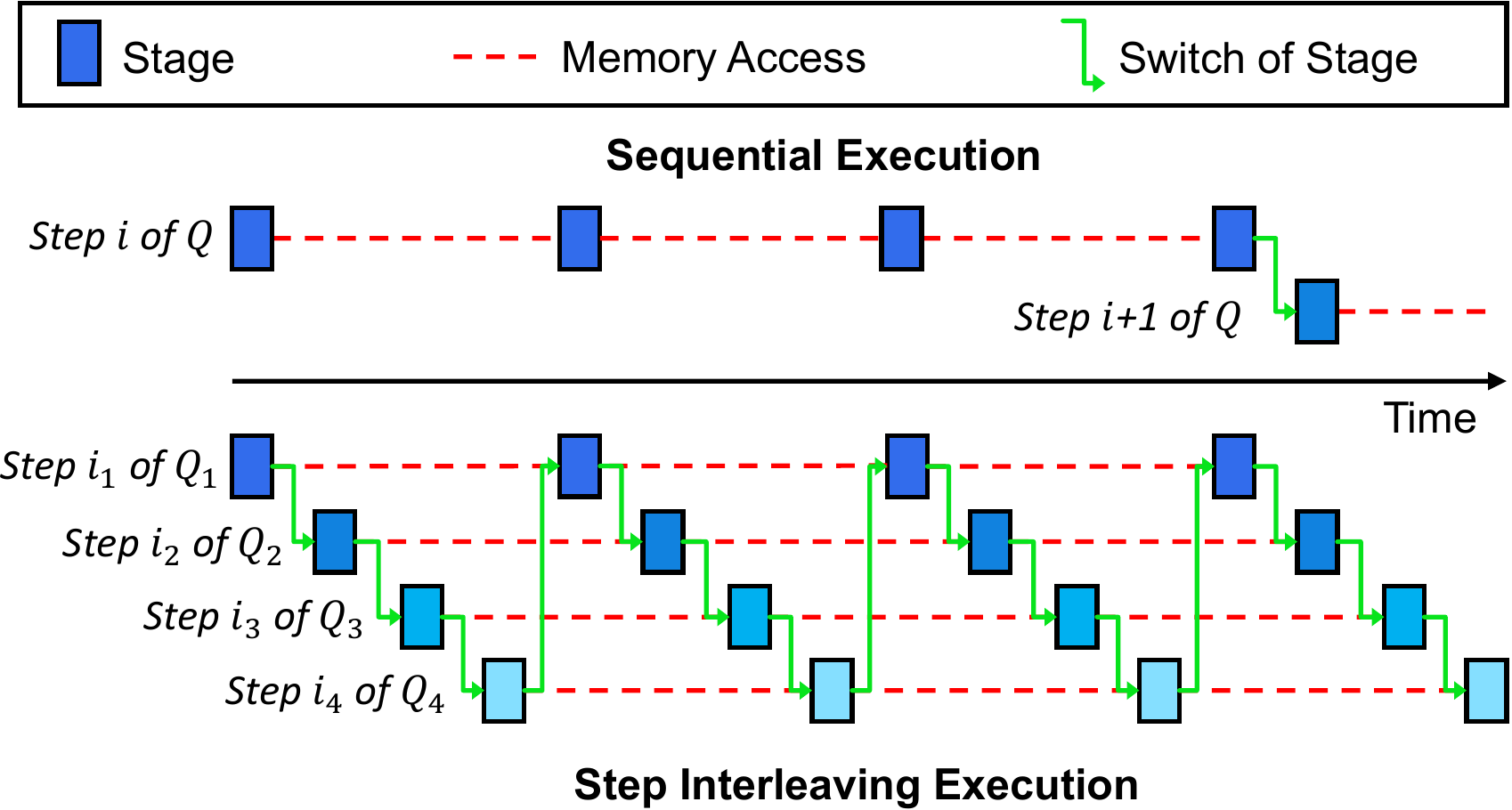}
    \caption{Sequential versus step interleaving.}
    \label{fig:sequential_vs_interleaving}
\end{figure}

\begin{example}
    Figure \ref{fig:sequential_vs_interleaving} presents an example where
    a step is divided into four stages. If executing a query step-by-step
    sequentially, then CPUs are frequently stalled due to memory access.
    Even with prefetching, the computation of a stage cannot hide the memory
    access latency. In contrast, the step interleaving hides the memory
    access latency by executing steps of different queries alternately.
\end{example}

\HBS{Let’s perform a simple back-of-envelop calculation on the performance gain of interleaving execution.} Given a group containing $k$ queries, we assume that \texttt{Move} of
each query executes the same number of stages and the cost $W_C$ of
each stage is the same for the ease of analysis. Suppose that there are
$m$ stages with memory access and $\overline{m}$ without.
$W_L$ denotes the latency of memory access. Then, the cost of
moving a step for the queries in sequential is equal to
$W_0 = k((m + \overline{m})W_C + mW_L$). Let $W_S$ denote the cost of switching. 
The cost of \texttt{Move} with step interleaving is
$W_1 = k((m + \overline{m})(W_C + W_S) + m(\max(W_L - kW_S - (k - 1)W_C, 0))$
where the last term calculates whether step interleaving hides memory access
latency. Therefore, the gain of step interleaving
for a step of $k$ queries can be estimated by Equation \ref{eq:gain}
where $W_{hide} = \max(W_L - kW_S - (k - 1)W_C, 0)$.

\begin{equation}\label{eq:gain}
\begin{aligned}
    W_{gain} &= (W_0 - W_1)/k \\
             &= mW_L - (m + \overline{m})W_S - mW_{hide}.
\end{aligned}
\end{equation}

From Equation \ref{eq:gain}, we can see that step interleaving requires
an efficient switch mechanism to reduce the overhead $W_S$ of performing
switching, and enough workload to overlap the memory access latency $W_{hide}$.

\subsection{Stage Dependency Graph}

We design the \emph{stage dependency graph} (SDG) to model stages of
a sequence of operations in a step. Each node in SDG is a stage
containing a set of operations and edges represent the dependency
relationship among them. Given the sequence of operations, we build
SDG in two steps, abstracting stages (nodes) and extracting
dependency relationships (edges).

\HBS{\textbf{Defining stages:}} As we hide memory access latency by switching the execution of queries,
the constraint on stages is that each stage contains at most
one memory access operation and the operations consuming the data are in subsequent
stages. Note that we view the
operation containing jump operation as a single stage for the ease of
the implementation of switching. We present an example in the following.

\begin{table}[t]
    \setlength{\abovecaptionskip}{0pt}
    \setlength{\belowcaptionskip}{0pt}
\footnotesize
\caption{Stages of \texttt{Move} with \texttt{ALIAS} and \texttt{REJ} ($v = Q.cur$).}
\label{tab:stages}
\begin{tabular}{c|l}
\hline
\textbf{Stage}         & \multicolumn{1}{c}{\textbf{\texttt{ALIAS}}}                                 \\ \hline\hline
$S_0$                  & $O_0$: Load $d_v$.                                                 \\ \hline
\multirow{3}{*}{$S_1$} & $O_1$: Generate an int random num $x$ in $[0, d_v)$.               \\ \cline{2-2} 
                       & $O_2$: Generate a real random num $y$ in $[0,1)$.                  \\ \cline{2-2} 
                       & $O_3$: Load $C[x] = (H[x], A[x])$.                                        \\ \hline
\multirow{2}{*}{$S_2$} & $O_4$: If $y < H[x]$,  $v' = A[x].first$; Else $v' = A[x].second$.   \\ \cline{2-2} 
                       & $O_5$: Add $v'$ to $Q$ and return $e(v, v')$.                                            \\ \hline\hline
\textbf{Stage}         & \multicolumn{1}{c}{\textbf{\texttt{REJ}}}                                   \\ \hline\hline
$S_0$                  & $O_0$: Load $d_v$.                                                 \\ \hline
$S_1$                  & $O_1$: Load the maximum value $p_v ^ *$.                           \\ \hline
\multirow{3}{*}{$S_2$} & $O_2$: Generate an int random num $x$ in $[0, d_v)$.               \\ \cline{2-2} 
                       & $O_3$: Generate a real random num $y$ in $[0, p_v ^ *)$.           \\ \cline{2-2} 
                       & $O_4$: Load $C[x] = p$.                                            \\ \hline
$S_3$                  & $O_5$: If $y > C[x]$, jump to $O_2$; Else jump to $O_6$.                                  \\ \hline
$S_4$                  & $O_6$: Load $e(v,v') = E_v[x]$.                                    \\ \hline
$S_5$                  & $O_7$: Add $v'$ to $Q$ and return $e(v, v')$.                                            \\ \hline
\end{tabular}
\end{table}

\begin{example} \label{exmp:stages}
    The right column of Table \ref{tab:stages} illustrates the sequence of operations
    in the \texttt{Move} function with the \texttt{ALIAS} and \texttt{REJ} sampling methods, respectively,
    to perform the neighbor selection. The left column lists stages. For example, $S_0$ of \texttt{ALIAS}
    loads $d_v$ consumed in $O_1$ of $S_1$. $O_5$ in \texttt{REJ}
    has the jump operation. Therefore, we regard it as a \HBS{separate} stage.
\end{example}

\HBS{\textbf{Defining edges:}} Next, we add edges among nodes in SDG based on their
dependency relationships. Given
stages $S$ and $S'$, if there is a dependency relationship
between $S$ and $S'$, we add an edge from $S$ to $S'$.
The edges are categorized into three types,
\emph{memory dependency}, \emph{computation dependency}
and \emph{control dependency}. We call the first two
relationship as \emph{data dependency}. More specifically,
if $S'$ consumes the data loaded from memory by $S$,
then the edge type is memory dependency. Otherwise,
$S'$ depends on the data computed by $S$ and the
edge type is computation dependency. The data leading to
the dependency is attached to each edge as properties.
Furthermore, if $S$ contains the operation jumping to $S'$,
we add the control dependency from $S$ to $S'$.
SDG allows that there are multiple edges (i.e., dependency
relationships) between nodes. \HBS{If we only consider data
dependency}, SDG is a directed acyclic graph (DAG), while the control dependency can
generate cycles in SDG. 

\begin{figure}[t]\small
    \setlength{\abovecaptionskip}{5pt}
    \setlength{\belowcaptionskip}{0pt}
    \centering
    \includegraphics[scale=0.49]{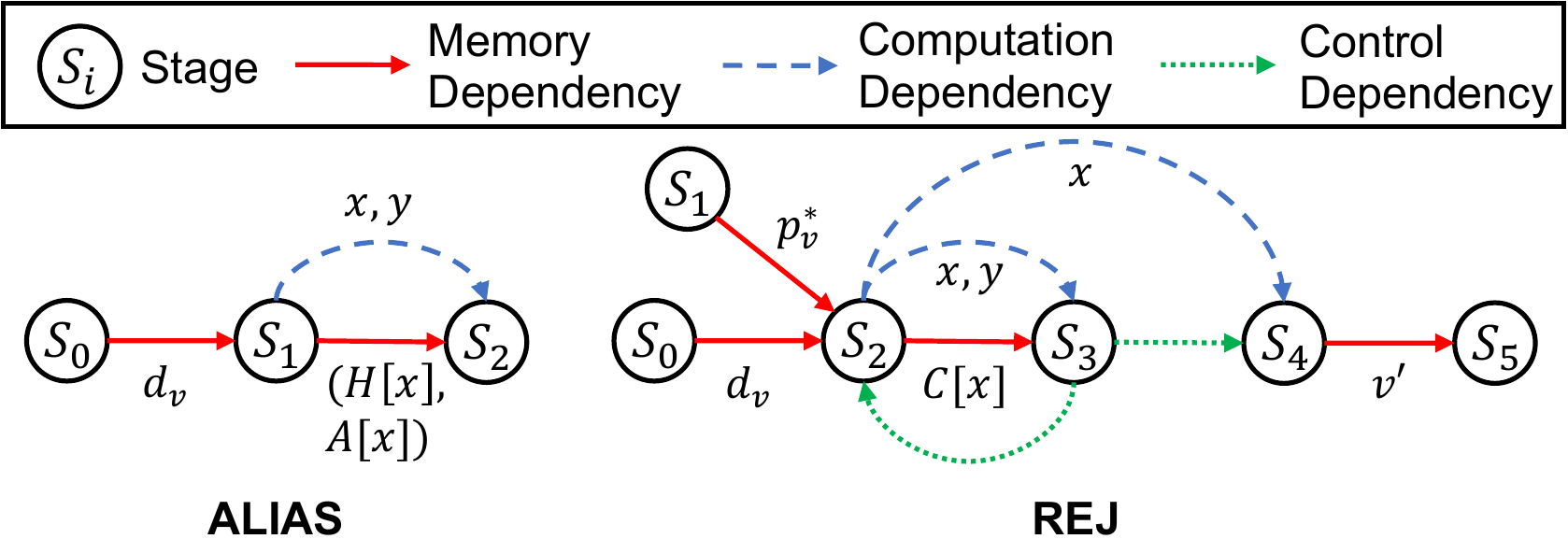}
    \caption{Stage dependency graph.}
    \label{fig:sdg}
\end{figure}

\begin{example} \label{exmp:sdg}
    Continuing with Example \ref{exmp:stages},
    Figure \ref{fig:sdg} shows SDGs.
    In SDG of \texttt{ALIAS}, $S_2$ relies on $x,y$,
    which are random numbers generated by $S_1$, while $(H[x],A[x])$
    is the data retrieved from memory. As such, $S_1$ and $S_2$ have
    both memory and computation dependency relationships. SDG of
    \texttt{ALIAS} is a DAG because there is no control dependency.
    In contrast, there is a cycle containing $S_2$ and $S_3$
    in SDG of \texttt{REJ} because of the control dependency. 
\end{example}

In summary, SDG is a methodology
to abstract stages from a sequence of operations in \texttt{Move}
and model the dependency relationship among them. Note that
the stage design of \texttt{MOVE} does not require user input but
it is implemented in the system space.

\subsection{State Switch Mechanism}

In this subsection, we introduce the implementation of step interleaving
under SDG. Based on Equation \ref{eq:gain}, we need an efficient switch mechanism.
For example, using multi-threading is forbidden because the overhead of context
switch among threads is in microseconds, whereas the main memory latency is in nanoseconds.
As each thread tends to take many RW queries, we switch the execution among stages in a single thread.

We categorize stages of a SDG into two classes
based on whether they belong to cycles in SDG, and efficiently handle them in different manners. For stages not in cycles (called \emph{non-cycle stages}),
a query visits them exactly once to complete \texttt{Move}. Given a group 
of queries $\mathbb{Q}'$, we execute them in a coupled manner. Particularly, once a query $Q_i \in \mathbb{Q}'$
completes a stage $S$, we switch to the next query $Q_{i + 1} \in \mathbb{Q}'$ to process $S$.
After all queries complete $S$, we move to the next stage. In contrast, stages in cycles (called \emph{cycle stages})
can be visited variant times for different queries. To deal with the irregularity, we process them
in a decoupled manner. Specifically, each query $Q$ records the stage $S$ to be executed. When switching to $Q$,
we execute $S$, set the next stage of $Q$ based on SDG, and switch to the next query after completing $S$.
As a result, each query executes asynchronous.

\sun{For data communication between different stages in a query, we
create two kinds of ring buffers based on SDG, in which the computation dependency edge indicates the information
requiring to be stored. In particular, the \emph{task ring} is used for data communication across all stages of a query,
while the search ring serves to process cycle stages. As we need to explicitly record states of cycle stages
and control the switch of them, processing cycle stages not only causes implementation complexities, but also incurs more overhead.
Note that the SDGs of \texttt{NAIVE} and \texttt{ALIAS} have no cycle stages because
there are no for loops in their generation phases, whereas that of \texttt{ITS}, \texttt{REJ}
and \texttt{O-REJ} have. The implementation details are introduced in the appendix.}

\subsection{Ring Size Tuning} \label{sec:tune_ring_size}

The task ring size $k$ and the search ring size $k'$ determine the group size of queries executed
simultaneously in a thread, and therefore control memory level parallelism of executing non-cycle
stages and cycle stages, respectively. According to Equation \ref{eq:gain}, we can
improve the performance by increasing $k$ to reduce $W_{hide}$.
However, $k$ is limited by hardware. Particularly,
modern CPUs can issue a limited number of outstanding memory requests,
and the L1 data cache size is only tens of kilobytes. Setting $k$ to
a large value can evict data before the usage. In ThunderRW,
we tune ring sizes by pre-executing a number of queries.
We start a RW query from each vertex with the target length as 10
and set the RW type as static. We first select the \texttt{NAIVE}
and \texttt{ALIAS} sampling methods, respectively and vary $k$ from
$1,2,...,512, 1024$ to pick an optimal value $k ^ *$. Next we fix $k$
to $k ^ *$ and vary $k'$ from $1,2,..., k^*$ to select optimal values
for \texttt{ITS}, \texttt{REJ} and \texttt{O-REJ}, respectively.

\setlength{\textfloatsep}{0pt}
\begin{algorithm}[t]
	\caption{ThunderRW using Step Interleaving}
	\label{algo:ThunderRW_framework_si}
	\footnotesize
	\SetKwProg{func}{Function}{}{}
	\SetKwFunction{Gather}{Gather}
	\SetKwFunction{Move}{Move}
	\SetKwRepeat{Do}{do}{while}
	 \KwIn{a graph $G$ and a set $\mathbb{Q}$ of random walk queries\;}
	 \KwOut{the walk sequences of each query in $\mathbb{Q}$\;}
	 Add the first $k$ queries in $\mathbb{Q}$ to $\mathbb{Q}'$\;
	 $completed \leftarrow 0$, $submitted \leftarrow k$\;
	 \While{$completed < |\mathbb{Q}|$}{
	    $\mathbb{C} \leftarrow \{\}$\;
	    \For{$Q \in \mathbb{Q}'$}{
	        $C \leftarrow$\Gather{$G, Q, \textbf{Weight}$}\;
	        Add $\mathbb{C}[Q]$ to $C$\;
	    }
	    
	    $\mathbb{U} \leftarrow$ \Move{$G, \mathbb{Q}', \mathbb{C}$}\;
	    
	    \For{$Q \in \mathbb{Q}'$}{
	        \If{\textbf{Update}($Q, \mathbb{U}[Q]$) is true}{
	            Remove $Q$ from $\mathbb{Q}'$\;
	            $completed \leftarrow completed + 1$\;
	            \If{$submitted < |\mathbb{Q}|$}{
	                Get next query $Q'$ from $\mathbb{Q}$ and add it to $\mathbb{Q}'$\;
	                $submitted \leftarrow submitted + 1$\;
	            }
	        }
	    }
	 }
\end{algorithm}

\subsection{Integration with ThunderRW}

Algorithm \ref{algo:ThunderRW_framework_si} illustrates
our ThunderRW framework using step interleaving. Line 1
adds the first $k$ queries in $\mathbb{Q}$ to $\mathbb{Q}'$ where
$k$ is the parameter setting the group size.
Lines 3-15 repeatedly execute GMU operations on queries in
$\mathbb{Q}'$ until all queries in $\mathbb{Q}$ complete. Specifically,
Lines 5-7 first execute the \texttt{Gather} operation on each query in $\mathbb{Q}'$.
Next, Line 8 invokes the \texttt{Move} operation using step interleaving to
process queries in $\mathbb{Q}'$. After that, Lines 9-15 apply the \texttt{Update}
operation to all queries in the group. If a query completes, then Lines 11-15 remove
it and submit the next query in $\mathbb{Q}$ to $\mathbb{Q}'$.
Thus, the step interleaving technique can be seamlessly integrated with ThunderRW
\HBS{without changing APIs}.

\textbf{Time and space.} The time complexity of Algorithm \ref{algo:ThunderRW_framework_si}
is the same with the analysis in Section \ref{sec:analysis} because the step interleaving
does not change the number of steps moved. Suppose that there are $n$ threads.
Then, the memory cost is $O(n \times k \times d_{max})$ in addition to the space storing
the graph and the output because each thread has at most $k$ queries in flight.

\section{Experiments} \label{sec:experiments}

We conduct experiments to evaluate the performance of ThunderRW in this section.

\subsection{Experimental Setup} \label{sec:experimental_setup}

We conduct experiments on a Linux server equipped with an Intel Xeon W-2155 CPU and 220GB RAM. The CPU has ten physical cores with hyper-threading disabled for consistent measurement.
The sizes of L1, L2 and L3 (last level cache, LLC) caches are 32KB, 1MB and 13.75MB, respectively.

\begin{table}[t]
\setlength{\abovecaptionskip}{0pt}
\setlength{\belowcaptionskip}{0pt}
\footnotesize
\caption{Properties of real-world datasets.}
\label{tab:datasets}
\begin{tabular}{cccclcc}
\hline
\textbf{Dataset} & \textbf{Name} & \textbf{$|V|$} & \textbf{$|E|$} & $d_{avg}$ & $d_{max}$ & \textbf{Memory} \\ \hline\hline
\sun{amazon}           & \sun{\textit{am}}   & \sun{0.55M}          & \sun{1.85M}          & \sun{3.38}    & \sun{549}  & \sun{0.01GB}          \\ \hline
youtube          & \textit{yt}   & 1.14M          & 2.99M          & 5.24    &28754  & 0.03GB          \\ \hline
us patents       & \textit{up}   & 3.78M          & 16.52M         & 8.74    &793  & 0.17GB          \\ \hline
eu-2005          & \textit{eu}   & 0.86M          & 19.24M         & 44.74   &68963  & 0.15GB          \\ \hline
\sun{amazon-clothing}  & \sun{\textit{ac}}   & \sun{15.16M}         & \sun{63.33M}          & \sun{4.18}   & \sun{12845} & \sun{0.35GB}         \\ \hline
\sun{amazon-book}      & \sun{\textit{ab}}   & \sun{18.29M}         & \sun{102.12M}          & \sun{5.58}  & \sun{58147} & \sun{0.52GB}  \\ \hline
livejournal      & \textit{lj}   & 4.85M          & 68.99M         & 28.45   &20333  & 0.54GB          \\ \hline
com-orkut        & \textit{ot}   & 3.07M          & 117.19M        & 76.34   &33313  & 0.89GB          \\ \hline
\sun{wikidata}         & \sun{\textit{wk}}   & \sun{40.96M}         & \sun{265.20M}          & \sun{6.47}   & \sun{8085513} & \sun{1.29GB}\\ \hline
uk-2002          & \textit{uk}   & 18.52M         & 298.11M        & 32.19   &194955  & 2.30GB          \\ \hline
twitter          & \textit{tw}   & 41.66M         & 1.21B          & 58.08   &2997487  & 9.27GB          \\ \hline
friendster       & \textit{fs}   & 65.61M         & 1.81B          & 55.17   &5214  & 13.71GB         \\ \hline
\end{tabular}
\end{table}

\textbf{Datasets.} \sun{Table \ref{tab:datasets} lists the statistics of the twelve real-world graphs in our experiments.
\emph{ab} and \emph{ac} are downloaded from \cite{amazon_review}, \emph{wk} is obtained from \cite{wikidata},
\emph{eu}, \emph{uk} and \emph{tw} are obtained from \cite{networkrepo}, and the other graphs are downloaded from
\cite{snapnets}.} The datasets are from different categories such as web, social and citation, and have different densities.
The number of vertices is ranged from hundreds of thousands to tens of millions, and the number of edges scales from millions to billions.
\sun{Except \emph{am}, all the graphs outsize LLC.}

\textbf{Workloads.} We study PPR, DeepWalk, Node2Vec and MetaPath to
evaluate the performance and generality of competing methods. The settings of the four algorithms are the same as that in
Section \ref{sec:workload_profiling}. \sun{\emph{ab} and \emph{ac} are weighted graphs where weights denote review ratings for products.
\emph{wk} has 1327 distinct labels, which represents the relationship between entities in a knowledge base. The other graphs are unweighted
and unlabeled.} Given a graph having no labels or weights, we set the weight and label of edges
with the same setting as previous work \cite{yang2019knightking}: (1) We choose a real number from [1, 5) uniformly at random, and assign
it to an edge as its weight; and (2) We set the edge label by randomly choosing a label from a set containing five distinct labels.

\textbf{Comparison.} \sun{We compare the performance of ThunderRW (called \emph{TRW} for short) with the following methods.}

\begin{itemize}[noitemsep,topsep=1pt,leftmargin=*]
    \item \sun{\emph{BL}: Baseline approaches that first load a graph entirely into memory and then execute random walks,
    the detail of which is presented in Section \ref{sec:workload_profiling}.}
    \item \sun{\emph{HG}: Our homegrown implementation optimizing \emph{BL} from two aspects: (1) select a suitable sampling method for each algorithm
    according to the recommendation in Section \ref{sec:analysis}; and (2) regard each query as a parallel task with OpenMP.}
    \item \emph{GW}: GraphWalker \cite{wang2020graphwalker}, the state-of-the-art RW framework in a single machine. \HBS{For the fair of comparison, we configure GraphWalker to execute in-memory, without any disk I/O.}
    \item \emph{KK}: KnightKing \cite{yang2019knightking}, the state-of-the-art distributed RW framework. \HBS{
    It supports to execute in a single machine. }
\end{itemize}

\begin{table*}[t]
    \setlength{\abovecaptionskip}{0pt}
    \setlength{\belowcaptionskip}{0pt}
\footnotesize
\caption{Overall performance comparison (seconds).}
\label{tab:overall_comparison}
\resizebox{0.96\textwidth}{!}{
\begin{tabular}{c|ccccc|cccc|cccc|ccc}
\hline
\textbf{}        & \multicolumn{5}{c|}{\sun{\textbf{PPR}}}                          & \multicolumn{4}{c|}{\textbf{DeepWalk}}                     & \multicolumn{4}{c|}{\textbf{Node2vec}}       & \multicolumn{3}{c}{\textbf{MetaPath}}        \\ \hline
\textbf{Dataset} &\sun{\textbf{\emph{BL}}} & \sun{\textbf{\emph{HG}}} & \sun{\textbf{\emph{GW}}} & \sun{\textbf{\emph{KK}}}  & \sun{\textbf{\emph{TRW}}} & \sun{\textbf{\emph{BL}}}& \textbf{\emph{HG}}   &\textbf{\emph{KK}} & \textbf{\emph{TRW}} & \sun{\textbf{\emph{BL}}}&\textbf{\emph{HG}} & \textbf{\emph{KK}} & \textbf{\emph{TRW}} & \sun{\textbf{\emph{BL}}}& \textbf{\emph{HG}} & \textbf{\emph{TRW}} \\ \hline\hline
\sun{\textit{am}}  &0.06     &0.008    &0.42     &0.012    &\textbf{0.007   } &2.16     &0.21     &0.44     &\textbf{0.07    } &9.97     &0.26     &2.08     &\textbf{0.14    } &0.22     &0.018    &\textbf{0.012   }\\ \hline
\textit{yt}  &0.33     &0.04     &1.68     &0.05     &\textbf{0.015   } &9.78     &0.98     &1.93     &\textbf{0.26    } &853.13   &1.30     &5.94     &\textbf{1.03    } &6.18     &\textbf{0.23    } &0.24    \\ \hline
\textit{up}  &1.24     &0.13     &7.19     &0.19     &\textbf{0.07    } &45.44    &4.33     &8.41     &\textbf{0.95    } &369.00   &6.20     &16.92    &\textbf{4.01    } &4.88     &0.40     &\textbf{0.24    }\\ \hline
\textit{eu}  &0.16     &0.02     &0.99     &0.03     &\textbf{0.011   } &8.16     &0.82     &1.56     &\textbf{0.20    } &2731.07  &1.47     &4.43     &\textbf{1.14    } &90.55    &\textbf{3.18    } &3.55    \\ \hline
\sun{\textit{ac}}  &4.84     &0.51     &19.31    &0.65     &\textbf{0.19    } &173.66   &17.86    &31.88    &\textbf{3.31    } &6951.12  &24.54    &87.86    &\textbf{6.26    } &45.01    &2.01     &\textbf{1.69    }\\ \hline
\sun{\textit{ab}}  &8.86     &0.94     &26.74    &1.09     &\textbf{0.26    } &212.80   &22.24    &40.07    &\textbf{4.01    } &26231.45 &32.04    &100.78   &\textbf{7.87    } &128.35   &5.06     &\textbf{4.47    }\\ \hline
\textit{lj}  &1.69     &0.19     &7.90     &0.23     &\textbf{0.06    } &55.63    &5.44     &10.67    &\textbf{1.19    } &2951.33  &9.09     &24.95    &\textbf{6.20    } &18.08    &0.94     &\textbf{0.75    }\\ \hline
\textit{ot}  &1.49     &0.16     &5.25     &0.19     &\textbf{0.04    } &38.54    &3.70     &7.97     &\textbf{0.80    } &5891.28  &7.28     &15.16    &\textbf{4.82    } &40.77    &1.72     &\textbf{1.57    }\\ \hline
\sun{\textit{wk}}  &21.86    &2.21     &47.05    &3.07     &\textbf{0.59    } &502.27   &49.67    &95.17    &\textbf{9.26    } &     \emph{OOT} &68.43    &216.24   &\textbf{27.68   } &5.98     &\textbf{0.54    } &0.55    \\ \hline
\textit{uk}  &6.47     &0.69     &27.72    &0.90     &\textbf{0.24    } &203.86   &20.42    &21.40    &\textbf{4.56    } &12630.01 &34.36    &94.69    &\textbf{28.68   } &322.66   &12.84    &\textbf{12.56   }\\ \hline
\textit{tw}  &26.42    &2.73     &77.12    &3.61     &\textbf{1.16    } &575.43   &61.18    &115.92   &\textbf{11.13   } &     \emph{OOT} &130.72   &232.41   &\textbf{91.00   } &     \emph{OOT} &12300.32 &\textbf{9780.20 }\\ \hline
\textit{fs}  &79.14    &8.20     &223.81   &10.72    &\textbf{4.10    } &1043.93  &108.23   &208.45   &\textbf{17.67   } &     \emph{OOT} &178.15   &364.51   &\textbf{120.16  } &683.05   &28.69    &\textbf{25.01   }\\ \hline
\end{tabular}
}
\vspace*{-10pt}
\end{table*}

We implement all our methods including \emph{BL}, \emph{HG} and \emph{TRW} in C++.
\emph{GW}\footnote{\url{https://github.com/ustcadsl/GraphWalker}, Last accessed on 2020/12/07.}
and \emph{KK}\footnote{\url{https://github.com/KnightKingWalk/KnightKing}, Last accessed on 2020/12/20.}
are programmed in C++ as well. All the source code is compiled by g++ 8.3.2 with -O3 enabled.
\sun{\emph{BL} executes in serial, while the other methods are running on all the cores of the single socket,
with one thread per core.}

We consider C-SAW \cite{pandey2020c}, the state-of-the-art RW framework
on GPUs, as well. However, its open source package\footnote{\url{https://github.com/concept-inversion/C-SAW}, Last accessed on 2020/12/07.}
supports 4000 queries at most, which cannot handle the workload containing massive queries in the experiment.
Previous experiment results \cite{yang2019knightking,wang2020graphwalker} show that
\emph{KK} and \emph{GW} significantly outperform generic graph computing frameworks
such as Gemini \cite{zhu2016gemini} on RW algorithms.
Therefore, our experiment does not involve C-SAW as well as any generic graph
computing frameworks. 

As for RW algorithms, \emph{GW} only supports unbiased RW. Thus, we execute PPR without considering edge weights,
and evaluate \emph{GW} on PPR only. Despite that \emph{KK} studies MetaPath in the original paper~\cite{yang2019knightking},
its open source package cannot handle labeled graphs. As such, it cannot execute MetaPath.
In contrast, \emph{TRW} supports all the four algorithms, which demonstrates its flexibility.

\sun{As for sampling methods, \emph{BL} uses \texttt{NAIVE} for PPR, while adopts \texttt{ALIAS} for the other three algorithms.
As discussed in Section \ref{sec:workload_profiling}, building alias tables for dynamic RW in an indexing phase can consume a
huge amount of memory. Therefore, in the experiments, \emph{BL} dynamically computes the alias table (i.e., perform the initialization of \texttt{ALIAS})
at each step of a query, which is the same as the computation flow of \emph{TRW} for dynamic RW.
Different from \emph{BL}, \emph{HG} adopts \texttt{O-REJ} for Node2Vec, and \texttt{ITS} for MetaPath. This is because (1)
the max value of transition probability of Node2Vec can be easily set as $\max (1, 1/a, 1/b)$, and \texttt{O-REJ}
can avoid scanning the neighbors of $Q.cur$ at each step; and (2) the probability distribution of MetaPath is skewed due to
filtering based on labels, which increases the generation cost of rejection sampling, and the initialization phase of \texttt{ITS}
is much faster than that of \texttt{ALIAS} in practice. \emph{TRW} adopts the same sampling method as \emph{HG} for each algorithm.}

\sun{\textbf{Ring Size Setting.} We tune the ring size with the method in Section
\ref{sec:tune_ring_size}. Despite that the graphs have variant structures, the optimal setting for them is close.
First, the optimal value for the graphs except \emph{am} is $k = 64$ and $k' = 32$ because the optimal ring size is
closely related to the instructions available for computation, the switch overhead, the memory access latency, and
the maximum number of outstanding memory requests, which are determined by the program and hardwares.
Second, the optimal value for \emph{am} is $k = 32$ and $k' = 32$ as \emph{am} fits in LLC and the memory access latency
is smaller than that of the other graphs.
Additionally, the tuning process is very
efficient, which takes less than one minute for most of the graphs. Even for \emph{fs} with more than 1.8 billion edges,
the tuning is completed with around four minutes.} 

\textbf{Metrics.} The \emph{total time} is the elapsed time on evaluating RW algorithms without counting the time on
loading data from the disk. For static random walk, the total time consists of the \emph{preprocessing time}, which is the time spent
on the preprocessing, and the \emph{execution time}, which is the time spent on executing queries. \sun{To complete experiments
in a reasonable time, we set the time limit for each algorithm as eight hours. If an algorithm cannot be completed within the limit,
we terminate it and record the execution time as \emph{OOT} (i.e., out-of-time).}
We measure the \emph{throughput} (steps per second) by
dividing the number of steps of all queries by the execution time. To provide more insights,
we adopt \emph{Intel Vtune Profiler} to examine the pipeline slot utilization and use \emph{Linux Perf}
to examine the \emph{instructions per step} and \emph{cycles per step}, which are the number of instructions and
the number of cycles on one step, respectively.

\sun{\textbf{Supplement experiments.} More experiment results including the impact of ring sizes,
memory bandwidth utilization, the effectiveness of prefetching data to different cache levels, the impact of the step interleaving on existing systems
and the comparison with AMAC \cite{kocberber2015asynchronous} are presented in the appendix.}

\subsection{Overall Comparison} \label{sec:overall_performance_comparison}

Table \ref{tab:overall_comparison} gives an overall comparison of competing methods
on the four RW algorithms. Although \emph{GW} is parallel, it runs slower than \emph{BL}, the sequential
baseline algorithm. \emph{KK} runs faster than \emph{GW} and \emph{BL}, but slower than
\emph{HG} because (1) the framework incurs extra overhead compared with \emph{HG}; and (2) \emph{HG} adopts an appropriate
sampling method for each algorithm. \emph{TRW} runs 54.6-131.7X and 1.7-14.6X faster
than \emph{GW} and \emph{KK}, respectively.

\sun{Benefiting from parallelization, \emph{HG} achieves 7.5-10.5X speedup over \emph{BL} on PPR and DeepWalk. Moreover,
\emph{HG} runs 38.3-1857.9X and 11.1-28.5X faster than \emph{BL} on Node2Vec and MetaPath, respectively, because \emph{HG}
adopts \texttt{O-REJ} sampling for Node2Vec, which avoids scanning the neighbors of $Q.cur$ at each step, and uses
\texttt{ITS} sampling for MetaPath, the initialization phase of which is more efficient than that of \texttt{ALIAS} in practice.
\emph{TRW} runs 8.6-3333.1X faster than \emph{BL}. Even compared with \emph{HG}, \emph{TRW} achives up to 6.1X speedup benefiting from
our step-centric model and step interleaving technique. As MetaPath is dynamic and both \emph{TRW} and \emph{HG} use \texttt{ITS} sampling,
the gather operation at each step dominates the cost. Still, MetaPath on ThunderRW outperforms
that on HG for nine out of twelve graphs, and is slightly slower on the other three graphs. \emph{tw} is
dense but highly skewed (as shown in Table~\ref{tab:datasets}) and the vertices with high degrees
are frequently visited. Consequently, the execution time on MetaPath against \emph{tw} is much longer
than that on other graphs.}

\sun{In summary, ThunderRW significantly outperforms
state-of-the-art frameworks and homegrown solutions (e.g., \emph{BL} takes more than eight hours for Node2Vec on \emph{tw}, while
\emph{TRW} completes the algorithm in two minutes). Furthermore, ThunderRW saves a lot of engineering effort on the
implementation and parallelization of RW algorithms compared with \emph{BL} and \emph{HG}.}

\begin{figure}[t]\small
    \setlength{\abovecaptionskip}{0pt}
    \setlength{\belowcaptionskip}{0pt}
    \captionsetup[subfigure]{aboveskip=0pt,belowskip=0pt}
    \centering
    \begin{subfigure}[t]{0.23\textwidth}
        \centering
        \includegraphics[scale=0.23]{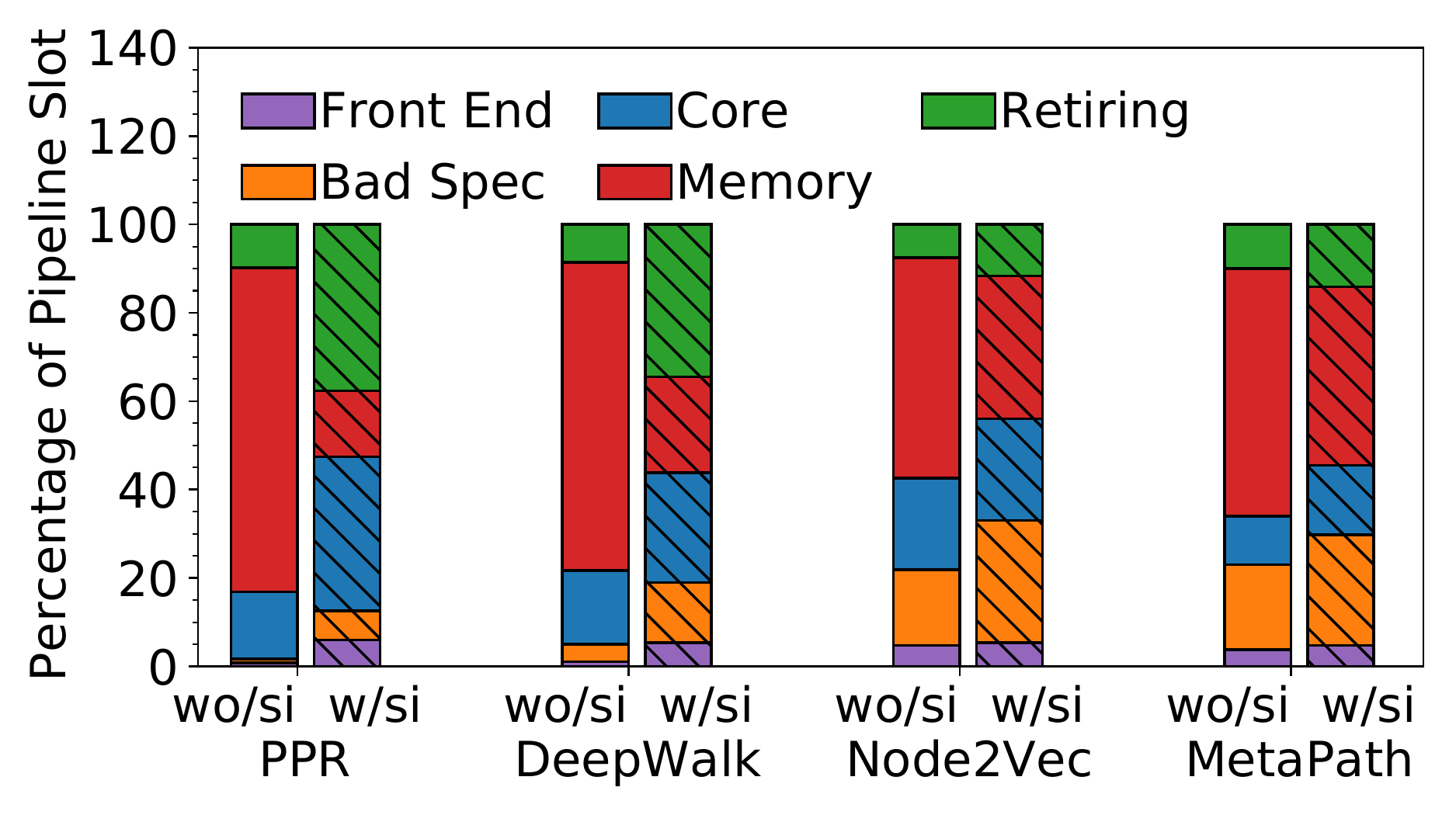}
        \caption{Pipeline slot breakdown.}
        \label{fig:lj_breakdown}
    \end{subfigure}
    \begin{subfigure}[t]{0.23\textwidth}
        \centering
        \includegraphics[scale=0.23]{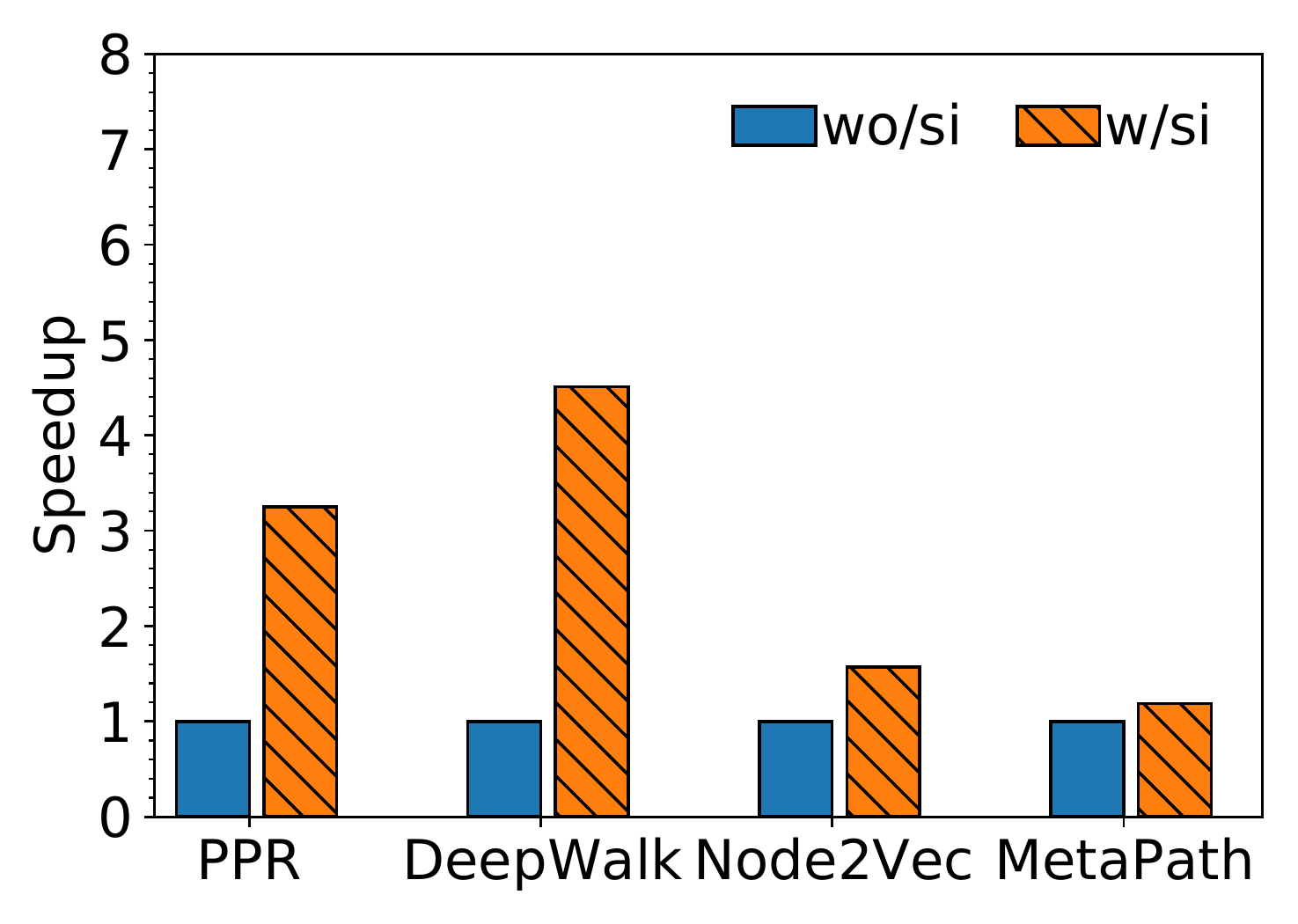}
        \caption{Speedup.}
        \label{fig:lj_speedup}
    \end{subfigure}
    \caption{\sun{Vary RW-algorithms on \emph{lj}.}}
    \label{fig:vary_rw_algorithms}
\end{figure}

\begin{figure}[t]\small
    \setlength{\abovecaptionskip}{0pt}
    \setlength{\belowcaptionskip}{0pt}
    \captionsetup[subfigure]{aboveskip=0pt,belowskip=0pt}
    \centering
    \begin{subfigure}[t]{0.23\textwidth}
        \centering
        \includegraphics[scale=0.23]{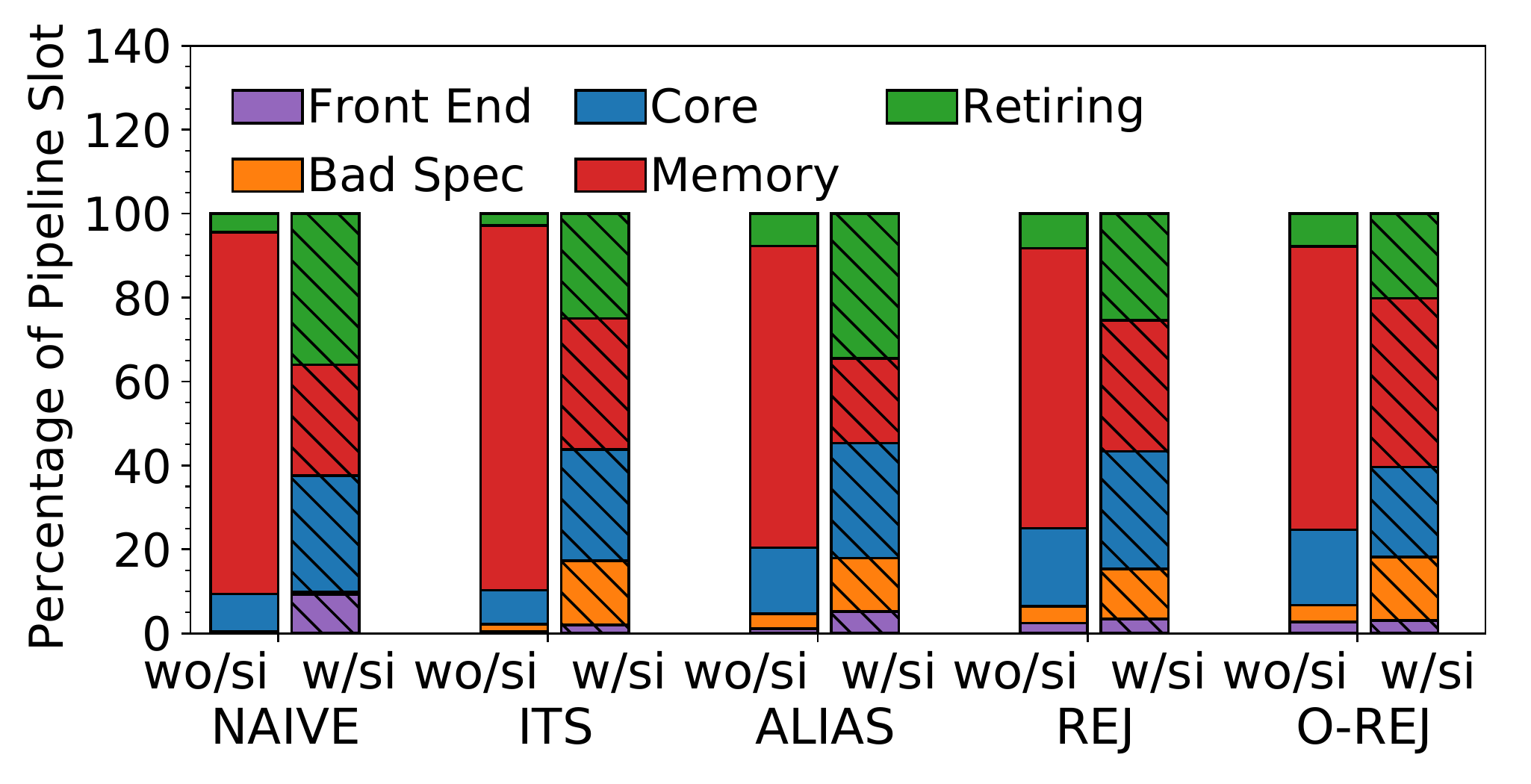}
        \caption{Pipeline slot breakdown.}
        \label{fig:lj_sampling_method_breakdown}
    \end{subfigure}
    \begin{subfigure}[t]{0.23\textwidth}
        \centering
        \includegraphics[scale=0.23]{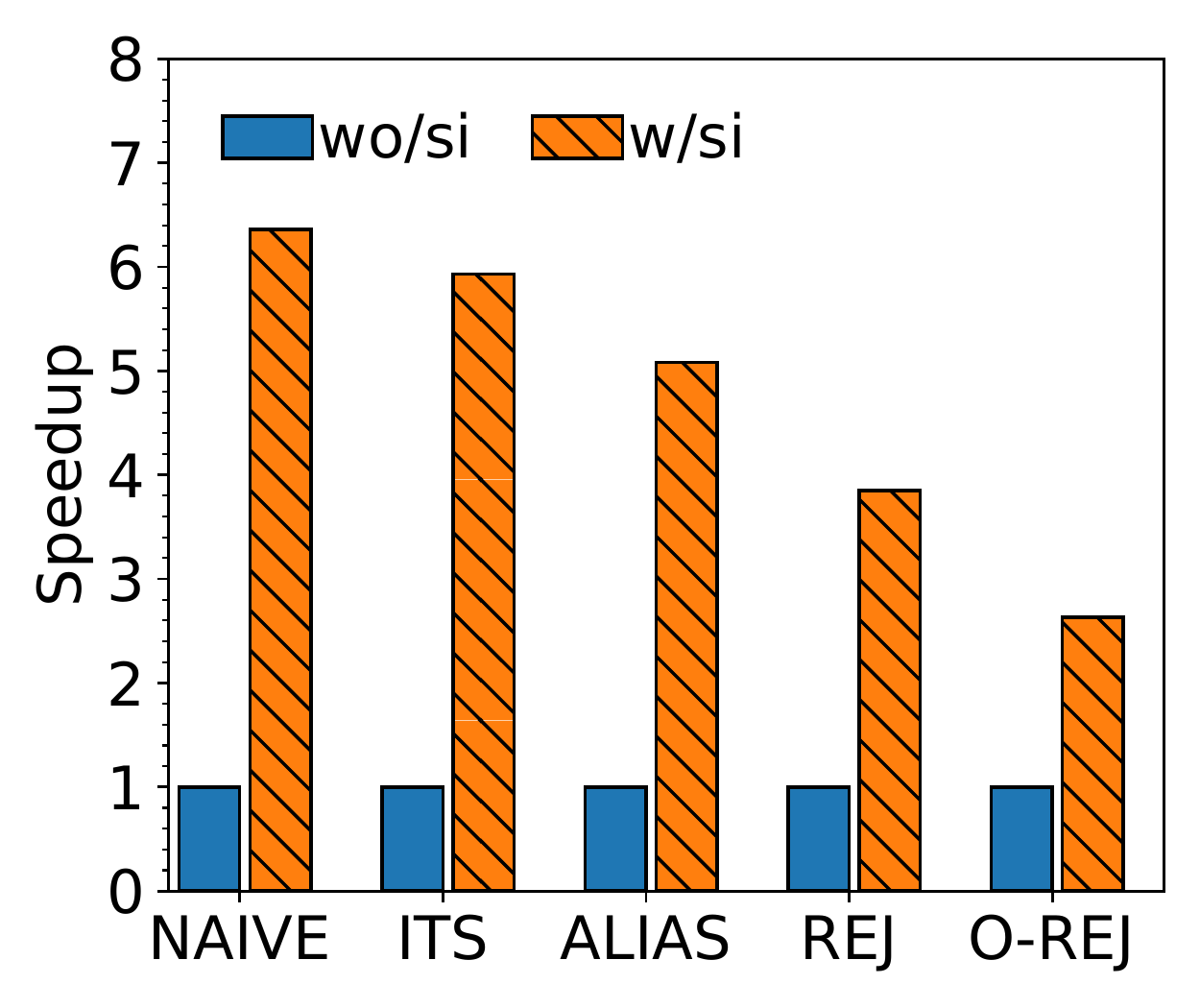}
        \caption{Speedup.}
        \label{fig:lj_sampling_method_speedup}
    \end{subfigure}
    \caption{Vary sampling methods on \emph{lj}.}
    \label{fig:vary_sampling_methods}
\end{figure}

\subsection{Evaluation of Step Interleaving} \label{sec:evaluate_individual_techniques}

We evaluate the effectiveness of step interleaving in this subsection. For brevity,
we use \emph{lj} as the representative graph by default.

\textbf{Varying RW algorithms.} We first evaluate the effectiveness of step interleaving
on different RW algorithms. Figure \ref{fig:vary_rw_algorithms} presents the pipeline slot breakdown
and speedup among the RW algorithms. wo/si and w/si denote ThunderRW without and with
the step interleaving technique, respectively. Enabling step interleaving drastically
reduces memory bound on PPR and DeepWalk, and improves the instruction retirement.
Correspondingly, w/si achieves significant speedup over wo/si in
Figure \ref{fig:lj_speedup}. \sun{The speedup on PPR is lower than that on DeepWalk because PPR issues all queries from a given
vertex and the expected length of a query is 5, which by default exhibits better memory locality than DeepWalk.}
The memory bound on Node2Vec is reduced from around 60\% to 40\%
because the \texttt{Weight} function checks whether two vertices are neighbors with a binary search,
which causes a number of random memory access. The speedup on MetaPath is small because MetaPath
is dynamic and the gather operation dominates the cost at each step.

\textbf{Varying sampling methods.} We next examine the performance of step interleaving on
variant sampling methods. As the gather operation dominates the cost on dynamic random walk,
we focus on unbiased and static random walk. Particularly, we use DeepWalk as the representative
RW algorithm and evaluate it with the five sampling methods in Section \ref{sec:sampling_methods}, respectively.
When adopting \texttt{NAIVE}, we regard DeepWalk as unbiased random walk (i.e., without considering edge weight).
Figure \ref{fig:vary_sampling_methods} presents the pipeline slot breakdown and speedup on \emph{lj} with
variant sampling methods. We can see that the step interleaving technique significantly reduces memory
bound on all the five sampling methods and achieves remarkable speedup. The results demonstrate both
the generality and effectiveness of the step interleaving technique.

\begin{figure}[t]\small
    \setlength{\abovecaptionskip}{0pt}
    \setlength{\belowcaptionskip}{0pt}
    \centering
    \includegraphics[scale=0.23]{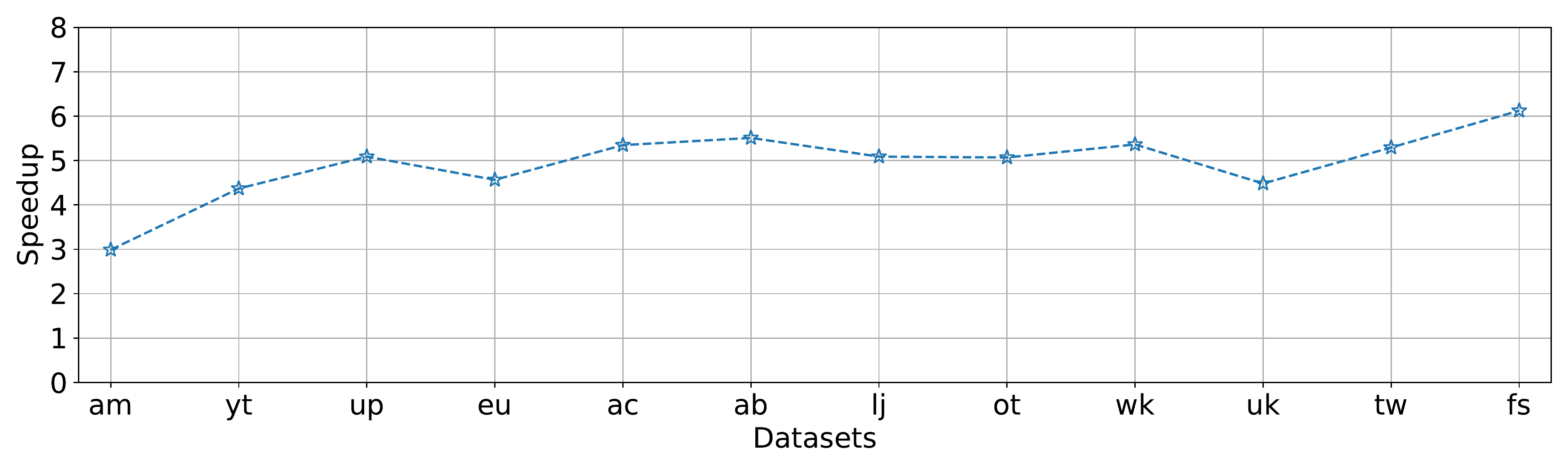}
    \caption{\sun{Vary datasets for DeepWalk.}}
    \label{fig:varying_dataset}
\end{figure}

\sun{\textbf{Varying datasets.} To explore the impact of graph structures on the performance, we evaluate the speedup of enabling
step interleaving for DeepWalk on different datasets. Figure \ref{fig:varying_dataset} presents the experiment results.
The speedup on \emph{am} and \emph{yt} is smaller than that on other graphs because \emph{am} can fit in LLC, and \emph{yt}
is only two times larger than LLC.  The speedup on
\emph{eu} and \emph{uk} is lower than the other graphs that are much larger than LLC since \emph{eu} and \emph{uk} have
dense communities (e.g., \emph{uk} has a clique containing around 1000 vertices \cite{chang2019efficient}), and RW queries
exhibit good memory locality. In contrast, the speedup on \emph{ac} and \emph{ab} is generally higher than the other graphs
because they are bipartite graphs and very sparse, and RW queries have poor memory locality. In summary, the optimization
tends to achieve higher speedup on large and sparse graphs than small graphs and graphs with dense community structures
because RW queries have poorer memory locality on the former one. Nevertheless, the optimization brings up to 3X speedup even on graphs
entirely fitting in LLC (i.e., \emph{am}) since L1 cache is only tens of kilobytes, but around ten times faster than LLC, and the
step interleaving directly fetches the data to L1 cache.} 

\subsection{Scalability Evaluation} \label{sec:scalability_evaluation}

In this section, we evaluate the scalability of ThunderRW. By default, we
execute $10 ^ 7$ RW queries on \emph{lj} with the target length
as 80. Each query starts from a vertex selected from the graph randomly.
We first evaluate the throughput in terms of steps per second
with the number of queries and the length of queries varying, respectively.
In that case, we set the RW as static and use the \texttt{ALIAS}
sampling method as the representative. Next, we evaluate the speedup
with the number of threads varying. When setting the RW as unbiased,
we use the \texttt{NAIVE} sampling method,  while we examine the speedup
on \texttt{ITS}, \texttt{ALIAS}, \texttt{REJ} and \texttt{O-REJ}, respectively,
when setting the RW as static and dynamic.

\begin{figure}[t]\small
    \setlength{\abovecaptionskip}{0pt}
    \setlength{\belowcaptionskip}{0pt}
    \captionsetup[subfigure]{aboveskip=0pt,belowskip=0pt}
    \centering
    \begin{subfigure}[t]{0.23\textwidth}
        \centering
        \includegraphics[scale=0.23]{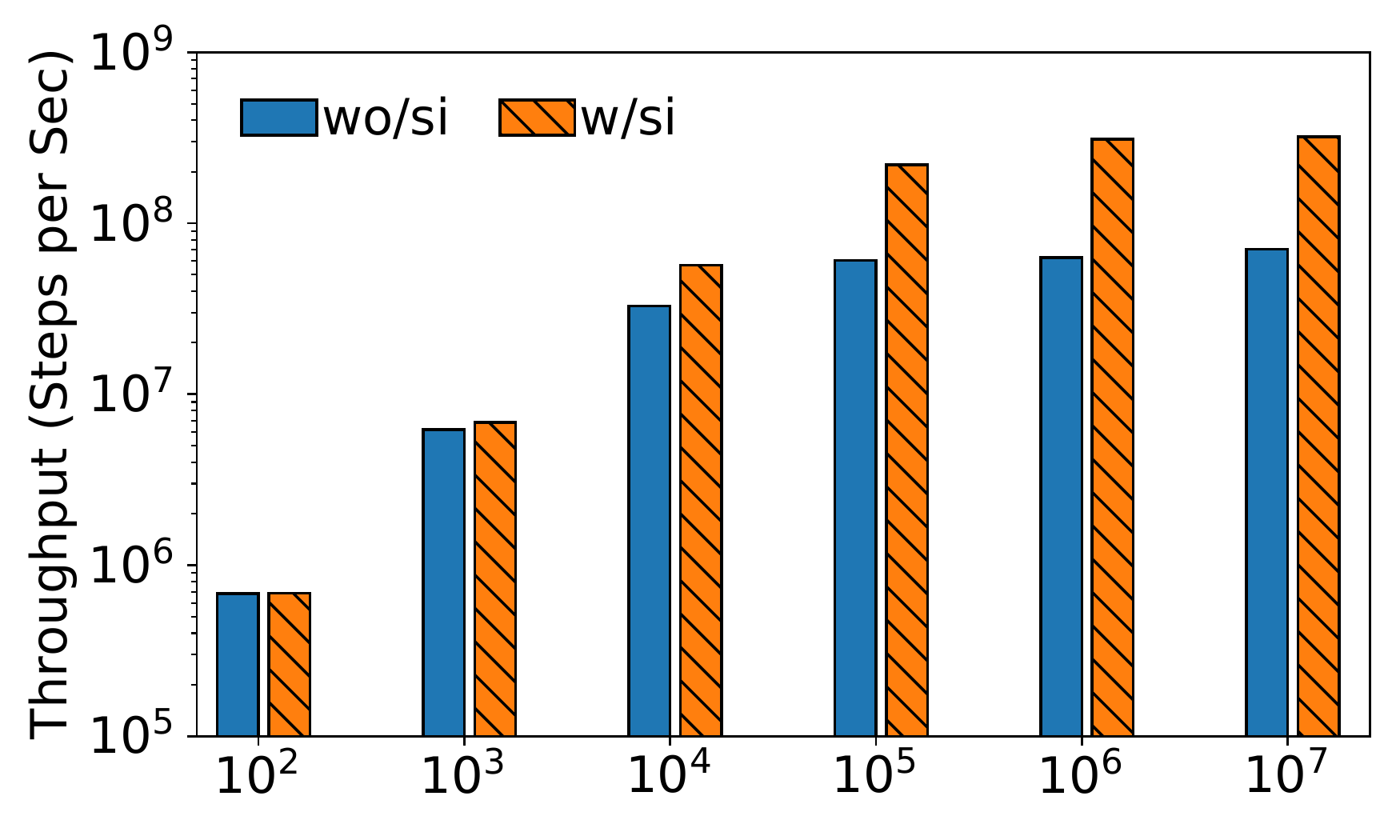}
        \caption{\sun{Varying number of queries.}}
        \label{fig:vary_num_queries}
    \end{subfigure}
    \begin{subfigure}[t]{0.23\textwidth}
        \centering
        \includegraphics[scale=0.23]{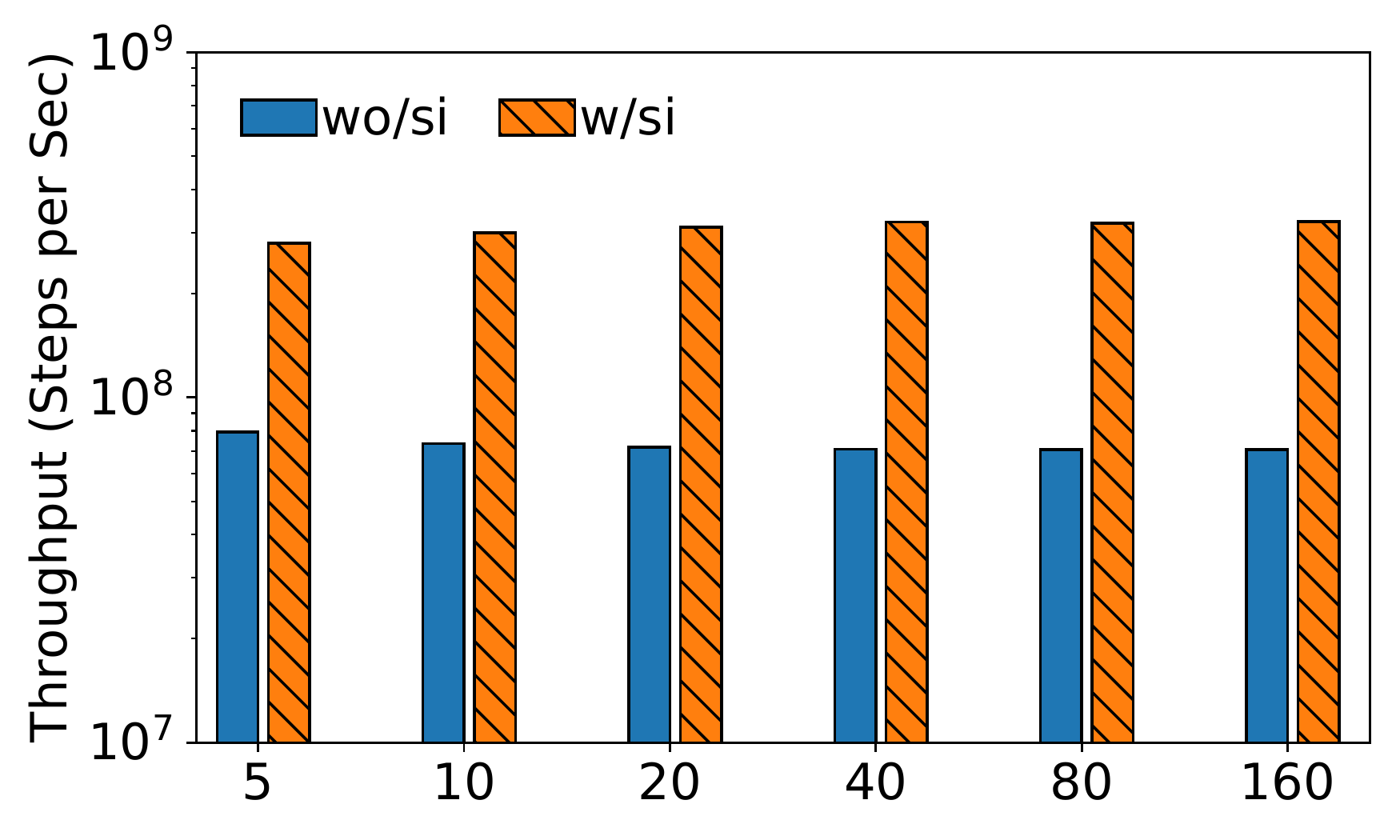}
        \caption{Varying length of queries.}
        \label{fig:vary_length_queries}
    \end{subfigure}
    \caption{Throughput on \emph{lj} with number and length of queries varying.}
    \label{fig:vary_number_length}
\end{figure}

\textbf{Varying number and length of queries.} \sun{Figure \ref{fig:vary_num_queries}
presents the throughput with the number of queries varying from $10 ^ 2$ to
$10 ^ 7$. For $10 ^ 2 - 10 ^ 4$ queries, the execution time
is very short and the start up and shut down time can dominate it. For example, for $10 ^ 2$ queries,
each thread spends less than 0.1 ms on performing random walks,
while the execution time is around 2 ms because of the cost on resource (e.g., memory and threads) initialization and release.
As a result, the benefit of the optimization is limited, and the throughput is lower than that with a large number of queries.}
The throughput is more than $3 \times 10 ^ 8$ and keeps
stable with the number of queries varying from $10 ^ 6$ to $10 ^ 7$.
Figure \ref{fig:vary_length_queries} presents the throughput with the length
of queries varying from 5 to 160. The throughput is steady. In summary,
ThunderRW has good scalability in terms of the number and length of queries.

\begin{figure}[t]\small
    \setlength{\abovecaptionskip}{0pt}
    \setlength{\belowcaptionskip}{0pt}
    \captionsetup[subfigure]{aboveskip=0pt,belowskip=0pt}
    \centering
    \begin{subfigure}[t]{0.23\textwidth}
        \centering
        \includegraphics[scale=0.23]{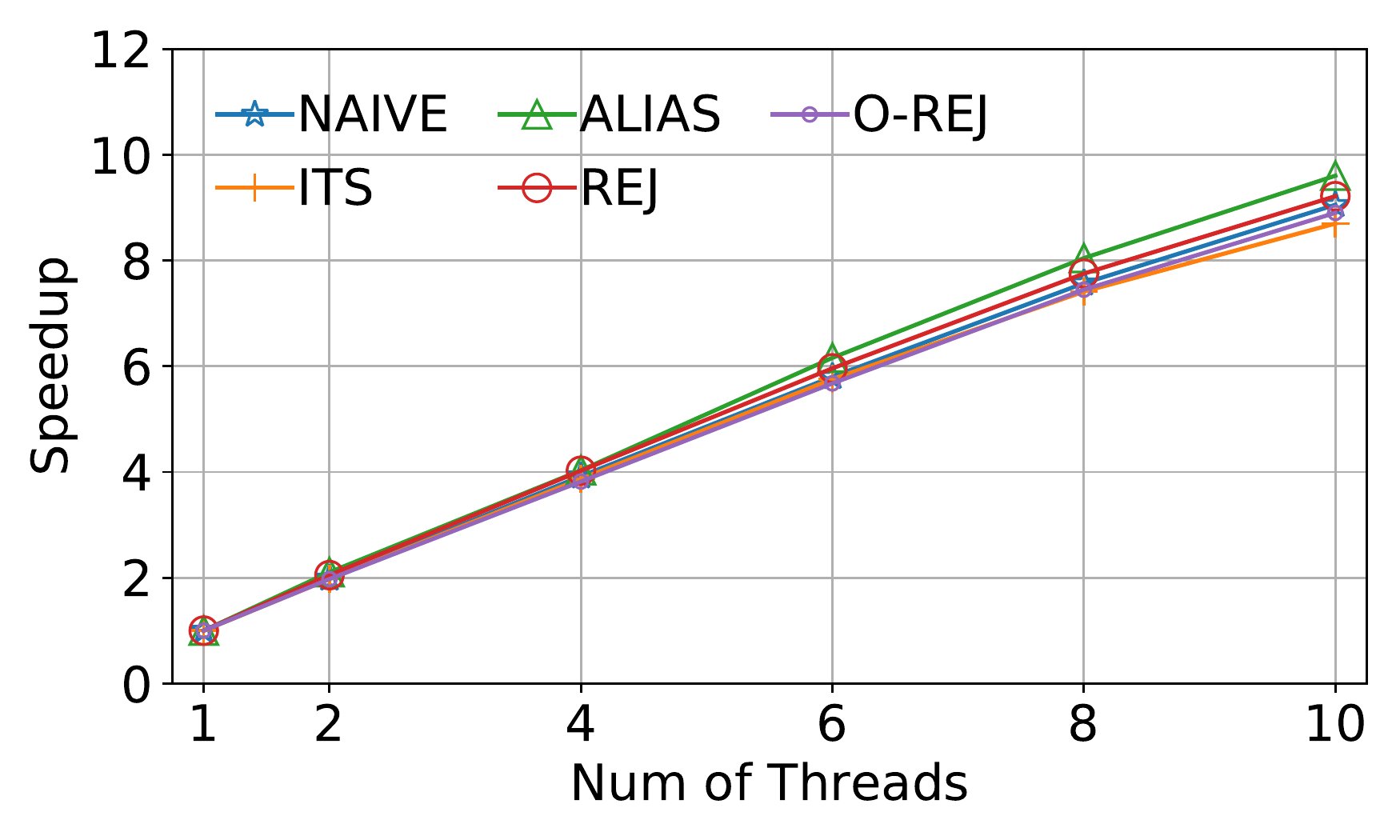}
        \caption{Unbiased/static RW.}
        \label{fig:scalability_on_static}
    \end{subfigure}
    \begin{subfigure}[t]{0.23\textwidth}
        \centering
        \includegraphics[scale=0.23]{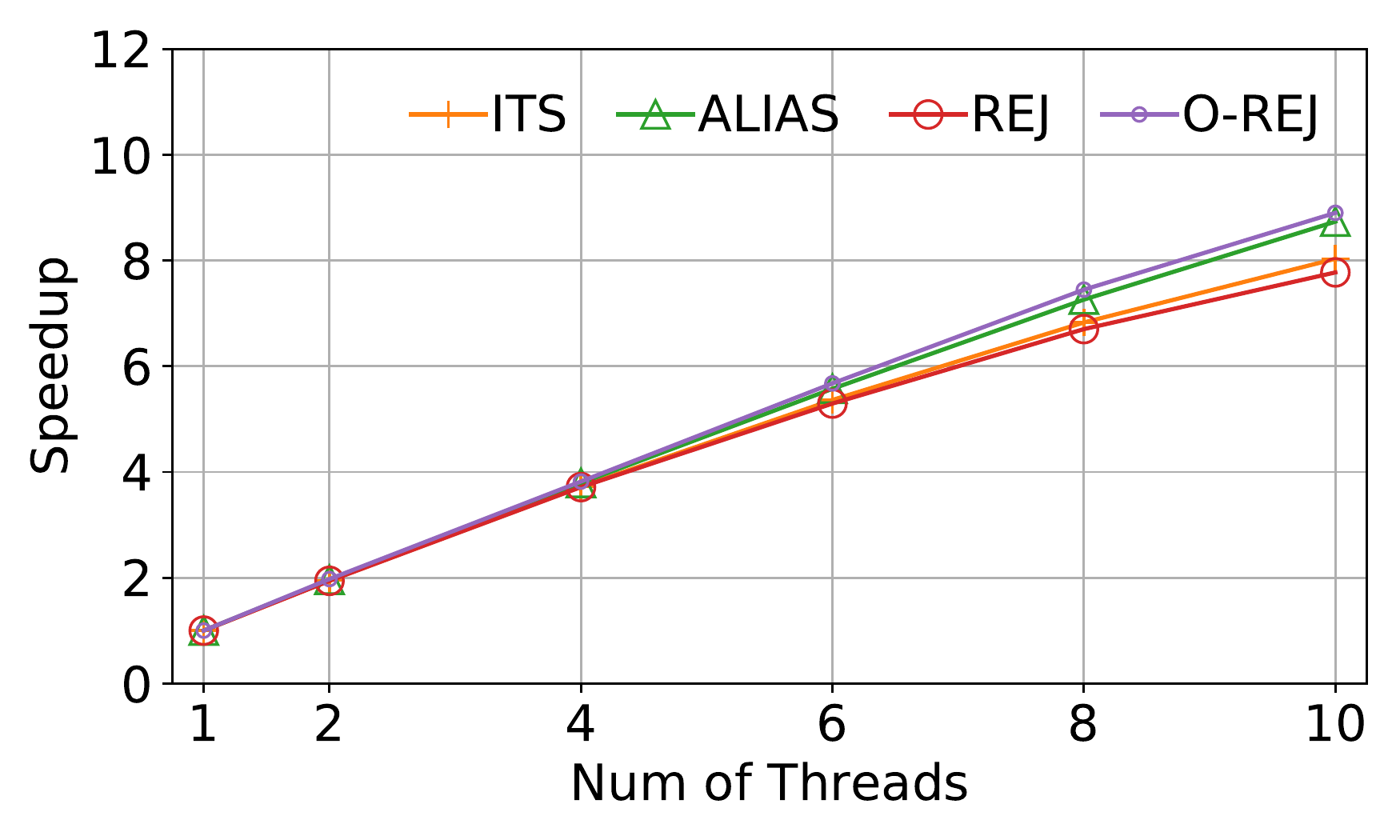}
        \caption{Dynamic RW.}
        \label{fig:scalability_on_dynamic}
    \end{subfigure}
    \caption{Speedup on \emph{lj} with number of threads varying.}
    \label{fig:scalability_with_thread_varying}
\end{figure}

\textbf{Varying number of threads.} Figure \ref{fig:scalability_with_thread_varying}
shows the speedup with the number of threads varying from 1 to 10 \HBS{(i.e., the number of cores in the machine)}. For all
the five sampling methods on unbiased/static RW, ThunderRW achieves
nearly linear speedup with the number of threads as shown in Figure \ref{fig:scalability_on_static}. Particularly, when the number of threads is 10,
the speedup is from 8.8X to 9.6X. Figure \ref{fig:scalability_on_dynamic}
presents the speedup on dynamic RW. The speedup is from
7.8X to 9.0X. Overall, ThunderRW achieves good scalability in terms
of the number of threads.

\subsection{\sun{Generality Evaluation}}

\begin{figure}[h]\small
    \setlength{\abovecaptionskip}{0pt}
    \setlength{\belowcaptionskip}{0pt}
    \captionsetup[subfigure]{aboveskip=0pt,belowskip=0pt}
    \centering
    \begin{subfigure}[t]{0.23\textwidth}
        \centering
        \includegraphics[scale=0.23]{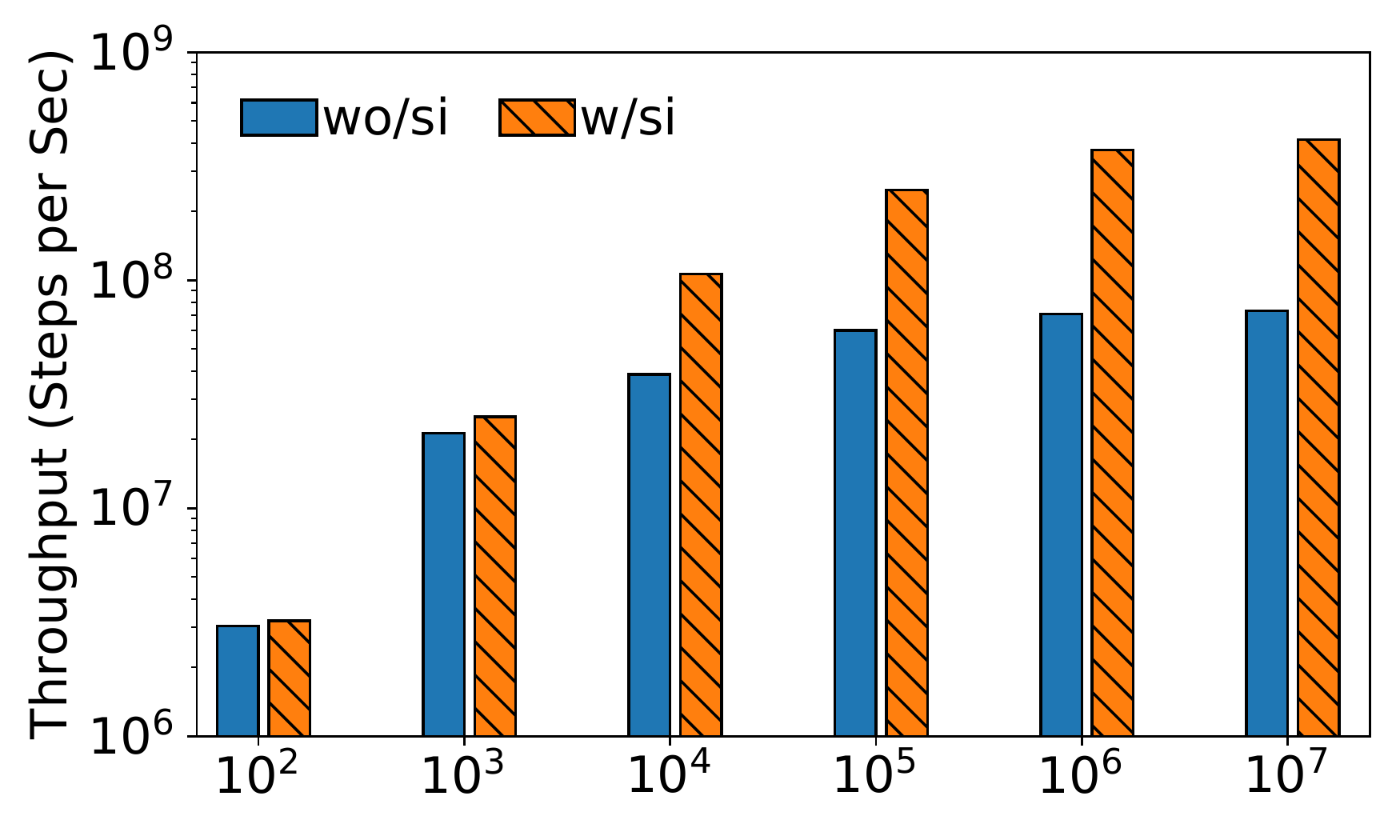}
        \caption{Varying number of queries.}
        \label{fig:vary_number_of_queries_throughput_generality}
    \end{subfigure}
    \begin{subfigure}[t]{0.23\textwidth}
        \centering
        \includegraphics[scale=0.23]{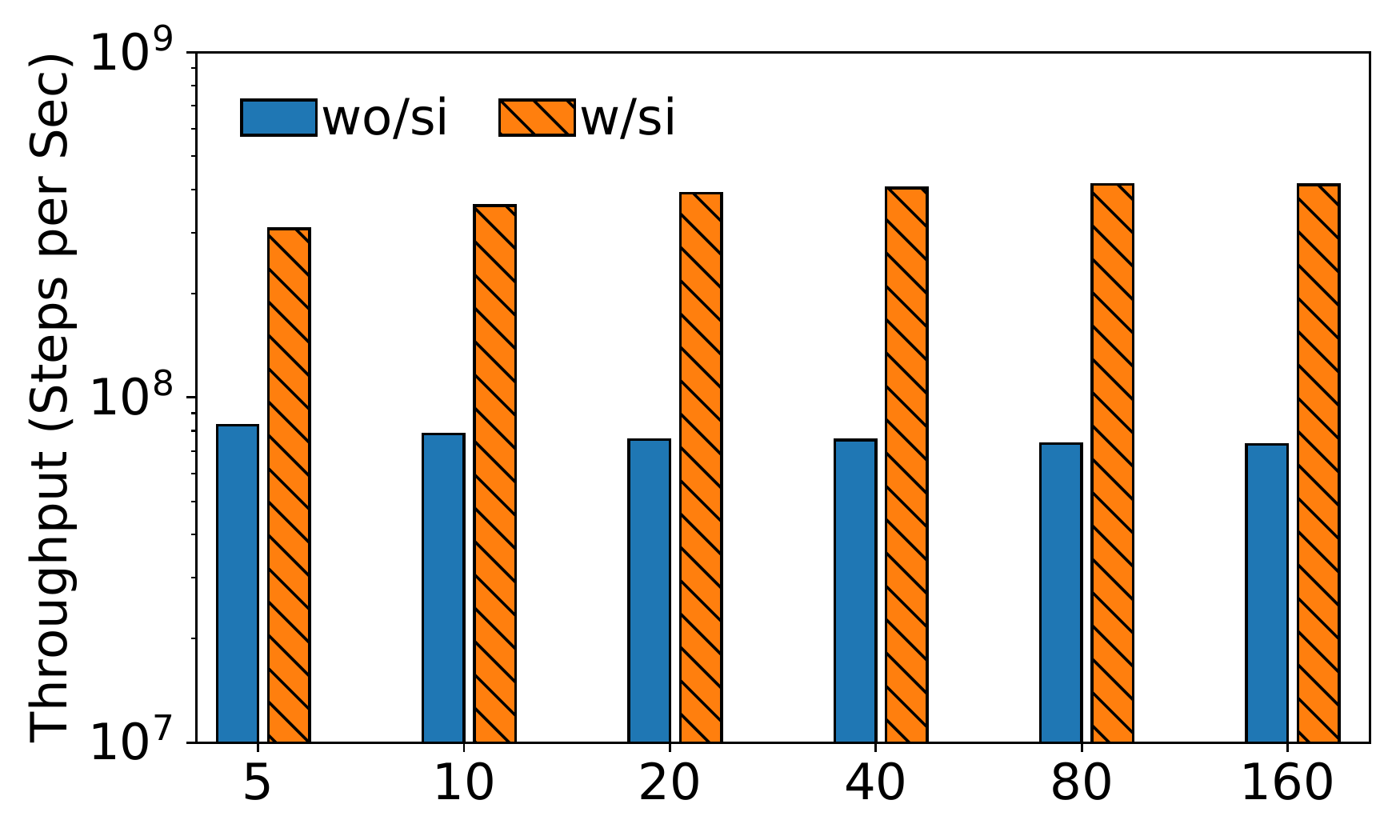}
        \caption{Varying length of queries.}
        \label{fig:vary_length_throughput_generality}
    \end{subfigure}
    \caption{\sun{Throughput on \emph{lj} with number and length of queries varying on processors with different architectures.}}
    \label{fig:throughput_generality}
\end{figure}

\sun{To evaluate the generality of ThunderRW, we repeat the first experiment in Section \ref{sec:scalability_evaluation} on a machine
equipped with an Intel Xeon Gold 6246R CPU, which has 16 physical cores. The sizes of L1, L2 and LLC caches are 32KB, 1MB and 35.75MB,
respectively. Additionally, the CPU is based on the \emph{Cascade Lake} microarchitecture, while that used in other experiments is
based on \emph{Skylake}. As the CPU has 16 physical cores, we set the number of workers as 16.
As shown in Figure \ref{fig:throughput_generality}, enabling the optimization significantly improves the throughput.
Moreover, using the new CPU increases the throughput, for example, when the length of queries is 160,
the throughput grows from $3 \times 10 ^ 8$ to $4.1 \times 10 ^ 8$. The experiment results show that
the techniques proposed in this paper are generic to different architectures.}

\subsection{\sun{Discussions}}

\sun{ThunderRW regards a step of a query as a parallel task unit, which parallelizes the computation from the perspective of queries
instead of the graph data. As RW algorithms consist of massive queries and the cost of moving a step is extremely small (e.g.,
around 34 ns for DeepWalk on \emph{lj}), there are a large number of small parallel tasks, which can be easily parallelized. As such,
the parallelization of ThunderRW can achieve significant speedup over the sequential despite that graph structures are
complex and flexible. Moreover, the sampling method has an important impact on the performance, and therefore providing variant
sampling methods is essential.}

\sun{The step interleaving technique executes different queries alternately to reduce memory bound incurred by random memory accesses.
Its effectiveness is closely related to the memory locality of workloads, which is determined by RW algorithms and graph structures.
In general, the optimization tends to achieve higher speedup on large and sparse graphs than small graphs and graphs with dense community
structures because RW queries have poorer memory locality on the former graphs. Nevertheless, the random memory access is a common issue
for RW algorithms since (1) graphs are much larger than cache sizes; and (2) RW queries wander randomly in the graph.
Thus, the step interleaving can achieve significant speedup even on graphs entirely fitting LLC.

However, the speedup achieved by the step interleaving on high order RW algorithms can be lower than that on first order algorithms.
First, the operations in user-defined functions can introduce random memory accesses.
Despite that, the optimization still brings 1.2-4.3X speedup on Node2Vec. Second,
the \texttt{Gather} operation dominates the cost at each step when performing it in run time.}

\section{Conclusion} \label{sec:conclusion}

In this paper, we propose ThunderRW, an efficient in-memory RW engine on which users can easily implement customized RW algorithms.
We design a step-centric model to abstract the computation from the local view of moving one step of a query.
Based on the model, we propose the step interleaving technique to hide memory access latency by executing multiple queries
alternately. We implement four representative RW algorithms including PPR, DeepWalk, Node2Vec and MetaPath with our framework.
Experimental results show that ThunderRW outperforms state-of-the-art RW frameworks by up to one order of magnitude
and the step interleaving reduces the memory bound from 73.1\% to 15.0\%. Currently, we implement the step interleaving technique in ThunderRW by explicitly and manually storing and restoring states
of each query. An interesting future work is to implement the method with \emph{coroutines}, which is an efficient technique supporting
interleaved execution~\cite{jonathan2018exploiting,psaropoulos2017interleaving,he2020corobase}.


\bibliographystyle{ACM-Reference-Format}
\bibliography{references}


\begin{thebibliography}{70}


\ifx \showCODEN    \undefined \def \showCODEN     #1{\unskip}     \fi
\ifx \showDOI      \undefined \def \showDOI       #1{#1}\fi
\ifx \showISBNx    \undefined \def \showISBNx     #1{\unskip}     \fi
\ifx \showISBNxiii \undefined \def \showISBNxiii  #1{\unskip}     \fi
\ifx \showISSN     \undefined \def \showISSN      #1{\unskip}     \fi
\ifx \showLCCN     \undefined \def \showLCCN      #1{\unskip}     \fi
\ifx \shownote     \undefined \def \shownote      #1{#1}          \fi
\ifx \showarticletitle \undefined \def \showarticletitle #1{#1}   \fi
\ifx \showURL      \undefined \def \showURL       {\relax}        \fi
\providecommand\bibfield[2]{#2}
\providecommand\bibinfo[2]{#2}
\providecommand\natexlab[1]{#1}
\providecommand\showeprint[2][]{arXiv:#2}

\bibitem[\protect\citeauthoryear{{Balkesen}, {Teubner}, {Alonso}, and
  {Özsu}}{{Balkesen} et~al\mbox{.}}{2013}]%
        {6544839}
\bibfield{author}{\bibinfo{person}{C. {Balkesen}}, \bibinfo{person}{J.
  {Teubner}}, \bibinfo{person}{G. {Alonso}}, {and} \bibinfo{person}{M.~T.
  {Özsu}}.} \bibinfo{year}{2013}\natexlab{}.
\newblock \showarticletitle{Main-memory hash joins on multi-core CPUs: Tuning
  to the underlying hardware}. In \bibinfo{booktitle}{\emph{2013 IEEE 29th
  International Conference on Data Engineering (ICDE)}}.
  \bibinfo{pages}{362--373}.
\newblock


\bibitem[\protect\citeauthoryear{Beamer, Asanovic, and Patterson}{Beamer
  et~al\mbox{.}}{2015}]%
        {beamer2015locality}
\bibfield{author}{\bibinfo{person}{Scott Beamer}, \bibinfo{person}{Krste
  Asanovic}, {and} \bibinfo{person}{David Patterson}.}
  \bibinfo{year}{2015}\natexlab{}.
\newblock \showarticletitle{Locality exists in graph processing: Workload
  characterization on an ivy bridge server}. In \bibinfo{booktitle}{\emph{2015
  IEEE International Symposium on Workload Characterization}}. IEEE,
  \bibinfo{pages}{56--65}.
\newblock


\bibitem[\protect\citeauthoryear{Chang}{Chang}{2019}]%
        {chang2019efficient}
\bibfield{author}{\bibinfo{person}{Lijun Chang}.}
  \bibinfo{year}{2019}\natexlab{}.
\newblock \showarticletitle{Efficient maximum clique computation over large
  sparse graphs}. In \bibinfo{booktitle}{\emph{Proceedings of the 25th ACM
  SIGKDD International Conference on Knowledge Discovery \& Data Mining}}.
  \bibinfo{pages}{529--538}.
\newblock


\bibitem[\protect\citeauthoryear{Chen, Ailamaki, Gibbons, and Mowry}{Chen
  et~al\mbox{.}}{2007}]%
        {chen2007improving}
\bibfield{author}{\bibinfo{person}{Shimin Chen}, \bibinfo{person}{Anastassia
  Ailamaki}, \bibinfo{person}{Phillip~B Gibbons}, {and} \bibinfo{person}{Todd~C
  Mowry}.} \bibinfo{year}{2007}\natexlab{}.
\newblock \showarticletitle{Improving hash join performance through
  prefetching}.
\newblock \bibinfo{journal}{\emph{ACM Transactions on Database Systems (TODS)}}
  \bibinfo{volume}{32}, \bibinfo{number}{3} (\bibinfo{year}{2007}),
  \bibinfo{pages}{17--es}.
\newblock


\bibitem[\protect\citeauthoryear{Chen, Gibbons, and Mowry}{Chen
  et~al\mbox{.}}{2001}]%
        {10.1145/376284.375688}
\bibfield{author}{\bibinfo{person}{Shimin Chen}, \bibinfo{person}{Phillip~B.
  Gibbons}, {and} \bibinfo{person}{Todd~C. Mowry}.}
  \bibinfo{year}{2001}\natexlab{}.
\newblock \showarticletitle{Improving Index Performance through Prefetching}.
\newblock \bibinfo{journal}{\emph{SIGMOD Rec.}} \bibinfo{volume}{30},
  \bibinfo{number}{2} (\bibinfo{year}{2001}), \bibinfo{pages}{235–246}.
\newblock


\bibitem[\protect\citeauthoryear{Cochez, Ristoski, Ponzetto, and
  Paulheim}{Cochez et~al\mbox{.}}{2017}]%
        {cochez2017biased}
\bibfield{author}{\bibinfo{person}{Michael Cochez}, \bibinfo{person}{Petar
  Ristoski}, \bibinfo{person}{Simone~Paolo Ponzetto}, {and}
  \bibinfo{person}{Heiko Paulheim}.} \bibinfo{year}{2017}\natexlab{}.
\newblock \showarticletitle{Biased graph walks for RDF graph embeddings}. In
  \bibinfo{booktitle}{\emph{Proceedings of the 7th International Conference on
  Web Intelligence, Mining and Semantics}}. \bibinfo{pages}{1--12}.
\newblock


\bibitem[\protect\citeauthoryear{Coorporation}{Coorporation}{2016}]%
        {coorporation2016intel}
\bibfield{author}{\bibinfo{person}{Intel Coorporation}.}
  \bibinfo{year}{2016}\natexlab{}.
\newblock \bibinfo{title}{Intel 64 and IA-32 architectures optimization
  reference manual}.
\newblock
\newblock


\bibitem[\protect\citeauthoryear{Dai, Li, Tang, and Wang}{Dai
  et~al\mbox{.}}{2018}]%
        {dai2018adversarial}
\bibfield{author}{\bibinfo{person}{Quanyu Dai}, \bibinfo{person}{Qiang Li},
  \bibinfo{person}{Jian Tang}, {and} \bibinfo{person}{Dan Wang}.}
  \bibinfo{year}{2018}\natexlab{}.
\newblock \showarticletitle{Adversarial network embedding}. In
  \bibinfo{booktitle}{\emph{Proceedings of the AAAI Conference on Artificial
  Intelligence}}, Vol.~\bibinfo{volume}{32}.
\newblock


\bibitem[\protect\citeauthoryear{Dhulipala}{Dhulipala}{[n.d.]}]%
        {dhulipalaprovably}
\bibfield{author}{\bibinfo{person}{Laxman Dhulipala}.}
  \bibinfo{year}{[n.d.]}\natexlab{}.
\newblock \showarticletitle{Provably Efficient and Scalable Shared-Memory Graph
  Processing}.
\newblock  (\bibinfo{year}{[n.\,d.]}).
\newblock


\bibitem[\protect\citeauthoryear{Fan, Yu, Xu, Zhou, Luo, Yin, Lu, Cao, and
  Xu}{Fan et~al\mbox{.}}{2018}]%
        {fan2018parallelizing}
\bibfield{author}{\bibinfo{person}{Wenfei Fan}, \bibinfo{person}{Wenyuan Yu},
  \bibinfo{person}{Jingbo Xu}, \bibinfo{person}{Jingren Zhou},
  \bibinfo{person}{Xiaojian Luo}, \bibinfo{person}{Qiang Yin},
  \bibinfo{person}{Ping Lu}, \bibinfo{person}{Yang Cao}, {and}
  \bibinfo{person}{Ruiqi Xu}.} \bibinfo{year}{2018}\natexlab{}.
\newblock \showarticletitle{Parallelizing sequential graph computations}.
\newblock \bibinfo{journal}{\emph{ACM Transactions on Database Systems (TODS)}}
  \bibinfo{volume}{43}, \bibinfo{number}{4} (\bibinfo{year}{2018}),
  \bibinfo{pages}{1--39}.
\newblock


\bibitem[\protect\citeauthoryear{Fogaras, R{\'a}cz, Csalog{\'a}ny, and
  Sarl{\'o}s}{Fogaras et~al\mbox{.}}{2005}]%
        {fogaras2005towards}
\bibfield{author}{\bibinfo{person}{D{\'a}niel Fogaras},
  \bibinfo{person}{Bal{\'a}zs R{\'a}cz}, \bibinfo{person}{K{\'a}roly
  Csalog{\'a}ny}, {and} \bibinfo{person}{Tam{\'a}s Sarl{\'o}s}.}
  \bibinfo{year}{2005}\natexlab{}.
\newblock \showarticletitle{Towards scaling fully personalized pagerank:
  Algorithms, lower bounds, and experiments}.
\newblock \bibinfo{journal}{\emph{Internet Mathematics}} \bibinfo{volume}{2},
  \bibinfo{number}{3} (\bibinfo{year}{2005}), \bibinfo{pages}{333--358}.
\newblock


\bibitem[\protect\citeauthoryear{Fortunato and Hric}{Fortunato and
  Hric}{2016}]%
        {fortunato2016community}
\bibfield{author}{\bibinfo{person}{Santo Fortunato} {and}
  \bibinfo{person}{Darko Hric}.} \bibinfo{year}{2016}\natexlab{}.
\newblock \showarticletitle{Community detection in networks: A user guide}.
\newblock \bibinfo{journal}{\emph{Physics reports}}  \bibinfo{volume}{659}
  (\bibinfo{year}{2016}), \bibinfo{pages}{1--44}.
\newblock


\bibitem[\protect\citeauthoryear{Fu, Lee, and Lei}{Fu et~al\mbox{.}}{2017}]%
        {fu2017hin2vec}
\bibfield{author}{\bibinfo{person}{Tao-yang Fu}, \bibinfo{person}{Wang-Chien
  Lee}, {and} \bibinfo{person}{Zhen Lei}.} \bibinfo{year}{2017}\natexlab{}.
\newblock \showarticletitle{Hin2vec: Explore meta-paths in heterogeneous
  information networks for representation learning}. In
  \bibinfo{booktitle}{\emph{Proceedings of the 2017 ACM on Conference on
  Information and Knowledge Management}}. \bibinfo{pages}{1797--1806}.
\newblock


\bibitem[\protect\citeauthoryear{Gonzalez, Low, Gu, Bickson, and
  Guestrin}{Gonzalez et~al\mbox{.}}{2012}]%
        {gonzalez2012powergraph}
\bibfield{author}{\bibinfo{person}{Joseph~E Gonzalez}, \bibinfo{person}{Yucheng
  Low}, \bibinfo{person}{Haijie Gu}, \bibinfo{person}{Danny Bickson}, {and}
  \bibinfo{person}{Carlos Guestrin}.} \bibinfo{year}{2012}\natexlab{}.
\newblock \showarticletitle{Powergraph: Distributed graph-parallel computation
  on natural graphs}. In \bibinfo{booktitle}{\emph{Presented as part of the
  10th USENIX Symposium on Operating Systems Design and Implementation (OSDI
  12)}}. \bibinfo{pages}{17--30}.
\newblock


\bibitem[\protect\citeauthoryear{Gonzalez, Xin, Dave, Crankshaw, Franklin, and
  Stoica}{Gonzalez et~al\mbox{.}}{2014}]%
        {gonzalez2014graphx}
\bibfield{author}{\bibinfo{person}{Joseph~E Gonzalez},
  \bibinfo{person}{Reynold~S Xin}, \bibinfo{person}{Ankur Dave},
  \bibinfo{person}{Daniel Crankshaw}, \bibinfo{person}{Michael~J Franklin},
  {and} \bibinfo{person}{Ion Stoica}.} \bibinfo{year}{2014}\natexlab{}.
\newblock \showarticletitle{Graphx: Graph processing in a distributed dataflow
  framework}. In \bibinfo{booktitle}{\emph{11th USENIX Symposium on Operating
  Systems Design and Implementation (OSDI 14)}}. \bibinfo{pages}{599--613}.
\newblock


\bibitem[\protect\citeauthoryear{Grover and Leskovec}{Grover and
  Leskovec}{2016}]%
        {grover2016node2vec}
\bibfield{author}{\bibinfo{person}{Aditya Grover} {and} \bibinfo{person}{Jure
  Leskovec}.} \bibinfo{year}{2016}\natexlab{}.
\newblock \showarticletitle{node2vec: Scalable feature learning for networks}.
  In \bibinfo{booktitle}{\emph{Proceedings of the 22nd ACM SIGKDD international
  conference on Knowledge discovery and data mining}}.
  \bibinfo{pages}{855--864}.
\newblock


\bibitem[\protect\citeauthoryear{Guo, Li, Sha, and Tan}{Guo
  et~al\mbox{.}}{2017}]%
        {guo2017parallel}
\bibfield{author}{\bibinfo{person}{Wentian Guo}, \bibinfo{person}{Yuchen Li},
  \bibinfo{person}{Mo Sha}, {and} \bibinfo{person}{Kian-Lee Tan}.}
  \bibinfo{year}{2017}\natexlab{}.
\newblock \showarticletitle{Parallel personalized pagerank on dynamic graphs}.
\newblock \bibinfo{journal}{\emph{Proceedings of the VLDB Endowment}}
  \bibinfo{volume}{11}, \bibinfo{number}{1} (\bibinfo{year}{2017}),
  \bibinfo{pages}{93--106}.
\newblock


\bibitem[\protect\citeauthoryear{He, Lu, and Wang}{He et~al\mbox{.}}{2020}]%
        {he2020corobase}
\bibfield{author}{\bibinfo{person}{Yongjun He}, \bibinfo{person}{Jiacheng Lu},
  {and} \bibinfo{person}{Tianzheng Wang}.} \bibinfo{year}{2020}\natexlab{}.
\newblock \showarticletitle{CoroBase: coroutine-oriented main-memory database
  engine}.
\newblock \bibinfo{journal}{\emph{Proceedings of the VLDB Endowment}}
  \bibinfo{volume}{14}, \bibinfo{number}{3} (\bibinfo{year}{2020}),
  \bibinfo{pages}{431--444}.
\newblock


\bibitem[\protect\citeauthoryear{Hu, Shi, Zhao, and Yu}{Hu
  et~al\mbox{.}}{2018}]%
        {hu2018leveraging}
\bibfield{author}{\bibinfo{person}{Binbin Hu}, \bibinfo{person}{Chuan Shi},
  \bibinfo{person}{Wayne~Xin Zhao}, {and} \bibinfo{person}{Philip~S Yu}.}
  \bibinfo{year}{2018}\natexlab{}.
\newblock \showarticletitle{Leveraging meta-path based context for top-n
  recommendation with a neural co-attention model}. In
  \bibinfo{booktitle}{\emph{Proceedings of the 24th ACM SIGKDD International
  Conference on Knowledge Discovery \& Data Mining}}.
  \bibinfo{pages}{1531--1540}.
\newblock


\bibitem[\protect\citeauthoryear{Jeh and Widom}{Jeh and Widom}{2002}]%
        {jeh2002simrank}
\bibfield{author}{\bibinfo{person}{Glen Jeh} {and} \bibinfo{person}{Jennifer
  Widom}.} \bibinfo{year}{2002}\natexlab{}.
\newblock \showarticletitle{SimRank: a measure of structural-context
  similarity}. In \bibinfo{booktitle}{\emph{Proceedings of the eighth ACM
  SIGKDD international conference on Knowledge discovery and data mining}}.
  \bibinfo{pages}{538--543}.
\newblock


\bibitem[\protect\citeauthoryear{Jha, He, Lu, Cheng, and Huynh}{Jha
  et~al\mbox{.}}{2015}]%
        {10.14778/2735703.2735704}
\bibfield{author}{\bibinfo{person}{Saurabh Jha}, \bibinfo{person}{Bingsheng
  He}, \bibinfo{person}{Mian Lu}, \bibinfo{person}{Xuntao Cheng}, {and}
  \bibinfo{person}{Huynh~Phung Huynh}.} \bibinfo{year}{2015}\natexlab{}.
\newblock \showarticletitle{Improving Main Memory Hash Joins on Intel Xeon Phi
  Processors: An Experimental Approach}.
\newblock \bibinfo{journal}{\emph{Proc. VLDB Endow.}} \bibinfo{volume}{8},
  \bibinfo{number}{6} (\bibinfo{year}{2015}), \bibinfo{pages}{642–653}.
\newblock


\bibitem[\protect\citeauthoryear{Jonathan, Minhas, Hunter, Levandoski, and
  Nishanov}{Jonathan et~al\mbox{.}}{2018}]%
        {jonathan2018exploiting}
\bibfield{author}{\bibinfo{person}{Christopher Jonathan},
  \bibinfo{person}{Umar~Farooq Minhas}, \bibinfo{person}{James Hunter},
  \bibinfo{person}{Justin Levandoski}, {and} \bibinfo{person}{Gor Nishanov}.}
  \bibinfo{year}{2018}\natexlab{}.
\newblock \showarticletitle{Exploiting coroutines to attack the" killer
  nanoseconds"}.
\newblock \bibinfo{journal}{\emph{Proceedings of the VLDB Endowment}}
  \bibinfo{volume}{11}, \bibinfo{number}{11} (\bibinfo{year}{2018}),
  \bibinfo{pages}{1702--1714}.
\newblock


\bibitem[\protect\citeauthoryear{Jun, Wright, Zhang, Xu, et~al\mbox{.}}{Jun
  et~al\mbox{.}}{2018}]%
        {jun2018grafboost}
\bibfield{author}{\bibinfo{person}{Sang-Woo Jun}, \bibinfo{person}{Andy
  Wright}, \bibinfo{person}{Sizhuo Zhang}, \bibinfo{person}{Shuotao Xu},
  {et~al\mbox{.}}} \bibinfo{year}{2018}\natexlab{}.
\newblock \showarticletitle{GraFBoost: Using accelerated flash storage for
  external graph analytics}. In \bibinfo{booktitle}{\emph{2018 ACM/IEEE 45th
  Annual International Symposium on Computer Architecture (ISCA)}}. IEEE,
  \bibinfo{pages}{411--424}.
\newblock


\bibitem[\protect\citeauthoryear{Khorasani, Vora, Gupta, and Bhuyan}{Khorasani
  et~al\mbox{.}}{2014}]%
        {khorasani2014cusha}
\bibfield{author}{\bibinfo{person}{Farzad Khorasani}, \bibinfo{person}{Keval
  Vora}, \bibinfo{person}{Rajiv Gupta}, {and} \bibinfo{person}{Laxmi~N
  Bhuyan}.} \bibinfo{year}{2014}\natexlab{}.
\newblock \showarticletitle{CuSha: vertex-centric graph processing on GPUs}. In
  \bibinfo{booktitle}{\emph{Proceedings of the 23rd international symposium on
  High-performance parallel and distributed computing}}.
  \bibinfo{pages}{239--252}.
\newblock


\bibitem[\protect\citeauthoryear{Kim, Kaldewey, Lee, Sedlar, Nguyen, Satish,
  Chhugani, Di~Blas, and Dubey}{Kim et~al\mbox{.}}{2009}]%
        {10.14778/1687553.1687564}
\bibfield{author}{\bibinfo{person}{Changkyu Kim}, \bibinfo{person}{Tim
  Kaldewey}, \bibinfo{person}{Victor~W. Lee}, \bibinfo{person}{Eric Sedlar},
  \bibinfo{person}{Anthony~D. Nguyen}, \bibinfo{person}{Nadathur Satish},
  \bibinfo{person}{Jatin Chhugani}, \bibinfo{person}{Andrea Di~Blas}, {and}
  \bibinfo{person}{Pradeep Dubey}.} \bibinfo{year}{2009}\natexlab{}.
\newblock \showarticletitle{Sort vs. Hash Revisited: Fast Join Implementation
  on Modern Multi-Core CPUs}.
\newblock  \bibinfo{volume}{2}, \bibinfo{number}{2} (\bibinfo{year}{2009}),
  \bibinfo{pages}{1378–1389}.
\newblock


\bibitem[\protect\citeauthoryear{Kocberber, Falsafi, and Grot}{Kocberber
  et~al\mbox{.}}{2015}]%
        {kocberber2015asynchronous}
\bibfield{author}{\bibinfo{person}{Onur Kocberber}, \bibinfo{person}{Babak
  Falsafi}, {and} \bibinfo{person}{Boris Grot}.}
  \bibinfo{year}{2015}\natexlab{}.
\newblock \showarticletitle{Asynchronous memory access chaining}.
\newblock \bibinfo{journal}{\emph{Proceedings of the VLDB Endowment}}
  \bibinfo{volume}{9}, \bibinfo{number}{4} (\bibinfo{year}{2015}),
  \bibinfo{pages}{252--263}.
\newblock


\bibitem[\protect\citeauthoryear{Kyrola, Blelloch, and Guestrin}{Kyrola
  et~al\mbox{.}}{2012}]%
        {kyrola2012graphchi}
\bibfield{author}{\bibinfo{person}{Aapo Kyrola}, \bibinfo{person}{Guy
  Blelloch}, {and} \bibinfo{person}{Carlos Guestrin}.}
  \bibinfo{year}{2012}\natexlab{}.
\newblock \showarticletitle{Graphchi: Large-scale graph computation on just a
  PC}. In \bibinfo{booktitle}{\emph{Presented as part of the 10th USENIX
  Symposium on Operating Systems Design and Implementation (OSDI 12)}}.
  \bibinfo{pages}{31--46}.
\newblock


\bibitem[\protect\citeauthoryear{Lao, Mitchell, and Cohen}{Lao
  et~al\mbox{.}}{2011}]%
        {lao2011random}
\bibfield{author}{\bibinfo{person}{Ni Lao}, \bibinfo{person}{Tom Mitchell},
  {and} \bibinfo{person}{William Cohen}.} \bibinfo{year}{2011}\natexlab{}.
\newblock \showarticletitle{Random walk inference and learning in a large scale
  knowledge base}. In \bibinfo{booktitle}{\emph{Proceedings of the 2011
  conference on empirical methods in natural language processing}}.
  \bibinfo{pages}{529--539}.
\newblock


\bibitem[\protect\citeauthoryear{Lee, Kim, and Vuduc}{Lee
  et~al\mbox{.}}{2012}]%
        {lee2012prefetching}
\bibfield{author}{\bibinfo{person}{Jaekyu Lee}, \bibinfo{person}{Hyesoon Kim},
  {and} \bibinfo{person}{Richard Vuduc}.} \bibinfo{year}{2012}\natexlab{}.
\newblock \showarticletitle{When prefetching works, when it doesn’t, and
  why}.
\newblock \bibinfo{journal}{\emph{ACM Transactions on Architecture and Code
  Optimization (TACO)}} \bibinfo{volume}{9}, \bibinfo{number}{1}
  (\bibinfo{year}{2012}), \bibinfo{pages}{1--29}.
\newblock


\bibitem[\protect\citeauthoryear{Leskovec and Krevl}{Leskovec and
  Krevl}{2014}]%
        {snapnets}
\bibfield{author}{\bibinfo{person}{Jure Leskovec} {and} \bibinfo{person}{Andrej
  Krevl}.} \bibinfo{year}{2014}\natexlab{}.
\newblock \bibinfo{title}{{SNAP Datasets}: {Stanford} Large Network Dataset
  Collection}.
\newblock \bibinfo{howpublished}{\url{http://snap.stanford.edu/data}}.
\newblock


\bibitem[\protect\citeauthoryear{Li, Yu, Huang, and Cheng}{Li
  et~al\mbox{.}}{2014}]%
        {li2014random}
\bibfield{author}{\bibinfo{person}{Rong-Hua Li}, \bibinfo{person}{Jeffrey~Xu
  Yu}, \bibinfo{person}{Xin Huang}, {and} \bibinfo{person}{Hong Cheng}.}
  \bibinfo{year}{2014}\natexlab{}.
\newblock \showarticletitle{Random-walk domination in large graphs}. In
  \bibinfo{booktitle}{\emph{2014 IEEE 30th International Conference on Data
  Engineering}}. IEEE, \bibinfo{pages}{736--747}.
\newblock


\bibitem[\protect\citeauthoryear{Liu and Huang}{Liu and Huang}{2017}]%
        {liu2017graphene}
\bibfield{author}{\bibinfo{person}{Hang Liu} {and} \bibinfo{person}{H~Howie
  Huang}.} \bibinfo{year}{2017}\natexlab{}.
\newblock \showarticletitle{Graphene: Fine-grained IO management for graph
  computing}. In \bibinfo{booktitle}{\emph{15th USENIX Conference on File and
  Storage Technologies (FAST 17)}}. \bibinfo{pages}{285--300}.
\newblock


\bibitem[\protect\citeauthoryear{Liu, Li, Lui, and Cheng}{Liu
  et~al\mbox{.}}{2016}]%
        {liu2016powerwalk}
\bibfield{author}{\bibinfo{person}{Qin Liu}, \bibinfo{person}{Zhenguo Li},
  \bibinfo{person}{John~CS Lui}, {and} \bibinfo{person}{Jiefeng Cheng}.}
  \bibinfo{year}{2016}\natexlab{}.
\newblock \showarticletitle{Powerwalk: Scalable personalized pagerank via
  random walks with vertex-centric decomposition}. In
  \bibinfo{booktitle}{\emph{Proceedings of the 25th ACM International on
  Conference on Information and Knowledge Management}}.
  \bibinfo{pages}{195--204}.
\newblock


\bibitem[\protect\citeauthoryear{Lofgren}{Lofgren}{2015}]%
        {lofgren2015efficient}
\bibfield{author}{\bibinfo{person}{Peter Lofgren}.}
  \bibinfo{year}{2015}\natexlab{}.
\newblock \showarticletitle{Efficient algorithms for personalized pagerank}.
\newblock \bibinfo{journal}{\emph{arXiv preprint arXiv:1512.04633}}
  (\bibinfo{year}{2015}).
\newblock


\bibitem[\protect\citeauthoryear{Lofgren, Banerjee, Goel, and
  Seshadhri}{Lofgren et~al\mbox{.}}{2014}]%
        {lofgren2014fast}
\bibfield{author}{\bibinfo{person}{Peter~A Lofgren},
  \bibinfo{person}{Siddhartha Banerjee}, \bibinfo{person}{Ashish Goel}, {and}
  \bibinfo{person}{C Seshadhri}.} \bibinfo{year}{2014}\natexlab{}.
\newblock \showarticletitle{FAST-PPR: scaling personalized pagerank estimation
  for large graphs}. In \bibinfo{booktitle}{\emph{Proceedings of the 20th ACM
  SIGKDD international conference on Knowledge discovery and data mining}}.
  \bibinfo{pages}{1436--1445}.
\newblock


\bibitem[\protect\citeauthoryear{Low, Gonzalez, Kyrola, Bickson, Guestrin, and
  Hellerstein}{Low et~al\mbox{.}}{2012}]%
        {low2012distributed}
\bibfield{author}{\bibinfo{person}{Yucheng Low}, \bibinfo{person}{Joseph
  Gonzalez}, \bibinfo{person}{Aapo Kyrola}, \bibinfo{person}{Danny Bickson},
  \bibinfo{person}{Carlos Guestrin}, {and} \bibinfo{person}{Joseph~M
  Hellerstein}.} \bibinfo{year}{2012}\natexlab{}.
\newblock \showarticletitle{Distributed graphlab: A framework for machine
  learning in the cloud}.
\newblock \bibinfo{journal}{\emph{arXiv preprint arXiv:1204.6078}}
  (\bibinfo{year}{2012}).
\newblock


\bibitem[\protect\citeauthoryear{Lu, Sun, Paul, Li, and He}{Lu
  et~al\mbox{.}}{2021}]%
        {lu2021cache}
\bibfield{author}{\bibinfo{person}{Shengliang Lu}, \bibinfo{person}{Shixuan
  Sun}, \bibinfo{person}{Johns Paul}, \bibinfo{person}{Yuchen Li}, {and}
  \bibinfo{person}{Bingsheng He}.} \bibinfo{year}{2021}\natexlab{}.
\newblock \showarticletitle{Cache-Efficient Fork-Processing Patterns on Large
  Graphs}. In \bibinfo{booktitle}{\emph{Proceedings of the 2021 International
  Conference on Management of Data}}. \bibinfo{pages}{1208--1221}.
\newblock


\bibitem[\protect\citeauthoryear{Lv, Gu, Han, Hou, Li, and Liu}{Lv
  et~al\mbox{.}}{2019}]%
        {lv2019adapting}
\bibfield{author}{\bibinfo{person}{Xin Lv}, \bibinfo{person}{Yuxian Gu},
  \bibinfo{person}{Xu Han}, \bibinfo{person}{Lei Hou}, \bibinfo{person}{Juanzi
  Li}, {and} \bibinfo{person}{Zhiyuan Liu}.} \bibinfo{year}{2019}\natexlab{}.
\newblock \showarticletitle{Adapting Meta Knowledge Graph Information for
  Multi-Hop Reasoning over Few-Shot Relations}. In
  \bibinfo{booktitle}{\emph{Proceedings of the 2019 Conference on Empirical
  Methods in Natural Language Processing and the 9th International Joint
  Conference on Natural Language Processing (EMNLP-IJCNLP)}}.
  \bibinfo{pages}{3367--3372}.
\newblock


\bibitem[\protect\citeauthoryear{Malewicz, Austern, Bik, Dehnert, Horn, Leiser,
  and Czajkowski}{Malewicz et~al\mbox{.}}{2010}]%
        {malewicz2010pregel}
\bibfield{author}{\bibinfo{person}{Grzegorz Malewicz},
  \bibinfo{person}{Matthew~H Austern}, \bibinfo{person}{Aart~JC Bik},
  \bibinfo{person}{James~C Dehnert}, \bibinfo{person}{Ilan Horn},
  \bibinfo{person}{Naty Leiser}, {and} \bibinfo{person}{Grzegorz Czajkowski}.}
  \bibinfo{year}{2010}\natexlab{}.
\newblock \showarticletitle{Pregel: a system for large-scale graph processing}.
  In \bibinfo{booktitle}{\emph{Proceedings of the 2010 ACM SIGMOD International
  Conference on Management of data}}. \bibinfo{pages}{135--146}.
\newblock


\bibitem[\protect\citeauthoryear{Marsaglia}{Marsaglia}{1963}]%
        {marsaglia1963generating}
\bibfield{author}{\bibinfo{person}{George Marsaglia}.}
  \bibinfo{year}{1963}\natexlab{}.
\newblock \showarticletitle{Generating discrete random variables in a
  computer}.
\newblock \bibinfo{journal}{\emph{Commun. ACM}} \bibinfo{volume}{6},
  \bibinfo{number}{1} (\bibinfo{year}{1963}), \bibinfo{pages}{37--38}.
\newblock


\bibitem[\protect\citeauthoryear{Mikolov, Chen, Corrado, and Dean}{Mikolov
  et~al\mbox{.}}{2013}]%
        {mikolov2013efficient}
\bibfield{author}{\bibinfo{person}{Tomas Mikolov}, \bibinfo{person}{Kai Chen},
  \bibinfo{person}{Greg Corrado}, {and} \bibinfo{person}{Jeffrey Dean}.}
  \bibinfo{year}{2013}\natexlab{}.
\newblock \showarticletitle{Efficient estimation of word representations in
  vector space}.
\newblock \bibinfo{journal}{\emph{arXiv preprint arXiv:1301.3781}}
  (\bibinfo{year}{2013}).
\newblock


\bibitem[\protect\citeauthoryear{n.d.}{n.d.}{[n.d.]}]%
        {wikidata}
\bibfield{author}{\bibinfo{person}{n.d.}} \bibinfo{year}{[n.d.]}\natexlab{}.
\newblock \bibinfo{title}{Wikimedia Downloads}.
\newblock \bibinfo{howpublished}{\url{https://dumps.wikimedia.org/}}.
\newblock


\bibitem[\protect\citeauthoryear{n.d.}{n.d.}{2018}]%
        {amazon_review}
\bibfield{author}{\bibinfo{person}{n.d.}} \bibinfo{year}{2018}\natexlab{}.
\newblock \bibinfo{title}{Amazon Review Data}.
\newblock
  \bibinfo{howpublished}{\url{https://nijianmo.github.io/amazon/index.html}}.
\newblock


\bibitem[\protect\citeauthoryear{Nguyen, Lenharth, and Pingali}{Nguyen
  et~al\mbox{.}}{2013}]%
        {nguyen2013lightweight}
\bibfield{author}{\bibinfo{person}{Donald Nguyen}, \bibinfo{person}{Andrew
  Lenharth}, {and} \bibinfo{person}{Keshav Pingali}.}
  \bibinfo{year}{2013}\natexlab{}.
\newblock \showarticletitle{A lightweight infrastructure for graph analytics}.
  In \bibinfo{booktitle}{\emph{Proceedings of the twenty-fourth ACM symposium
  on operating systems principles}}. \bibinfo{pages}{456--471}.
\newblock


\bibitem[\protect\citeauthoryear{Page, Brin, Motwani, and Winograd}{Page
  et~al\mbox{.}}{1999}]%
        {page1999pagerank}
\bibfield{author}{\bibinfo{person}{Lawrence Page}, \bibinfo{person}{Sergey
  Brin}, \bibinfo{person}{Rajeev Motwani}, {and} \bibinfo{person}{Terry
  Winograd}.} \bibinfo{year}{1999}\natexlab{}.
\newblock \bibinfo{booktitle}{\emph{The PageRank citation ranking: Bringing
  order to the web.}}
\newblock \bibinfo{type}{{T}echnical {R}eport}. \bibinfo{institution}{Stanford
  InfoLab}.
\newblock


\bibitem[\protect\citeauthoryear{Pandey, Li, Hoisie, Li, and Liu}{Pandey
  et~al\mbox{.}}{[n.d.]}]%
        {pandey2020c}
\bibfield{author}{\bibinfo{person}{Santosh Pandey}, \bibinfo{person}{Lingda
  Li}, \bibinfo{person}{Adolfy Hoisie}, \bibinfo{person}{Xiaoye Li}, {and}
  \bibinfo{person}{Hang Liu}.} \bibinfo{year}{[n.d.]}\natexlab{}.
\newblock \showarticletitle{C-SAW: A Framework for Graph Sampling and Random
  Walk on GPUs}. In \bibinfo{booktitle}{\emph{2020 SC20: International
  Conference for High Performance Computing, Networking, Storage and Analysis
  (SC)}}. IEEE Computer Society, \bibinfo{pages}{780--794}.
\newblock


\bibitem[\protect\citeauthoryear{Perozzi, Al-Rfou, and Skiena}{Perozzi
  et~al\mbox{.}}{2014}]%
        {perozzi2014deepwalk}
\bibfield{author}{\bibinfo{person}{Bryan Perozzi}, \bibinfo{person}{Rami
  Al-Rfou}, {and} \bibinfo{person}{Steven Skiena}.}
  \bibinfo{year}{2014}\natexlab{}.
\newblock \showarticletitle{Deepwalk: Online learning of social
  representations}. In \bibinfo{booktitle}{\emph{Proceedings of the 20th ACM
  SIGKDD international conference on Knowledge discovery and data mining}}.
  \bibinfo{pages}{701--710}.
\newblock


\bibitem[\protect\citeauthoryear{Pr{\v{z}}ulj}{Pr{\v{z}}ulj}{2007}]%
        {prvzulj2007biological}
\bibfield{author}{\bibinfo{person}{Nata{\v{s}}a Pr{\v{z}}ulj}.}
  \bibinfo{year}{2007}\natexlab{}.
\newblock \showarticletitle{Biological network comparison using graphlet degree
  distribution}.
\newblock \bibinfo{journal}{\emph{Bioinformatics}} \bibinfo{volume}{23},
  \bibinfo{number}{2} (\bibinfo{year}{2007}), \bibinfo{pages}{e177--e183}.
\newblock


\bibitem[\protect\citeauthoryear{Psaropoulos, Legler, May, and
  Ailamaki}{Psaropoulos et~al\mbox{.}}{2017}]%
        {psaropoulos2017interleaving}
\bibfield{author}{\bibinfo{person}{Georgios Psaropoulos},
  \bibinfo{person}{Thomas Legler}, \bibinfo{person}{Norman May}, {and}
  \bibinfo{person}{Anastasia Ailamaki}.} \bibinfo{year}{2017}\natexlab{}.
\newblock \showarticletitle{Interleaving with coroutines: a practical approach
  for robust index joins}.
\newblock \bibinfo{journal}{\emph{Proceedings of the VLDB Endowment}}
  \bibinfo{volume}{11}, \bibinfo{number}{CONF} (\bibinfo{year}{2017}),
  \bibinfo{pages}{230--242}.
\newblock


\bibitem[\protect\citeauthoryear{Robert and Casella}{Robert and
  Casella}{2013}]%
        {robert2013monte}
\bibfield{author}{\bibinfo{person}{Christian Robert} {and}
  \bibinfo{person}{George Casella}.} \bibinfo{year}{2013}\natexlab{}.
\newblock \bibinfo{booktitle}{\emph{Monte Carlo statistical methods}}.
\newblock \bibinfo{publisher}{Springer Science \& Business Media}.
\newblock


\bibitem[\protect\citeauthoryear{Rossi and Ahmed}{Rossi and Ahmed}{2015}]%
        {networkrepo}
\bibfield{author}{\bibinfo{person}{Ryan~A. Rossi} {and}
  \bibinfo{person}{Nesreen~K. Ahmed}.} \bibinfo{year}{2015}\natexlab{}.
\newblock \showarticletitle{The Network Data Repository with Interactive Graph
  Analytics and Visualization}. In \bibinfo{booktitle}{\emph{AAAI}}.
\newblock
\urldef\tempurl%
\url{http://networkrepository.com}
\showURL{%
\tempurl}


\bibitem[\protect\citeauthoryear{Schwarz}{Schwarz}{2011}]%
        {schwarz2011darts}
\bibfield{author}{\bibinfo{person}{Keith Schwarz}.}
  \bibinfo{year}{2011}\natexlab{}.
\newblock \showarticletitle{Darts, dice, and coins: Sampling from a discrete
  distribution}.
\newblock \bibinfo{journal}{\emph{Retrieved}} \bibinfo{volume}{3},
  \bibinfo{number}{28} (\bibinfo{year}{2011}), \bibinfo{pages}{2012}.
\newblock


\bibitem[\protect\citeauthoryear{Shao, Huang, Miao, Cui, and Chen}{Shao
  et~al\mbox{.}}{2020}]%
        {shao2020memory}
\bibfield{author}{\bibinfo{person}{Yingxia Shao}, \bibinfo{person}{Shiyue
  Huang}, \bibinfo{person}{Xupeng Miao}, \bibinfo{person}{Bin Cui}, {and}
  \bibinfo{person}{Lei Chen}.} \bibinfo{year}{2020}\natexlab{}.
\newblock \showarticletitle{Memory-Aware Framework for Efficient Second-Order
  Random Walk on Large Graphs}. In \bibinfo{booktitle}{\emph{Proceedings of the
  2020 ACM SIGMOD International Conference on Management of Data}}.
  \bibinfo{pages}{1797--1812}.
\newblock


\bibitem[\protect\citeauthoryear{Shi, Yang, Jin, Xiao, and Yang}{Shi
  et~al\mbox{.}}{2019}]%
        {shi2019realtime}
\bibfield{author}{\bibinfo{person}{Jieming Shi}, \bibinfo{person}{Renchi Yang},
  \bibinfo{person}{Tianyuan Jin}, \bibinfo{person}{Xiaokui Xiao}, {and}
  \bibinfo{person}{Yin Yang}.} \bibinfo{year}{2019}\natexlab{}.
\newblock \showarticletitle{Realtime top-k personalized pagerank over large
  graphs on gpus}.
\newblock \bibinfo{journal}{\emph{Proceedings of the VLDB Endowment}}
  \bibinfo{volume}{13}, \bibinfo{number}{1} (\bibinfo{year}{2019}),
  \bibinfo{pages}{15--28}.
\newblock


\bibitem[\protect\citeauthoryear{Shun and Blelloch}{Shun and Blelloch}{2013}]%
        {shun2013ligra}
\bibfield{author}{\bibinfo{person}{Julian Shun} {and} \bibinfo{person}{Guy~E
  Blelloch}.} \bibinfo{year}{2013}\natexlab{}.
\newblock \showarticletitle{Ligra: a lightweight graph processing framework for
  shared memory}. In \bibinfo{booktitle}{\emph{Proceedings of the 18th ACM
  SIGPLAN symposium on Principles and practice of parallel programming}}.
  \bibinfo{pages}{135--146}.
\newblock


\bibitem[\protect\citeauthoryear{Sun and Han}{Sun and Han}{2013}]%
        {sun2013mining}
\bibfield{author}{\bibinfo{person}{Yizhou Sun} {and} \bibinfo{person}{Jiawei
  Han}.} \bibinfo{year}{2013}\natexlab{}.
\newblock \showarticletitle{Mining heterogeneous information networks: a
  structural analysis approach}.
\newblock \bibinfo{journal}{\emph{Acm Sigkdd Explorations Newsletter}}
  \bibinfo{volume}{14}, \bibinfo{number}{2} (\bibinfo{year}{2013}),
  \bibinfo{pages}{20--28}.
\newblock


\bibitem[\protect\citeauthoryear{Sundaram, Satish, Patwary, Dulloor, Vadlamudi,
  Das, and Dubey}{Sundaram et~al\mbox{.}}{2015}]%
        {sundaram2015graphmat}
\bibfield{author}{\bibinfo{person}{Narayanan Sundaram},
  \bibinfo{person}{Nadathur~Rajagopalan Satish},
  \bibinfo{person}{Md~Mostofa~Ali Patwary}, \bibinfo{person}{Subramanya~R
  Dulloor}, \bibinfo{person}{Satya~Gautam Vadlamudi}, \bibinfo{person}{Dipankar
  Das}, {and} \bibinfo{person}{Pradeep Dubey}.}
  \bibinfo{year}{2015}\natexlab{}.
\newblock \showarticletitle{Graphmat: High performance graph analytics made
  productive}.
\newblock \bibinfo{journal}{\emph{arXiv preprint arXiv:1503.07241}}
  (\bibinfo{year}{2015}).
\newblock


\bibitem[\protect\citeauthoryear{Walker}{Walker}{1977}]%
        {walker1977efficient}
\bibfield{author}{\bibinfo{person}{Alastair~J Walker}.}
  \bibinfo{year}{1977}\natexlab{}.
\newblock \showarticletitle{An efficient method for generating discrete random
  variables with general distributions}.
\newblock \bibinfo{journal}{\emph{ACM Transactions on Mathematical Software
  (TOMS)}} \bibinfo{volume}{3}, \bibinfo{number}{3} (\bibinfo{year}{1977}),
  \bibinfo{pages}{253--256}.
\newblock


\bibitem[\protect\citeauthoryear{Wang, Li, Xie, Xu, and Lui}{Wang
  et~al\mbox{.}}{2020}]%
        {wang2020graphwalker}
\bibfield{author}{\bibinfo{person}{Rui Wang}, \bibinfo{person}{Yongkun Li},
  \bibinfo{person}{Hong Xie}, \bibinfo{person}{Yinlong Xu}, {and}
  \bibinfo{person}{John~CS Lui}.} \bibinfo{year}{2020}\natexlab{}.
\newblock \showarticletitle{GraphWalker: An I/O-Efficient and Resource-Friendly
  Graph Analytic System for Fast and Scalable Random Walks}. In
  \bibinfo{booktitle}{\emph{2020 USENIX Annual Technical Conference (USENIX ATC
  20)}}. \bibinfo{pages}{559--571}.
\newblock


\bibitem[\protect\citeauthoryear{Wang, Yang, Xiao, Wei, and Yang}{Wang
  et~al\mbox{.}}{2017}]%
        {wang2017fora}
\bibfield{author}{\bibinfo{person}{Sibo Wang}, \bibinfo{person}{Renchi Yang},
  \bibinfo{person}{Xiaokui Xiao}, \bibinfo{person}{Zhewei Wei}, {and}
  \bibinfo{person}{Yin Yang}.} \bibinfo{year}{2017}\natexlab{}.
\newblock \showarticletitle{FORA: simple and effective approximate
  single-source personalized pagerank}. In
  \bibinfo{booktitle}{\emph{Proceedings of the 23rd ACM SIGKDD International
  Conference on Knowledge Discovery and Data Mining}}.
  \bibinfo{pages}{505--514}.
\newblock


\bibitem[\protect\citeauthoryear{Wang, Davidson, Pan, Wu, Riffel, and
  Owens}{Wang et~al\mbox{.}}{2016}]%
        {wang2016gunrock}
\bibfield{author}{\bibinfo{person}{Yangzihao Wang}, \bibinfo{person}{Andrew
  Davidson}, \bibinfo{person}{Yuechao Pan}, \bibinfo{person}{Yuduo Wu},
  \bibinfo{person}{Andy Riffel}, {and} \bibinfo{person}{John~D Owens}.}
  \bibinfo{year}{2016}\natexlab{}.
\newblock \showarticletitle{Gunrock: A high-performance graph processing
  library on the GPU}. In \bibinfo{booktitle}{\emph{Proceedings of the 21st ACM
  SIGPLAN Symposium on Principles and Practice of Parallel Programming}}.
  \bibinfo{pages}{1--12}.
\newblock


\bibitem[\protect\citeauthoryear{Wei, He, Xiao, Wang, Shang, and Wen}{Wei
  et~al\mbox{.}}{2018}]%
        {wei2018topppr}
\bibfield{author}{\bibinfo{person}{Zhewei Wei}, \bibinfo{person}{Xiaodong He},
  \bibinfo{person}{Xiaokui Xiao}, \bibinfo{person}{Sibo Wang},
  \bibinfo{person}{Shuo Shang}, {and} \bibinfo{person}{Ji-Rong Wen}.}
  \bibinfo{year}{2018}\natexlab{}.
\newblock \showarticletitle{Topppr: top-k personalized pagerank queries with
  precision guarantees on large graphs}. In
  \bibinfo{booktitle}{\emph{Proceedings of the 2018 International Conference on
  Management of Data}}. \bibinfo{pages}{441--456}.
\newblock


\bibitem[\protect\citeauthoryear{Williams, Waterman, and Patterson}{Williams
  et~al\mbox{.}}{2009}]%
        {williams2009roofline}
\bibfield{author}{\bibinfo{person}{Samuel Williams}, \bibinfo{person}{Andrew
  Waterman}, {and} \bibinfo{person}{David Patterson}.}
  \bibinfo{year}{2009}\natexlab{}.
\newblock \showarticletitle{Roofline: an insightful visual performance model
  for multicore architectures}.
\newblock \bibinfo{journal}{\emph{Commun. ACM}} \bibinfo{volume}{52},
  \bibinfo{number}{4} (\bibinfo{year}{2009}), \bibinfo{pages}{65--76}.
\newblock


\bibitem[\protect\citeauthoryear{Yan, Cheng, Lu, and Ng}{Yan
  et~al\mbox{.}}{2014}]%
        {yan2014blogel}
\bibfield{author}{\bibinfo{person}{Da Yan}, \bibinfo{person}{James Cheng},
  \bibinfo{person}{Yi Lu}, {and} \bibinfo{person}{Wilfred Ng}.}
  \bibinfo{year}{2014}\natexlab{}.
\newblock \showarticletitle{Blogel: A block-centric framework for distributed
  computation on real-world graphs}.
\newblock \bibinfo{journal}{\emph{Proceedings of the VLDB Endowment}}
  \bibinfo{volume}{7}, \bibinfo{number}{14} (\bibinfo{year}{2014}),
  \bibinfo{pages}{1981--1992}.
\newblock


\bibitem[\protect\citeauthoryear{Yang, Zhang, Chen, Ma, Bai, and Jiang}{Yang
  et~al\mbox{.}}{2019}]%
        {yang2019knightking}
\bibfield{author}{\bibinfo{person}{Ke Yang}, \bibinfo{person}{MingXing Zhang},
  \bibinfo{person}{Kang Chen}, \bibinfo{person}{Xiaosong Ma},
  \bibinfo{person}{Yang Bai}, {and} \bibinfo{person}{Yong Jiang}.}
  \bibinfo{year}{2019}\natexlab{}.
\newblock \showarticletitle{Knightking: a fast distributed graph random walk
  engine}. In \bibinfo{booktitle}{\emph{Proceedings of the 27th ACM Symposium
  on Operating Systems Principles}}. \bibinfo{pages}{524--537}.
\newblock


\bibitem[\protect\citeauthoryear{Ye, Zhao, Zhang, Zhu, Xiao, and Wang}{Ye
  et~al\mbox{.}}{2019}]%
        {ye2019improved}
\bibfield{author}{\bibinfo{person}{Zhonglin Ye}, \bibinfo{person}{Haixing
  Zhao}, \bibinfo{person}{Ke Zhang}, \bibinfo{person}{Yu Zhu},
  \bibinfo{person}{Yuzhi Xiao}, {and} \bibinfo{person}{Zhaoyang Wang}.}
  \bibinfo{year}{2019}\natexlab{}.
\newblock \showarticletitle{Improved DeepWalk Algorithm Based on Preference
  Random Walk}. In \bibinfo{booktitle}{\emph{CCF International Conference on
  Natural Language Processing and Chinese Computing}}. Springer,
  \bibinfo{pages}{265--276}.
\newblock


\bibitem[\protect\citeauthoryear{Zhang, Yang, Baghdadi, Kamil, Shun, and
  Amarasinghe}{Zhang et~al\mbox{.}}{2018}]%
        {zhang2018graphit}
\bibfield{author}{\bibinfo{person}{Yunming Zhang}, \bibinfo{person}{Mengjiao
  Yang}, \bibinfo{person}{Riyadh Baghdadi}, \bibinfo{person}{Shoaib Kamil},
  \bibinfo{person}{Julian Shun}, {and} \bibinfo{person}{Saman Amarasinghe}.}
  \bibinfo{year}{2018}\natexlab{}.
\newblock \showarticletitle{Graphit: A high-performance graph dsl}.
\newblock \bibinfo{journal}{\emph{Proceedings of the ACM on Programming
  Languages}} \bibinfo{volume}{2}, \bibinfo{number}{OOPSLA}
  (\bibinfo{year}{2018}), \bibinfo{pages}{1--30}.
\newblock


\bibitem[\protect\citeauthoryear{Zhong and He}{Zhong and He}{2013}]%
        {zhong2013medusa}
\bibfield{author}{\bibinfo{person}{Jianlong Zhong} {and}
  \bibinfo{person}{Bingsheng He}.} \bibinfo{year}{2013}\natexlab{}.
\newblock \showarticletitle{Medusa: Simplified graph processing on GPUs}.
\newblock \bibinfo{journal}{\emph{IEEE Transactions on Parallel and Distributed
  Systems}} \bibinfo{volume}{25}, \bibinfo{number}{6} (\bibinfo{year}{2013}),
  \bibinfo{pages}{1543--1552}.
\newblock


\bibitem[\protect\citeauthoryear{Zhou, Niu, and Chen}{Zhou
  et~al\mbox{.}}{2018}]%
        {zhou2018efficient}
\bibfield{author}{\bibinfo{person}{Dongyan Zhou}, \bibinfo{person}{Songjie
  Niu}, {and} \bibinfo{person}{Shimin Chen}.} \bibinfo{year}{2018}\natexlab{}.
\newblock \showarticletitle{Efficient graph computation for Node2Vec}.
\newblock \bibinfo{journal}{\emph{arXiv preprint arXiv:1805.00280}}
  (\bibinfo{year}{2018}).
\newblock


\bibitem[\protect\citeauthoryear{Zhu, Chen, Zheng, and Ma}{Zhu
  et~al\mbox{.}}{2016}]%
        {zhu2016gemini}
\bibfield{author}{\bibinfo{person}{Xiaowei Zhu}, \bibinfo{person}{Wenguang
  Chen}, \bibinfo{person}{Weimin Zheng}, {and} \bibinfo{person}{Xiaosong Ma}.}
  \bibinfo{year}{2016}\natexlab{}.
\newblock \showarticletitle{Gemini: A computation-centric distributed graph
  processing system}. In \bibinfo{booktitle}{\emph{12th USENIX Symposium on
  Operating Systems Design and Implementation (OSDI 16)}}.
  \bibinfo{pages}{301--316}.
\newblock


\end{thebibliography}

\appendix

\section{Other Profiling Results}

In this section, we evaluate the impact of varying the length and the number of queries, respectively.
We use a micro benchmark that assembles the access pattern of RWs and also we can control the parameters easily.
Particularly, we set the number of queries as $10 ^ 7$ and configure the target length as 80 by default. Each query starts from a vertex
randomly selected from the graph. We use the \texttt{ALIAS} sampling method to perform the queries.
We first evaluate the impact of varying the length from 5 to 160, and then examine the performance of varying the number of queries
from \sun{$10^2$ to $10^8$}.

Tables \ref{tab:vary_length} and \ref{tab:vary_num_queries} present the results with the length of queries varying from 5 to 160 and the
number of queries varying from \sun{$10 ^ 2$ to $10 ^ 8$} on the \emph{livejournal} graph, respectively. We can see that the memory bound
is consistently above 60\% despite the variance in the length and number of queries. With the length (or the number of queries) increasing,
the memory bound grows slightly. The memory bandwidth utilization is also far from the maximum bandwidth in all test cases.
In summary the in-memory computation of RW algorithms suffers severe performance
issues due to memory stalls caused by cache misses and under-utilizes the memory bandwidth
regardless of the length and number of queries.

\begin{table}[h]
\footnotesize
    \setlength{\abovecaptionskip}{0pt}
    \setlength{\belowcaptionskip}{0pt}
\caption{Pipeline slot breakdown and memory bandwidth with the length of queries varying.}
\label{tab:vary_length}
\begin{tabular}{c|c|c|c|c|c|c}
\hline
\textbf{\begin{tabular}[c]{@{}c@{}}Length of\\ Queries\end{tabular}} & \textbf{\begin{tabular}[c]{@{}c@{}}Front\\ End\end{tabular}} & \textbf{\begin{tabular}[c]{@{}c@{}}Bad\\ Spec\end{tabular}} & \textbf{Core} & \textbf{Memory} & \textbf{Retiring} & \textbf{\begin{tabular}[c]{@{}c@{}}Memory\\ Bandwidth\end{tabular}} \\ \hline\hline
5               & 3.6\%                                                        & 5.5\%                                                       & 16.6\%        & 61.3\%          & 13.0\%            & 7.7GB/s                                                             \\ \hline
10              & 2.7\%                                                        & 4.0\%                                                       & 18.5\%        & 63.4\%          & 11.2\%            & 6.6GB/s                                                             \\ \hline
20              & 2.7\%                                                        & 4.1\%                                                       & 18.1\%        & 64.0\%          & 11.1\%            & 6.0GB/s                                                             \\ \hline
40              & 2.5\%                                                        & 4.0\%                                                       & 18.1\%        & 64.5\%          & 10.9\%            & 5.8GB/s                                                             \\ \hline
80              & 2.3\%                                                        & 3.7\%                                                       & 18.6\%        & 64.8\%          & 10.6\%            & 5.6GB/s                                                             \\ \hline
160             & 2.3\%                                                        & 3.6\%                                                       & 18.5\%        & 65\%            & 10.5\%            & 5.6GB/s                                                             \\ \hline
\end{tabular}
\end{table}

\begin{table}[h]
\footnotesize
    \setlength{\abovecaptionskip}{0pt}
    \setlength{\belowcaptionskip}{0pt}
\caption{Pipeline slot breakdown and memory bandwidth with the number of queries varying.}
\label{tab:vary_num_queries}
\begin{tabular}{c|c|c|c|c|c|c}
\hline
\textbf{\begin{tabular}[c]{@{}c@{}}Num of\\ Queries\end{tabular}} & \textbf{\begin{tabular}[c]{@{}c@{}}Front\\ End\end{tabular}} & \textbf{\begin{tabular}[c]{@{}c@{}}Bad\\ Spec\end{tabular}} & \textbf{Core} & \textbf{Memory} & \textbf{Retiring} & \textbf{\begin{tabular}[c]{@{}c@{}}Memory\\ Bandwidth\end{tabular}} \\ \hline\hline
\sun{$10 ^ 2$}                          & \sun{4.1\%}         & \sun{2.6\%}         & \sun{16.5\%}           & \sun{66.4\%}     & \sun{10.4\%}   & \sun{5.9GB/s} \\ \hline
$10 ^ 3$                                                          & 4.5\%                                                        & 7.4\%                                                       & 12.1\%        & 63.8\%          & 12.2\%            & 8.0GB/s                                                             \\ \hline
$10 ^ 4$                                                          & 4.4\%                                                        & 6.9\%                                                       & 12.7\%        & 64.3\%          & 11.8\%            & 6.6GB/s                                                             \\ \hline
$10 ^ 5$                                                          & 4.0\%                                                        & 6.2\%                                                       & 16.5\%        & 60.9\%          & 12.4\%            & 6.0GB/s                                                             \\ \hline
$10 ^ 6$                                                          & 2.7\%                                                        & 4.1\%                                                       & 19.0\%        & 63.2\%          & 11.0\%            & 5.8GB/s                                                             \\ \hline
$10 ^ 7$                                                          & 2.3\%                                                        & 3.7\%                                                       & 18.6\%        & 64.8\%          & 10.6\%            & 5.6GB/s                                                             \\ \hline
$10 ^ 8$                                                          & 2.3\%                                                        & 3.6\%                                                       & 18.5\%        & 65.1\%          & 10.5\%            & 5.6GB/s                                                             \\ \hline
\end{tabular}
\end{table}

\section{Other Implementation Details}

In this section, we present the implementation details of the stage switch mechanism,
the graph storage, the walker management and the input/output of the framework.

\textbf{Stage switch.} Continuing with Example \ref{exmp:sdg}, we use \texttt{Move} with
the \texttt{REJ} sampling method to demonstrate the implementation of stage switch.
Algorithm \ref{algo:move_with_rej} presents the details where $\mathbb{Q}'$
is a group of queries and $\mathbb{C}$ maintains the transition probability
$C$ for each $Q \in \mathbb{Q}'$. Line 2 creates a task ring $TR$ with
$|\mathbb{Q}'|$ slots. Each slot records states of a query $Q \in \mathbb{Q}'$.
The load operations are replaced with the \texttt{PREFETCH} operations. We
process a non-cycle stage $S$ in SDG with a for loop where all queries evaluate
$S$ one by one. For example, Lines 4-5 deal with $S_0$ in which we
fetch the degree of $Q.cur$ for each $Q \in \mathbb{Q}'$. The \texttt{Search}
function handles cycle stages. Line 16 first creates a search ring $SR$ with
$k'$ slots to process cycle stages. If a slot $R \in SR$ is empty and
there are queries in $TR$ not submitted to $SR$ (Line 20),
Lines 21-23 submit a query $Q$ to $SR$ and initialize the slot $R$.
If the stage is $S_2$, Lines 25-27 perform the operations in $S_2$.
Moreover, Line 28 sets $R.S$ to $S_3$ and stores $x,y$ because
the next stage is $S_3$ and $S_3$ depends on the value of $x, y$ according to
SDG. When $S_3$ is completed, we write $x$ to $TR$ because
$S_4$ consumes it as shown in SDG. Lines 18-34 repeat the process until
all queries jump out the cycle. Lines 9-13 continue the computation
with values generated by \texttt{Search}. Line 14 returns $\mathbb{U}$ that
maintains the selected edge for each query.

\setlength{\textfloatsep}{0pt}
\begin{algorithm}[t]
	\caption{\texttt{Move} with \texttt{REJ} using Step Interleaving}
	\label{algo:move_with_rej}
	\footnotesize
	\SetKwProg{func}{Function}{}{}
	\SetKwFunction{Move}{Move}
	\SetKwFunction{Search}{Search}
	\func{\Move{$G, \mathbb{Q}', \mathbb{C}$}}{
	    Initialize a task ring $TR$ each slot of which corresponds to $Q \in \mathbb{Q}'$\;
	    Initialize $\mathbb{U}$ as $\{\}$ to store the selected edge for $Q \in \mathbb{Q}'$\;
	    \tcc{Stage $S_0$.}
	    \ForEach{$Q \in \mathbb{Q}'$}{
	        \texttt{PREFETCH} $d_v$ where $v = Q.cur$\;
	    }
	    \tcc{Stage $S_1$.}
	    \ForEach{$Q \in \mathbb{Q}'$}{
	        \texttt{PREFETCH} $p_v ^ *$ where $v = Q.cur$\;
	    }
	    \Search{$\mathbb{C}, TR$}\;
	    \tcc{Stage $S_4$.}
	    \ForEach{$Q \in \mathbb{Q}'$}{
	        \texttt{PREFETCH} $E_v[TR[Q].x]$ where $v = Q.cur$\;
	    }
	    \tcc{Stage $S_5$.}
	    \ForEach{$Q \in \mathbb{Q}'$}{
	        Add $v'$ to $Q$ where $v = Q.cur$ and $e(v, v') = E_v[TR[Q].x]$\;
	        Set $\mathbb{U}[Q]$ to $e(v, v')$\;
	    }
	    
	    \KwRet $\mathbb{U}$\;
	}
	
	\func{\Search{$\mathbb{C}, TR$}}{
	    Initialize a search ring $SR$ with $k'$ slots where $k' \leqslant |TR|$\;
	    $submitted, completed, index \leftarrow 0$\;
	    \While{$completed < |TR|$}{
	        $R \leftarrow SR[index]$\;
	        \If{$R.S = null$ and $submitted < |TR|$} {
	            Get next slot $R' \in TR$ and set $v$ to $R'.Q.cur$\;
	            Set $R.Q$, $R.S$, $R.d$ and $R.p ^ *$ to $R'.Q$, $S_2$, $d_v$ and $p_v ^ *$, respectively\;
	            $submitted \leftarrow submitted + 1$\;
	        }
	        \tcc{Stage $S_2$.}
	        \ElseIf{$R.S = S_2$}{
	             Generate an int random number $x$ in $[0, R.d)$\;
	             Generate a real random number $y$ in $[0, R.p ^ *)$\;
	             \texttt{PREFETCH} $C[x]$ where $C = \mathbb{C}[R.Q]$\;
	             Set $R.S$, $R.x$ and $R.y$ to $S_3$, $x$ and $y$, respectively\;
	        }
	        \tcc{Stage $S_3$.}
	        \ElseIf{$R.S = S_3$}{
	            \lIf{$R.y > C[R.x]$}{Set $R.S$ to $S_2$}
	            \Else{
	                Set $R.S$ and $TR[R.Q].x$ to \emph{null} and $R.x$, respectively\;
	                $completed \leftarrow completed + 1$\;
	            }
	        }
	        $index \leftarrow (index + 1) \mod k'$\;
	    }
	}
\end{algorithm}

\textbf{Graph storage.} We store the graph $G$ in \emph{compressed sparse row} (CSR)
where $G$ consists of an array of vertices and an array of edges. Each vertex in CSR
points to the start of its adjacent edges in the edge array. Moreover, we associate
the edge label and edge weight to each edge and store them as two arrays, respectively.

\textbf{Walker management.} Given a set $\mathbb{Q}$ of random walk queries,
we assign an unique ID from $0$ to $|\mathbb{Q}| - 1$ to each of them. For a
query $Q \in \mathbb{Q}$, we maintain the query ID, the source vertex,
the length of $Q$ and a pointer linking to the payload (e.g., the walk path).
In addition, user can customize the data associating with each query.

\sun{\textbf{Input and output.} ThunderRW provides APIs for users to specify the source
vertices of RW queries and the number of queries from each source. For example, we can
start a RW query from each vertex in $G$ for DeepWalk, while issue a number of queries
from a given vertex for single source PPR.}

\sun{ThunderRW outputs the walk path for each RW query. The output can be either consumed by down streaming tasks on the fly or
stored for the future usage. The former case consumes a small amount of memory space, whereas the memory cost of the latter
can be $O(\sum_{Q \in \mathbb{Q}}|Q|)$. Fortunately, it is unnecessary to maintain all walks in memory in practical implementation.
Instead, we can use the classic double buffering mechanism to efficiently dump the output to the disk in batch..}

\sun{Specifically, one is used to write results to the disk, while the other records new results generated
by the engine. When the second one is full, we swap the role of the two buffers. In this way, the I/O cost can be easily and seamlessly
overlapped by the computation because (1) modern computers support direct memory access (DMA), which transfers data independent of CPUs,
and operating systems provide simple APIs for async I/O programming (e.g., \emph{aio\_write} in Linux); and (2) the time on filling a
buffer is much longer than that on writing to disks because of the rapid advancement of storage hardwares.
For example, the time on filling 2 GB buffer by the engine is around 1.79 second in
our test bed (equipped with Samsung PM981 NVMe SSD), which can be completely stored to disk in around 1.20 second.
Moreover, the 980 PRO series with PCIe-4.0 achieve up to 5100MB/second sequential write speed, which
can output 2 GB data in around 0.4 second.}

\section{Supplement Experiments}

\subsection{Tuning Ring Sizes}

\sun{\textbf{Time on tuning ring sizes.} Table \ref{tab:tuning_ring_size} presents the time on tuning the ring size.
We can see that the tuning process is very
efficient. Even for \emph{fs} having more than 1.8 billion edges, the tuning takes around four minutes, whereas the tuning on most of the graphs
takes less than one minute.}

\begin{table}[h]
\footnotesize
    \setlength{\abovecaptionskip}{0pt}
    \setlength{\belowcaptionskip}{0pt}
\caption{\sun{The time on tuning ring sizes (seconds).}}
\label{tab:tuning_ring_size}
\begin{tabular}{c|cccccc}
\hline
\textbf{Dataset} & \textit{am} & \textit{yt} & \textit{up} & \textit{eu} & \textit{ac} & \textit{ab} \\ \hline
\textbf{Time}    & 0.87        & 2.67        & 9.45        & 2.55        & 35.12       & 39.23       \\ \hline\hline
\textbf{Dataset} & \textit{lj} & \textit{ot} & \textit{wk} & \textit{uk} & \textit{tw} & \textit{fs} \\ \hline
\textbf{Time}    & 13.19       & 9.82        & 132.4       & 51.86       & 156.37      & 241.44      \\ \hline
\end{tabular}
\end{table}

\begin{figure}[t]\small
    \setlength{\abovecaptionskip}{0pt}
    \setlength{\belowcaptionskip}{0pt}
    \captionsetup[subfigure]{aboveskip=0pt,belowskip=0pt}
    \centering
    \begin{subfigure}[t]{0.23\textwidth}
        \centering
        \includegraphics[scale=0.23]{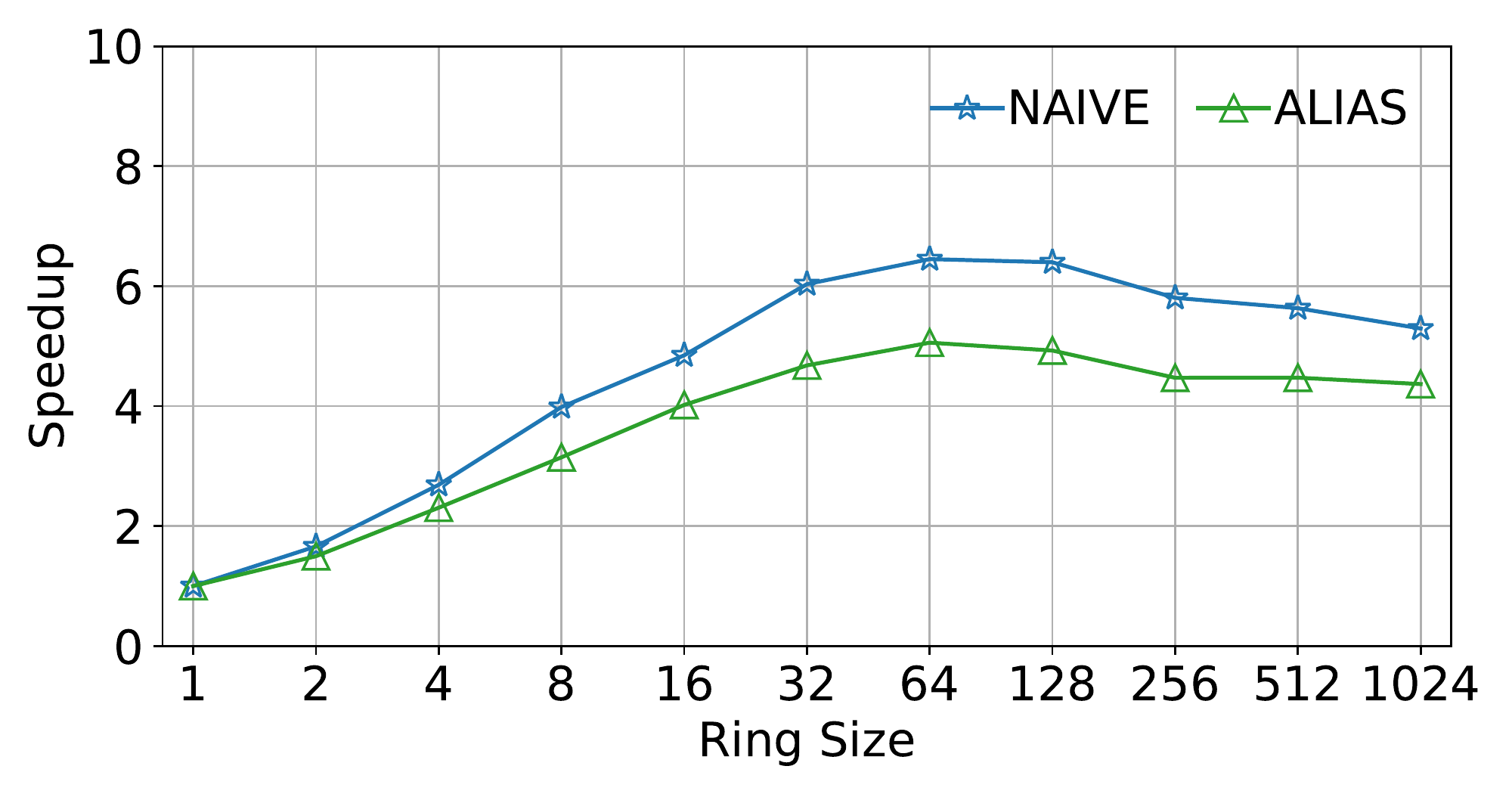}
        \caption{Task ring size ($k$).}
        \label{fig:vary_task_ring}
    \end{subfigure}
    \begin{subfigure}[t]{0.23\textwidth}
        \centering
        \includegraphics[scale=0.23]{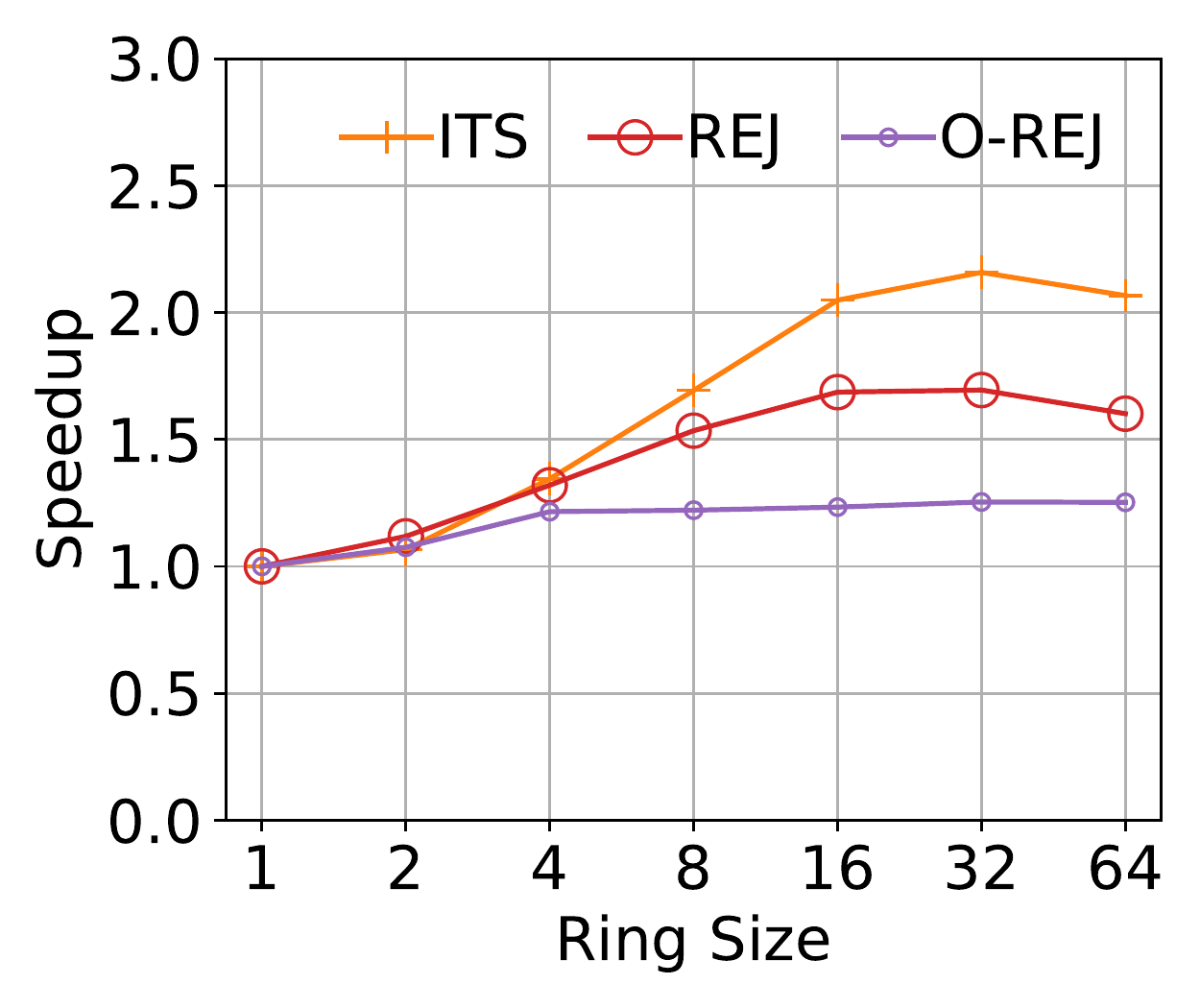}
        \caption{Search ring size ($k'$).}
        \label{fig:vary_search_ring}
    \end{subfigure}
    \caption{Speedup with ring size varying on \emph{lj}.}
    \label{fig:vary_ring_size}
\end{figure}

\textbf{Impact of ring sizes.} We evaluate the impact of ring sizes on
the performance. Based on our parameter tuning method, we first vary the
task ring size from 1 to 1024 on \texttt{NAIVE} and \texttt{ALIAS} to
pick the optimal value $k^*$, and then fix the task ring size to $k^*$ and
vary the search ring size from 1 to $k^*$ on \texttt{ITS}, \texttt{REJ} and \texttt{O-REJ}
to determine the search ring size. 
As shown in Figure \ref{fig:vary_task_ring},
the speedup first increases quickly with $k$ varying from 1 to 8 because
one core in our CPUs can support ten L1-D outstanding misses as it has ten
MSHRs. The optimal speedup is achieved when $k=64$ because we need to introduce
enough computation workload between the data request and the data usage to hide
memory access latency. Further increasing $k$ degrades the performance
as the L1-D cache size is limited and the request data can be evicted. 
Next, we fix the task ring size and vary the search ring size. When $k' = 32$, ThunderRW achieves the highest speedup.

\subsection{Prefetching Data to Different Cache Levels}

We use the intrinsic \texttt{\_mm\_prefetch(PTR, HINT)}
to prefetch the data. The intrinsic fetches the line of data from memory containing
address \texttt{PTR} to a location in the cache hierarchy specified by locality hint
\texttt{HINT} \cite{lee2012prefetching}. The intrinsic can load the data to L1, L2 or
L3 cache based on the hint. When fetching the data to L1 or L2,
it loads the data to the higher cache level as well. Moreover, we can specify the
data as non-temporal with \texttt{\_MM\_HINT\_NTA}. Then, the intrinsic will
load the data to L1 cache, mark it as non-temporal and bypass L2 and L3 caches. We set \texttt{HINT} to
\texttt{\_MM\_HINT\_T0} to fetch the data to L1 cache with respect to all level caches,
which has good performance based on our experiment.

We evaluate the effectiveness of prefetching the data to L1, L2, and L3 cache, respectively.
Table \ref{tab:cache_level} lists the experiment results on the \emph{livejournal} graph. The performance of fetching
data to L1/L2/L3 cache is close. In contrast, marking the data as non-temporal degrades the performance. This is because
the penalty of L3 cache miss is much more than that of L1/L2 cache misses and bypassing L3 cache results in
more L3 cache misses. Thus, ThunderRW uses the \texttt{\_MM\_HINT\_T0} cache locality hint to fetch
the data to L1 cache.

\begin{table}[h]
\setlength{\abovecaptionskip}{0pt}
\setlength{\belowcaptionskip}{0pt}
\footnotesize
\caption{Effectiveness of prefetching data to different cache levels (Speedup over loading data to L1 Cache).}
\label{tab:cache_level}
\begin{tabular}{c|c|c|c|c}
\hline
\textbf{Method} & \textbf{L1 Cache} & \textbf{L2 Cache} & \textbf{L3 Cache} & \textbf{Non-temporal Data} \\ \hline\hline
\texttt{NAVIE}  &1.00     &0.97     &0.95     &0.79    \\ \hline
\texttt{ITS}  &1.00     &1.01     &1.00     &0.95    \\ \hline
\texttt{ALIAS}  &1.00     &0.95     &0.95     &0.80    \\ \hline
\texttt{REJ}  &1.00     &1.00     &0.99     &0.92    \\ \hline
\texttt{O-REJ}  &1.00     &1.01     &1.01     &0.96    \\ \hline
\end{tabular}
\end{table}

\subsection{Pipeline Slot Breakdown and Memory Bandwidth of ThunderRW}

Tables \ref{tab:vary_length_thunderrw} and \ref{tab:vary_num_queries_thunderrw} present the pipleline
slot breakdown and memory bandwidth of ThunderRW with the length of queries and the number of queries varying,
respectively. Compared with results in Tables \ref{tab:vary_length} and \ref{tab:vary_num_queries},
ThunderRW dramatically reduces the memory bound, while significantly increases the retiring. Moreover,
the memory bandwidth utilization is improved. \sun{The memory bound for $10 ^ 2$ queries is higher than other settings
because each thread has only 10 queries, whereas the optimal task ring size $k$ is 64.}

\begin{table}[h]
\footnotesize
    \setlength{\abovecaptionskip}{0pt}
    \setlength{\belowcaptionskip}{0pt}
\caption{Pipeline slot breakdown and memory bandwidth of ThunderRW with the length of queries varying.}
\label{tab:vary_length_thunderrw}
\begin{tabular}{c|c|c|c|c|c|c}
\hline
\textbf{\begin{tabular}[c]{@{}c@{}}Length of\\ Queries\end{tabular}} & \textbf{\begin{tabular}[c]{@{}c@{}}Front\\ End\end{tabular}} & \textbf{\begin{tabular}[c]{@{}c@{}}Bad\\ Spec\end{tabular}} & \textbf{Core} & \textbf{Memory} & \textbf{Retiring} & \textbf{\begin{tabular}[c]{@{}c@{}}Memory\\ Bandwidth\end{tabular}} \\ \hline\hline
5               & 5.0\%                                                        & 10.8\%                                                       & 25.7\%        & 27.0\%          & 31.5\%            & 29.4GB/s                                                             \\ \hline
10              & 6.4\%                                                        & 10.3\%                                                       & 29.9\%        & 18.0\%          & 36.1\%            & 29.8GB/s                                                             \\ \hline
20              & 6.8\%                                                        & 10.6\%                                                       & 30.6\%        & 12.4\%          & 40.1\%            & 30.8GB/s                                                             \\ \hline
40              & 6.8\%                                                        & 10.7\%                                                       & 31.0\%        & 9.2\%          & 42.3\%            & 31.1GB/s                                                             \\ \hline
80              & 6.9\%                                                        & 10.8\%                                                       & 31.2\%        & 7.9\%          & 43.2\%            & 31.1GB/s                                                             \\ \hline
160             & 7.0\%                                                        & 10.8\%                                                       & 31.3\%        & 7.3\%            & 43.7\%            & 31.2GB/s                                                             \\ \hline
\end{tabular}
\vspace*{-10pt}
\end{table}

\begin{table}[h]
\footnotesize
    \setlength{\abovecaptionskip}{0pt}
    \setlength{\belowcaptionskip}{0pt}
\caption{Pipeline slot breakdown and memory bandwidth of ThunderRW with the number of queries varying.}
\label{tab:vary_num_queries_thunderrw}
\begin{tabular}{c|c|c|c|c|c|c}
\hline
\textbf{\begin{tabular}[c]{@{}c@{}}Num of\\ Queries\end{tabular}} & \textbf{\begin{tabular}[c]{@{}c@{}}Front\\ End\end{tabular}} & \textbf{\begin{tabular}[c]{@{}c@{}}Bad\\ Spec\end{tabular}} & \textbf{Core} & \textbf{Memory} & \textbf{Retiring} & \textbf{\begin{tabular}[c]{@{}c@{}}Memory\\ Bandwidth\end{tabular}} \\ \hline\hline
\sun{$10 ^ 2$}                          & \sun{5.3\%}         & \sun{6.5\%}         & \sun{28.1\%}             & \sun{27.3\%}             & \sun{32.8\%}     & \sun{26.1GB/s} \\ \hline
$10 ^ 3$                                                          & 6.3\%                                                        & 10.4\%                                                       & 30.7\%        & 9.8\%          & 42.8\%            & 30.1GB/s                                                             \\ \hline
$10 ^ 4$                                                          & 7.2\%                                                        & 11.1\%                                                       & 32.2\%        & 7.7\%          & 43.9\%            & 29.0GB/s                                                             \\ \hline
$10 ^ 5$                                                          & 6.9\%                                                        & 10.8\%                                                       & 31.1\%        & 7.9\%          & 43.2\%            & 31.5GB/s                                                             \\ \hline
$10 ^ 6$                                                          & 6.9\%                                                        & 10.8\%                                                       & 31.0\%        & 8.0\%          & 43.3\%            & 31.4GB/s                                                             \\ \hline
$10 ^ 7$                                                          & 6.9\%                                                        & 10.7\%                                                       & 31.4\%        & 8.2\%          & 42.8\%            & 31.1GB/s                                                             \\ \hline
$10 ^ 8$                                                          & 6.8\%                                                        & 10.7\%                                                       & 31.4\%        & 8.4\%          & 42.7\%            & 31.0GB/s                                                             \\ \hline
\end{tabular}
\end{table}

\subsection{Impact on Existing Systems}

\sun{In principle, the step interleaving technique is a generic optimization for RW
algorithms because it accelerates in-memory computation by hiding memory access latency in a single query via executing
a group of queries alternately, and RW algorithms generally consist of a number of random walks. However,
directly implementing it in the code base of GraphWalker and KnightKing is difficult because (1) their walker-centric model
regards each query as a task unit, which cannot support to execute steps of different queries alternately; and (2)
their source code does not consider the extensibility to support further enhancement. As such, we emulate the execution paradigm
of the two systems to study the impact of our optimization on their in-memory computation.}

\sun{Specifically, the in-memory computation of KnightKing adopts the BSP model, which executes random walks iteratively and moves one step
for all queries at each iteration. We implement this procedure, and integrate SI into it as follows: (1) divide queries into a
number of groups; (2) run queries in a group with the step interleaving; and (3) execute queries group by group at each iteration.
The implementation without/with the step interleaving is denoted by \emph{KK}/\emph{KK-si}.
The in-memory computation of GraphWalker adopts the ASP model, which assigns a query to each core and executes it independently.
We implement the procedure, and integrate the step interleaving into it as follows: assign a group of random walks to each core
and execute them with the step interleaving. The implementation without/with the step interleaving is denoted by \emph{GW}/\emph{GW-si}.}

\begin{figure}[t]\small
    \setlength{\abovecaptionskip}{0pt}
    \setlength{\belowcaptionskip}{-10pt}
    \captionsetup[subfigure]{aboveskip=0pt,belowskip=0pt}
    \centering
    \begin{subfigure}[t]{0.23\textwidth}
        \centering
        \includegraphics[scale=0.23]{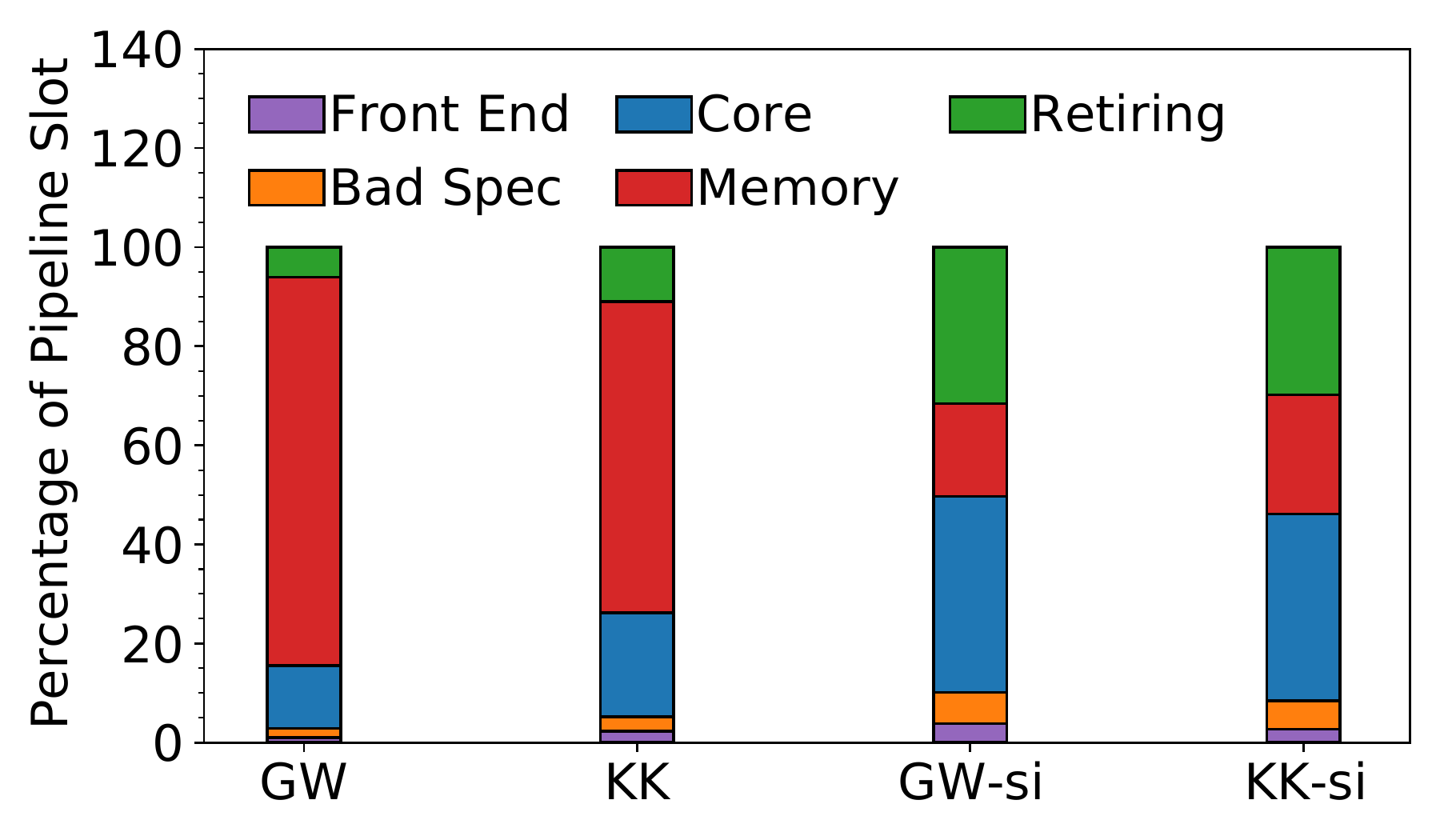}
        \caption{Pipeline slot breakdown.}
        \label{fig:impact_pipeline_slot}
    \end{subfigure}
    \begin{subfigure}[t]{0.23\textwidth}
        \centering
        \includegraphics[scale=0.23]{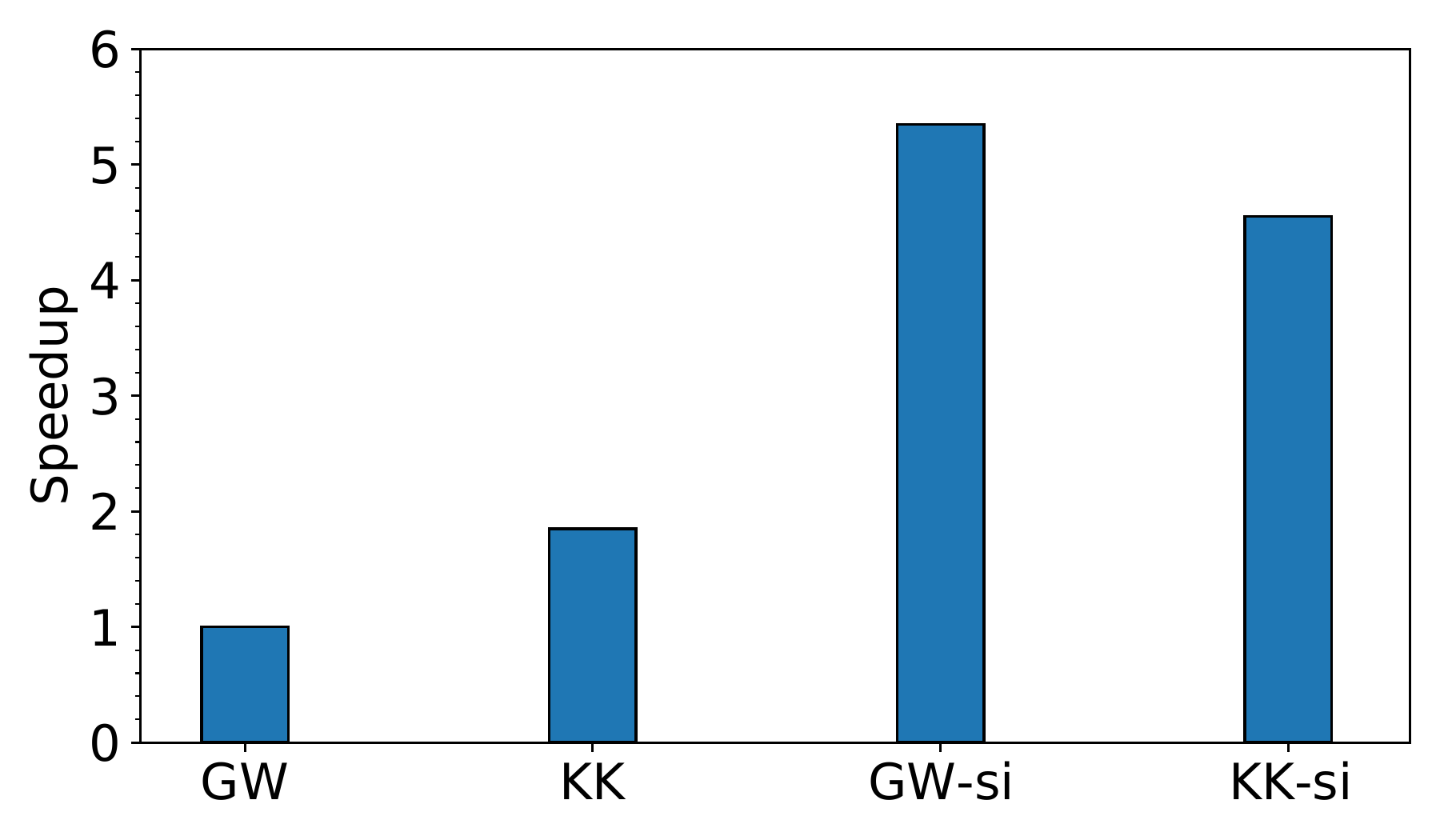}
        \caption{Speedup.}
        \label{fig:impact_speedup}
    \end{subfigure}
    \caption{\sun{Impact on in-memory computation of GraphWalker (GW) and KnightKing (KK).}}
    \label{fig:impact}
\end{figure}

\sun{Figure \ref{fig:impact} presents experiment results of DeepWalk on \emph{lj} with \texttt{ALIAS} sampling.
We set the group size as 64, which is the same as the optimal ring size. Enabling step interleaving
significantly reduces memory bound for both \emph{GW} and \emph{KK}, and improves the instruction retirement.
Figure \ref{fig:impact_speedup} shows the speedup over \emph{GW}. We find that \emph{KK}, which uses BSP, runs 1.8X faster than \emph{GW},
which utilizes ASP, because modern CPUs execute instructions out-of-order and steps of different queries at each iteration are independent of each other,
which benefits from this feature. After adopting the step interleaving, both \emph{GW} and \emph{KK} achieve a significant speedup. \emph{GW-si}
runs faster than \emph{KK-si} since \emph{KK-si} executes each query at one iteration and the context switch of each query incurs overhead.}

\subsection{Comparison with AMAC \cite{kocberber2015asynchronous}}

To compare with prefetching techniques designed for index lookups in database systems,
we implement the \texttt{Move} operation with the stage switch mechanism in AMAC \cite{kocberber2015asynchronous}. AMAC explicitly maintains states of all
stages in a SDG and performs the stage transition, which is similar to the method processing cycle stages.

Table~\ref{tab:detailed_metrics} presents instructions per step and cycles per step of wo/si, w/si and AMAC.
Enabling step interleaving leads to more instructions per step due to the overhead of prefetching
and stage transitions. The overhead on \texttt{NAIVE} and \texttt{ALIAS} is smaller than that on the
other three methods because all stages in \texttt{NAIVE} and \texttt{ALIAS} are non-cycle stages.
The benefit of hiding memory access latency offsets the overhead of executing
extra instructions. Therefore, the step interleaving technique significantly reduces
cycles per step. As \texttt{NAIVE} and \texttt{ALIAS} have only a few stages, instructions per step
of AMAC is close to that of w/si on the two methods. However, AMAC takes 1.57-2.03X more instructions
per step than w/si on \texttt{ITS}, \texttt{REJ} and \texttt{O-REJ}, which consist of several stages.
AMAC incurs more overhead because it explicitly maintains states of all stages in SDG and controls
the stage transition. In contrary, our stage switch mechanism processes cycle stages and non-cycle
stages with different methods, and controls the stage transition for cycle stages only.
Consequently, AMAC spends 1.18-1.64X more cycles per step than w/si. The results demonstrate the effectiveness
of our stage switch mechanism. 

\begin{table}[t]
\setlength{\abovecaptionskip}{0pt}
\setlength{\belowcaptionskip}{0pt}
    \captionsetup{aboveskip=0pt}
    \captionsetup{belowskip=0pt}
\footnotesize
    \caption{Detailed metrics with sampling method varying.}
    \label{tab:detailed_metrics}
\begin{tabular}{c|ccc|ccc}
\hline
   & \multicolumn{3}{c|}{\textbf{Instructions per Step}} & \multicolumn{3}{c}{\textbf{Cycles per Step}}\\ \hline
\textbf{Method}   &\textbf{wo/si} & \textbf{w/si}     & \textbf{AMAC}   &\textbf{wo/si}   & \textbf{w/si}               & \textbf{AMAC}      \\ \hline\hline
\texttt{NAIVE}  &131.24  &132.32  &137.42  &596.12  &111.26  &112.55 \\ \hline
\texttt{ITS}  &157.06  &335.75  &681.05  &1716.52 &327.65  &537.09 \\ \hline
\texttt{ALIAS}  &134.56  &139.17  &179.54  &740.73  &139.14  &140.26 \\ \hline
\texttt{REJ}  &187.87  &260.83  &464.78  &940.75  &273.44  &352.84   \\ \hline
\texttt{O-REJ}  &180.14  &264.56  &414.27  &1000.66 &333.21  &392.21 \\ \hline
\end{tabular}
\end{table}

\subsection{\sun{Future Extension}}

\sun{In case for extremely large graphs that cannot fit into the main memory of a single machine, we consider two approaches. First, we can develop
external memory graph systems to host the graph in the hard disk. With the recent advent of emerging storage such as Intel DCPMM persistent memory,
the I/O cost can be largely overlapped by in-memory processing (where ThunderRW can be leveraged and adopted for performance improvement).
Second, we plan to develop distributed systems such as KnightKing, where our ThunderRW can be leveraged as a single-node engine.
We leave the extension of ThunderRW in the future work.}

\end{document}